\documentclass[apj]{emulateapj}

\usepackage{longtable}

\def\cm{{\rm\thinspace cm}}

\def\erg{{\rm\thinspace erg}}

\def\km{{\rm\thinspace km}}

\def\Lsun{\hbox{$\rm\thinspace L_{\odot}$}}

\def\Mpc{{\rm\thinspace Mpc}}

\def\s{{\rm\thinspace s}}

\def\ergpcmsqps{\hbox{$\erg\cm^{-2}\s^{-1}\,$}}

\def\ergps{\hbox{$\erg\s^{-1}\,$}}

\def\kmpspMpc{\hbox{$\km\s^{-1}\Mpc^{-1}\,$}}

\def\pcmsq{\hbox{$\cm^{-2}\,$}}

\shorttitle{\sc{XDEEP2: the {\it Chandra} X-ray point-source catalog}}
\shortauthors{{\sc A.D.~Goulding et al.}}

\begin{document}

\title{The {\it Chandra} X-ray point-source catalog in the DEEP2 Galaxy
  Redshift Survey fields}


\author{A. D. Goulding\altaffilmark{1}, W. R. Forman\altaffilmark{1},
  R. C. Hickox\altaffilmark{1,2,3},  C. Jones\altaffilmark{1},
  R. Kraft\altaffilmark{1}, S. S. Murray\altaffilmark{1,4},
  A. Vikhlinin\altaffilmark{1}, \\ A.~L.~Coil\altaffilmark{5},
  M. C. Cooper\altaffilmark{6,\dag}, M. Davis\altaffilmark{7},
  J. A. Newman\altaffilmark{8}}
\email{E-mail:- agoulding@cfa.harvard.edu}

\altaffiltext{1}{Harvard-Smithsonian Center for Astrophysics, 60
  Garden St., Cambridge, MA 02138, USA}
\altaffiltext{2}{Department of Physics and Astronomy, Dartmouth
  College, Hanover, NH 03755, USA}
\altaffiltext{3}{Department of Physics, University of Durham, South Road, Durham DH1 3LE, UK}
\altaffiltext{4}{Department of Physics and Astronomy, Johns Hopkins
  University, 3400 North Charles Street, Baltimore, MD 21218, USA}
\altaffiltext{5}{Department of Physics, Center for Astrophysics and Space Sciences,
  University of California at San Diego, 9500 Gilman Dr., La Jolla,
  San Diego, CA 92093}
\altaffiltext{6}{Center for Galaxy Evolution, Department of Physics
  and Astronomy, University of California, Irvine, 4129 Frederick
  Reines Hall, Irvine, CA 92697, USA}
\altaffiltext{\dag}{Hubble Fellow}
\altaffiltext{7}{Department of Astronomy, University of California, Berkeley, Hearst
  Field Annex B, Berkeley, CA 94720, USA}
\altaffiltext{8}{Department of Physics and Astronomy, University of Pittsburgh, 3941
  O'Hara Street, Pittsburgh, PA 15260, USA}

\begin{abstract}
  We present the X-ray point-source catalog produced from the {\it
    Chandra} Advanced CCD Imaging Spectrometer (ACIS-I) observations
  of the combined $\sim3.2$~deg$^2$ DEEP2 (XDEEP2) survey fields,
  which consist of four $\sim 0.7$--1.1~deg$^2$ fields. The combined
  total exposures across all four XDEEP2 fields range from $\sim
  10$ks--1.1Ms. We detect X-ray point-sources in both the individual
  ACIS-I observations and the overlapping regions in the merged
  (stacked) images. We find a total of 2976 unique X-ray sources
  within the survey area with an expected false-source contamination
  of $\approx 30$ sources ($\lesssim 1$\%). We present the combined
  log N -- log S distribution of sources detected across the XDEEP2
  survey fields and find good agreement with the Extended {\it
    Chandra} Deep Field and {\it Chandra}-COSMOS fields to $f_{\rm
    X,0.5-2keV} \sim 2 \times 10^{-16} \ergpcmsqps$. Given the large
  survey area of XDEEP2, we additionally place relatively strong
  constraints on the log N -- log S distribution at high fluxes
  ($f_{\rm X,0.5-2keV} \sim 3 \times 10^{-14} \ergpcmsqps$), and find
  a small systematic offset (a factor $\sim 1.5$) towards lower source
  numbers in this regime, when compared to smaller area surveys. The
  number counts observed in XDEEP2 are in close agreement with those
  predicted by X-ray background synthesis models. Additionally, we
  present a Bayesian-style method for associating the X-ray sources
  with optical photometric counterparts in the DEEP2 catalog (complete
  to $R_{\rm AB} < 25.2$) and find that 2126 ($\approx 71.4 \pm
  2.8$\%) of the 2976 X-ray sources presented here have a secure
  optical counterpart with a $\lesssim 6$\% contamination fraction. We
  provide the DEEP2 optical source properties (e.g., magnitude,
  redshift) as part of the X-ray--optical counterpart catalog.
\end{abstract}

\keywords{galaxies: active -- surveys -- X-rays: galaxies}

\section{Introduction} \label{sec:intro}

Understanding the role of active galactic nuclei (AGN) in galaxy
evolution is a major focus in present day astrophysics.  It is now
becoming increasingly clear that, despite their vastly differing
size-scales, the evolution of massive host galaxies and the growth of
their central supermassive black holes (SMBHs) may not be independent
events (e.g.,
\citealt{boyle98,hopkins06,silverman09,Hopkins08,smolic09}). Indeed,
AGN activity and galaxy properties, such as luminosity, color and
morphology, are shown to evolve with time. The redshift range $z \sim
1$--2 is a crucial epoch: (1) galaxies are evolving strongly as a
function of stellar mass (e.g.,
\citealt{Zheng09,Franceschini99,Serjeant10}); (2) AGN activity is
prevalent (e.g.,
\citealt{ueda03,hasinger05,lafranca05,barger05,Richards06}); (3)
massive clusters are forming (e.g.,
\citealt{lidman08,hilton09,papovich10,fassbender11,bauer11,mehrtens12,nastasi11})
and (4) the red sequence is becoming established (e.g.,
\citealt{bell04,faber07,willmer06,brand05,dominguez11}). To
unambiguously determine the dominant physical processes that are
driving the growth and evolution of galaxies and their central black
holes requires sensitive, wide-field spectroscopic surveys of AGN.

Sensitive blank-field X-ray surveys arguably provide the most
efficient selection of AGN that is unbiased to moderate-to-high
obscuration, and in general is not readily contaminated by host-galaxy
emission. Indeed, as star-formation is relatively weak at X-ray
energies ($L_{\rm X,0.5-8keV} \lesssim 10^{42} \ergps$;
\citealt{moran99,lira02}), selection of AGN at these wavelengths can
identify many of the most low-luminosity and/or obscured systems
(e.g.,
\citealt{fukazawa01,done96,risaliti99,matt96,maiolino98,georgantopoulos09}). By
harnessing the unprecedented angular resolution provided by the {\it
  Chandra} X-ray Observatory, both deep and wide-field X-ray surveys
have been instrumental in our current understanding of AGN evolution
(e.g.,
\citealt{kenter05,nandra05,worsley05,brandt05,brand06,hasinger07,laird09}). To
date, the two deepest X-ray surveys are the pencil-beam ($\sim
0.1$~deg$^2$) $\sim 4$~Ms {\it Chandra} Deep Field South (CDF-S;
\citealt{giacconi02,Luo08,xue11}) and the $\sim 2$~Ms {\it Chandra}
Deep Field North (CDF-N; \citealt{dma03b}) which have successfully
identified AGN across more than 95\% of cosmic time (out to $z \sim
7$). Complementary to the highest redshift sources detected in the
deep fields, nearby ($z < 0.8$) AGN, identified in the relatively
shallow contiguous wide-field surveys, such as the 5~ks $\sim
9.3$~deg$^2$ XBootes field (\citealt{murray05,kenter05,brand06}), have
provided the ability to measure {\it environment}, a key component in
galaxy and AGN evolution (e.g.,
\citealt{cooper05,cooper06,georgakakis08,coil09,hickox09,cappelluti10,gilli09}). Furthermore,
these wide-field X-ray surveys serendipitously detect significant
numbers of rare, extremely luminous AGN and dozens of extended groups
and clusters, which allow for a more complete understanding of the
most massive SMBHs and cosmic structures in the Universe.

However, the peak of AGN activity, both in total luminosity and
relative abundance is believed to occur at $z \sim 1$--2 (e.g.,
\citealt{hopkins07,Zheng09,Serjeant10}). The $3.6$~deg$^2$ DEEP2
Galaxy Redshift Survey (\citealt{davis03,madgwick03}) provides one of
the most detailed censuses of the $z \sim 1$ Universe. DEEP2 is
currently one of the widest area and most complete spectroscopic
surveys of $z > 1$ galaxies, making it the ideal survey to target
large numbers of AGN at $z \sim 1$--2. Indeed, the fourth data release
(DR4) of the survey contains spectra for $\sim 50,300$ distant
galaxies (with $R_{\rm AB} < 24.1$) within four $\sim
0.7$--1.1~deg$^2$ fields, which are primarily in the redshift range $z
\sim 0.75$--1.4; these were collected using the DEIMOS spectrograph
($R \sim 5000$ in the wavelength range $6400 < \lambda < 9200$\AA) on
the Keck II telescope. A complete description of the DEEP2 DR4
spectroscopic catalog is available in \citet{newman12}.

We have used the {\it Chandra} Advanced CCD Imaging Spectrometer
(ACIS-I) to provide high-angular resolution X-ray coverage across
almost the entire $\sim 3.6$~deg$^2$ survey area covered by the four
DEEP2 fields (Field 1 - PI:K.Nandra; Fields 2, 3 and 4 -
PI:S.Murray). Here we present the X-ray source catalog for our {\it
  Chandra} ACIS-I observations of the combined $\sim 3.2$~deg$^2$
DEEP2 (XDEEP2) survey. The four contiguous XDEEP2 fields have combined
total exposures ranging from $\sim 10$ks -- 1Ms. In section 2 we
present a brief introduction to the construction of the survey fields
and the data reduction and processing of the X-ray observations. In
section 3 we provide an in-depth methodology for the detection of the
point sources and the building of the final XDEEP2 catalog. In section
4, we compare our new catalog of Field 1 (the Extended Groth strip),
which now includes three recent $\sim 600$~ks ACIS-I observations, to
the previous catalog of Laird et al. (2009), and compare the XDEEP2
catalog to the {\it Chandra} Source Catalog (\citealt{evans10}). We
further present the flux band ratio and density of number count
distributions for the XDEEP2 catalog. In section 5, we outline the
optical--X-ray source matching technique used to compare our new X-ray
catalog with the Fourth Data Release of the optical DEEP2 photometric
catalog. Finally, in section 6 we present a summary of our
findings. Throughout the manuscript we adopt a standard flat
$\Lambda$CDM cosmology with $H_0 = 71\kmpspMpc$ and $\Omega_M = 0.3$.

When combined, the redshift and galaxy property information
established using the DEEP2 optical spectra and the AGN identified
using the new {\it Chandra} X-ray observations provide one of the most
complete views of AGN activity and the growth of large scale structure
at $z \sim 1$--2. In forthcoming papers we will present a
statistically complete and obscuration-independent view of the
evolution of AGN and their host-galaxies identified across the entire
electromagnetic spectrum, in the epoch $z \sim 1.5$ to the
present-day.

\begin{table*}
\begin{center}
\caption{XDEEP2 Field Properties\label{tbl_field_props}}
\begin{tabular}{ccccccc}
\tableline\tableline
\multicolumn{1}{c}{\textbf{Field \#}\tablenotemark{a}} &
\multicolumn{1}{c}{\textbf{Pointings}\tablenotemark{b}} &
\multicolumn{1}{c}{$\alpha_{\rm center}$\tablenotemark{c}} &
\multicolumn{1}{c}{$\delta_{\rm center}$\tablenotemark{c}} &
\multicolumn{1}{c}{\textbf{Total Area}\tablenotemark{d}} &
\multicolumn{1}{c}{\textbf{Exp$_{\rm Eff,20\%}$}\tablenotemark{e}} &
\multicolumn{1}{c}{\textbf{Exp$_{\rm Eff,80\%}$}\tablenotemark{f}} \\
\multicolumn{1}{c}{} &
\multicolumn{1}{c}{} &
\multicolumn{1}{c}{(deg)} &
\multicolumn{1}{c}{(deg)} &
\multicolumn{1}{c}{(deg$^2$)} &
\multicolumn{1}{c}{(ks)} &
\multicolumn{1}{c}{(ks)} \\
\tableline
1 & 96 & 214.7388 & +52.7838 & 0.66 & 662.3 & 139.2 \\
2 & 12 & 252.4470 & +34.9300 & 0.74 & 15.8 & 8.1 \\
3 & 17 & 352.4711 & +0.1869 & 1.13 & 10.1 & 8.1 \\
4 & 12 & 37.2497  & +0.5916 & 0.75 & 15.9 & 8.2 \\
\tableline
\end{tabular}
\end{center}
\footnotesize
$^{a}$XDEEP2 field number \\
$^{b}$Number of {\it Chandra} pointings within field \\
$^{c}$Center co-ordinates of field in degrees as projected onto the
  sky in J2000 system \\
$^{d}$Total projected area of field in square degrees \\
$^{e}$Effective exposure in kilo-seconds at 20\% of total
  field area \\
$^{f}$Effective exposure in kilo-seconds at 80\% of total 
  field area
\end{table*}

\section{{\it Chandra} X-ray observations} \label{sec:obs}
\subsection{Construction of the XDEEP2 fields} \label{sec:deep2}

\begin{figure}[b]
\centering
\includegraphics[width=0.99\linewidth]{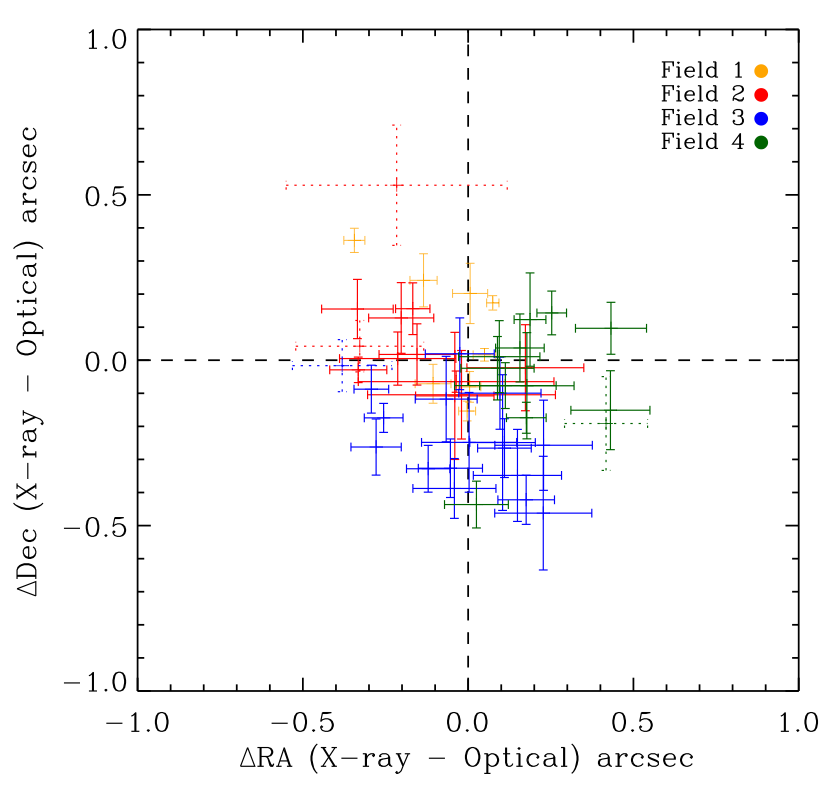}
\caption{Median offsets between optical and X-ray source positions
  with associated rms uncertainties in arc-seconds are plotted for the
  eight merged sub-fields in XDEEP2 Field 1 and the individual ObsIDs
  for Fields 2, 3 and 4. These median offsets were used to calculate
  astrometric corrections for the sub-fields in Field 1 and the
  individual X-ray observations in Fields 2--4. {\it Chandra} ObsIDs
  containing five or fewer X-ray--optical counterparts within 5
  arc-minutes of the observation aim-point are shown with dotted error
  bars.}
\label{fig:astrometry}
\end{figure}

The XDEEP2 survey region consists of four contiguous $\sim
0.7$--1.1~deg$^2$ fields covered by {\it Chandra} ACIS-I
observations. The field positions, arrangements and main properties
are outlined in Table 1. The total area covered by XDEEP2 is $\approx
3.2$~deg$^2$. X-ray catalogs for the previous 200~ks observations in
Field 1, also referred to as the Extended Groth Strip, have been
presented in \citet{nandra05} and \citet{laird09}. For consistency and
ease of reference to the previous Field 1 catalogs, we adopt the same
sub-field naming convention defined in \citet{laird09} (see column 2
of Table \ref{tbl_obslog}). Additionally, in this manuscript we
include the more recent 600~ks ACIS-I observations within three
sub-fields (EGS-3; EGS-4; EGS-5) of the Groth Strip centered at
$\alpha=215.0733^o$, $\delta= +53.008^o$; 214.808$^o$, +52.806$^o$;
214.527$^o$, +52.622$^o$, which for distinction between this and the
previous catalogs, we rename as AEGIS-1, AEGIS-2 and AEGIS-3,
respectively.

The catalog presented here was derived from multi-epoch observations
taken during AO3 (PI~K.~Nandra), AO6 and AO9, combined with Guaranteed
Time Observations (PI~S.~Murray; AO9). All {\it Chandra} observations
for XDEEP2 are publicly available through the {\it Chandra} X-ray
Center Archive. XDEEP2 consists of 126 separate pointings with varying
individual exposures ($\sim 3$--85~ks). With the exception of three
exposures, all XDEEP2 observations were performed in {\sc vfaint} mode
to allow for the best possible background rejection. ObsIDs 3305, 4537
and 4365 were taken in {\sc faint} mode.  In Table \ref{tbl_obslog} we
provide the individual pointing details for each observation and
field.

\begin{longtable*}{lcccrccrc}
\caption{Observation Log.\label{tbl_obslog}} \\
\hline\hline \multicolumn{1}{l}{\textbf{Field}$^a$} &
\multicolumn{1}{c}{\textbf{ObsID}$^{b,c}$} &
\multicolumn{1}{c}{\textbf{Sub-field}$^{d}$} &
\multicolumn{1}{c}{\textbf{Obs. Start}$^e$} &
\multicolumn{1}{c}{\textbf{Exp}$^f$} &
\multicolumn{1}{c}{$\alpha_{\rm J2000}$$^g$} &
\multicolumn{1}{c}{$\delta_{\rm J2000}$$^g$} &
\multicolumn{1}{c}{\textbf{Roll}$^h$} &
\multicolumn{1}{c}{\textbf{Mode}$^i$} \\
\multicolumn{1}{c}{} &
\multicolumn{1}{c}{} &
\multicolumn{1}{c}{} &
\multicolumn{1}{c}{(UT)} &
\multicolumn{1}{c}{(ks)} &
\multicolumn{1}{c}{(deg)} &
\multicolumn{1}{c}{(deg)} &
\multicolumn{1}{c}{(deg)} &
\multicolumn{1}{c}{} \\ \hline
\endfirsthead

\multicolumn{9}{c}%
{\tablename\ \thetable{} -- continued from previous page} \\
\hline\hline \multicolumn{1}{l}{\textbf{Field}} &
\multicolumn{1}{c}{\textbf{ObsID}} &
\multicolumn{1}{c}{\textbf{Sub-field}} &
\multicolumn{1}{c}{\textbf{Obs. Start (UT)}} &
\multicolumn{1}{c}{\textbf{Exp (ks)}} &
\multicolumn{1}{c}{$\alpha_{\rm J2000}$} &
\multicolumn{1}{c}{$\delta_{\rm J2000}$} &
\multicolumn{1}{c}{\textbf{Roll Angle}} &
\multicolumn{1}{c}{\textbf{Mode}} \\
\multicolumn{1}{c}{} &
\multicolumn{1}{c}{} &
\multicolumn{1}{c}{(UT)} &
\multicolumn{1}{c}{(ks)} &
\multicolumn{1}{c}{(deg)} &
\multicolumn{1}{c}{(deg)} &
\multicolumn{1}{c}{(deg)} &
\multicolumn{1}{c}{} \\ \hline
\endhead

\hline \multicolumn{9}{r}{{Continued on next page...}} \\
\endfoot

\hline\hline
\multicolumn{9}{r}{
\begin{minipage}[h!]{\textwidth}
{\footnotesize
{\sc Notes}-- \\
$^a$ XDEEP2 Field number. \\
$^b$ {\it Chandra} observation identification number. \\
$^c$ Due to missing gain files within the CALDB, ObsID 6221 is not
included in the analyses of Field 1 which are presented here. The
exposure time of 6221 is only 4.15~ks, hence, its rejection is
relatively insignificant compared to the total exposure time within
Field 1 and will have a negligible effect on our conclusions. \\
$^d$ Sub-field name for observations in Field 1, adopted from Laird et
al. (2009). \\
$^e$ Observing date and start time in UT. \\
$^f$ Exposure time in kiloseconds after appropriate screening. \\
$^g$ Aim point position of observation in degrees in J2000
coordinates. \\
$^h$ Spacecraft roll angle in degrees in standard north-east
co-ordinate system. \\
$^i$ {\it Chandra} observing mode.}
\end{minipage}
}
\endlastfoot

1 & 3305 & EGS-8 & 2002-08-11 21:43:57 & 29.40 & 214.42932 & 52.47367 & 84.74 & FAINT  \\
1 & 4357 & EGS-8 & 2002-08-12 22:32:00 & 84.36 & 214.42932 & 52.47367 & 84.74 & FAINT  \\
1 & 4365 & EGS-8 & 2002-08-21 10:56:53 & 83.75 & 214.42933 & 52.47367 & 84.74 & FAINT  \\
1 & 5841 & EGS-1 & 2005-03-14 00:04:09 & 44.45 & 215.67386 & 53.43149 & 229.11 & VFAINT \\
1 & 5842 & EGS-1 & 2005-03-16 15:54:34 & 46.42 & 215.67348 & 53.43141 & 226.09 & VFAINT \\
1 & 5843 & EGS-2 & 2005-03-19 17:13:09 & 44.46 & 215.38301 & 53.22857 & 222.38 & VFAINT \\
1 & 5844 & EGS-2 & 2005-03-21 22:37:40 & 45.85 & 215.38274 & 53.22848 & 219.86 & VFAINT \\
1 & 5845 & AEGIS-1 (EGS-3) & 2005-03-24 14:33:31 & 48.40 & 215.11277 & 53.03769 & 216.61 & VFAINT \\
1 & 5846 & AEGIS-1 (EGS-3) & 2005-03-27 04:51:15 & 49.40 & 215.11244 & 53.03755 & 213.67 & VFAINT \\
1 & 5847 & AEGIS-2 (EGS-4) & 2005-04-06 20:01:09 & 44.55 & 214.84408 & 52.84563 & 200.99 & VFAINT \\
1 & 5848 & AEGIS-2 (EGS-4) & 2005-04-07 21:03:59 & 44.45 & 214.84409 & 52.84563 & 200.99 & VFAINT \\
1 & 5849 & AEGIS-3 (EGS-5) & 2005-10-11 12:47:43 & 49.46 & 214.59049 & 52.64737 & 19.89 & VFAINT \\
1 & 5850 & AEGIS-3 (EGS-5) & 2005-10-14 05:15:37 & 45.55 & 214.59075 & 52.64755 & 16.94 & VFAINT \\
1 & 5851 & EGS-6 & 2005-10-15 03:03:18 & 35.68 & 214.10808 & 52.33121 & 14.80 & VFAINT \\
1 & 5852 & EGS-6 & 2005-12-03 13:00:33 & 10.62 & 214.10950 & 52.33506 & 324.80 & VFAINT \\
1 & 5853 & EGS-7 & 2005-10-16 20:16:24 & 42.57 & 213.84975 & 52.13788 & 14.04 & VFAINT \\
1 & 5854 & EGS-7 & 2005-09-30 23:52:23 & 50.07 & 213.84816 & 52.13692 & 31.03 & VFAINT \\
1 & 6210 & EGS-1 & 2005-10-03 14:56:50 & 45.94 & 215.68094 & 53.42341 & 29.70 & VFAINT \\
1 & 6211 & EGS-1 & 2005-10-12 11:43:28 & 35.64 & 215.68196 & 53.42394 & 19.80 & VFAINT \\
1 & 6212 & EGS-2 & 2005-10-04 22:56:06 & 46.28 & 215.39109 & 53.22076 & 28.00 & VFAINT \\
1 & 6213 & EGS-2 & 2005-10-06 06:52:13 & 47.51 & 215.39126 & 53.22084 & 26.47 & VFAINT \\
1 & 6214 & AEGIS-1 (EGS-3) & 2005-09-28 08:09:03 & 47.50 & 215.12071 & 53.02977 & 35.13 & VFAINT \\
1 & 6215 & AEGIS-1 (EGS-3) & 2005-09-29 15:58:09 & 48.63 & 215.12088 & 53.02982 & 33.67 & VFAINT \\
1 & 6216 & AEGIS-2 (EGS-4) & 2005-09-20 09:35:13 & 49.48 & 214.85259 & 52.83816 & 43.79 & VFAINT \\
1 & 6217 & AEGIS-2 (EGS-4) & 2005-09-23 01:34:59 & 49.50 & 214.85259 & 52.83816 & 43.79 & VFAINT \\
1 & 6218 & AEGIS-3 (EGS-5) & 2005-10-07 05:31:36 & 40.58 & 214.59005 & 52.64709 & 24.71 & VFAINT \\
1 & 6219 & AEGIS-3 (EGS-5) & 2005-09-25 15:57:04 & 49.48 & 214.58864 & 52.64648 & 37.70 & VFAINT \\
1 & 6220 & EGS-6 & 2005-09-13 09:17:01 & 37.63 & 214.10431 & 52.32963 & 49.79 & VFAINT \\
1 & 6222 & EGS-7 & 2005-08-28 17:20:24 & 34.69 & 213.84478 & 52.13607 & 59.04 & VFAINT \\
1 & 6223 & EGS-7 & 2005-08-31 05:06:47 & 49.51 & 213.84477 & 52.13608 & 59.04 & VFAINT \\
1 & 6366 & EGS-7 & 2005-09-03 06:30:11 & 14.58 & 213.84476 & 52.13605 & 59.04 & VFAINT \\
1 & 6391 & EGS-6 & 2005-09-16 20:43:01 & 8.45 & 214.10440 & 52.32956 & 49.79 & VFAINT \\
1 & 7169 & EGS-6 & 2005-12-06 02:29:46 & 16.03 & 214.10951 & 52.33509 & 324.25 & VFAINT \\
1 & 7180 & EGS-1 & 2005-10-13 05:16:04 & 20.43 & 215.68191 & 53.42394 & 19.80 & VFAINT \\
1 & 7181 & EGS-6 & 2005-10-15 21:17:21 & 15.98 & 214.10803 & 52.33122 & 14.80 & VFAINT \\
1 & 7187 & EGS-7 & 2005-10-17 19:07:08 & 6.59 & 213.84962 & 52.13785 & 14.04 & VFAINT \\
1 & 7188 & EGS-6 & 2005-12-05 04:50:40 & 2.58 & 214.10909 & 52.33499 & 324.80 & VFAINT \\
1 & 7236 & EGS-6 & 2005-11-30 19:29:34 & 20.37 & 214.10952 & 52.33505 & 324.80 & VFAINT \\
1 & 7237 & EGS-6 & 2005-12-04 05:26:20 & 16.93 & 214.10947 & 52.33504 & 324.80 & VFAINT \\
1 & 7238 & EGS-6 & 2005-12-03 10:02:10 & 9.53 & 214.10956 & 52.33502 & 324.80 & VFAINT \\
1 & 7239 & EGS-6 & 2005-12-11 08:31:06 & 16.03 & 214.10932 & 52.33545 & 319.59 & VFAINT \\
1 & 9450 & AEGIS-1 & 2007-12-11 04:24:07 & 28.78 & 215.07183 & 53.00951 & 319.80 & VFAINT \\
1 & 9451 & AEGIS-1 & 2007-12-16 10:52:06 & 25.21 & 215.07180 & 53.00951 & 319.80 & VFAINT \\
1 & 9452 & AEGIS-1 & 2007-12-18 05:45:49 & 13.29 & 215.07001 & 53.01006 & 311.30 & VFAINT \\
1 & 9453 & AEGIS-1 & 2008-06-15 21:28:03 & 44.69 & 215.05924 & 52.99529 & 130.79 & VFAINT \\
1 & 9454 & AEGIS-2 & 2008-09-11 04:47:10 & 59.35 & 214.81134 & 52.80632 & 49.30 & VFAINT \\
1 & 9455 & AEGIS-2 & 2008-09-13 19:38:46 & 99.72 & 214.81134 & 52.80633 & 49.30 & VFAINT \\
1 & 9456 & AEGIS-2 & 2008-09-24 08:15:30 & 58.35 & 214.81276 & 52.80818 & 34.80 & VFAINT \\
1 & 9457 & AEGIS-2 & 2008-06-27 07:08:38 & 32.74 & 214.79607 & 52.80288 & 124.29 & VFAINT \\
1 & 9458 & AEGIS-3 & 2009-03-18 12:20:16 & 6.65 & 214.52536 & 52.62140 & 223.14 & VFAINT \\
1 & 9459 & AEGIS-3 & 2008-09-30 19:20:28 & 69.55 & 214.55046 & 52.61607 & 30.30 & VFAINT \\
1 & 9460 & AEGIS-3 & 2008-10-10 06:17:49 & 21.36 & 214.55050 & 52.61613 & 29.80 & VFAINT \\
1 & 9461 & AEGIS-3 & 2009-06-26 09:30:12 & 23.73 & 214.53241 & 52.61042 & 129.79 & VFAINT \\
1 & 9720 & AEGIS-1 & 2008-06-17 05:14:02 & 27.79 & 215.05922 & 52.99527 & 130.79 & VFAINT \\
1 & 9721 & AEGIS-1 & 2008-06-12 08:09:14 & 16.55 & 215.05741 & 52.99587 & 139.79 & VFAINT \\
1 & 9722 & AEGIS-1 & 2008-06-13 07:02:28 & 19.89 & 215.05735 & 52.99589 & 139.79 & VFAINT \\
1 & 9723 & AEGIS-1 & 2008-06-18 13:42:40 & 34.47 & 215.05923 & 52.99528 & 130.79 & VFAINT \\
1 & 9724 & AEGIS-1 & 2007-12-22 13:37:26 & 14.08 & 215.07007 & 53.01006 & 311.30 & VFAINT \\
1 & 9725 & AEGIS-1 & 2008-03-31 05:21:42 & 31.13 & 215.05145 & 53.00445 & 209.78 & VFAINT \\
1 & 9726 & AEGIS-1 & 2008-06-05 08:45:04 & 39.62 & 215.05737 & 52.99587 & 139.79 & VFAINT \\
1 & 9727 & AEGIS-2 & 2008-09-12 16:44:12 & 34.94 & 214.81132 & 52.80634 & 49.30 & VFAINT \\
1 & 9729 & AEGIS-2 & 2008-07-09 16:47:58 & 48.09 & 214.79710 & 52.80272 & 119.79 & VFAINT \\
1 & 9730 & AEGIS-2 & 2008-09-25 16:50:54 & 53.72 & 214.81277 & 52.80817 & 34.80 & VFAINT \\
1 & 9731 & AEGIS-2 & 2008-07-03 10:58:47 & 21.38 & 214.79688 & 52.80275 & 120.79 & VFAINT \\
1 & 9733 & AEGIS-2 & 2008-09-27 01:15:33 & 58.36 & 214.81275 & 52.80818 & 34.80 & VFAINT \\
1 & 9734 & AEGIS-3 & 2008-09-16 11:01:21 & 49.47 & 214.54931 & 52.61415 & 44.80 & VFAINT \\
1 & 9735 & AEGIS-3 & 2008-09-19 03:14:15 & 49.47 & 214.54930 & 52.61415 & 44.80 & VFAINT \\
1 & 9736 & AEGIS-3 & 2008-09-20 11:07:10 & 49.48 & 214.54930 & 52.61416 & 44.80 & VFAINT \\
1 & 9737 & AEGIS-3 & 2008-09-21 17:53:00 & 49.48 & 214.54931 & 52.61415 & 44.80 & VFAINT \\
1 & 9738 & AEGIS-3 & 2008-10-02 06:56:22 & 61.39 & 214.55047 & 52.61607 & 30.30 & VFAINT \\
1 & 9739 & AEGIS-3 & 2008-10-05 11:28:12 & 42.59 & 214.55049 & 52.61614 & 29.80 & VFAINT \\
1 & 9740 & AEGIS-3 & 2009-03-09 22:24:18 & 20.37 & 214.52625 & 52.62221 & 229.78 & VFAINT \\
1 & 9793 & AEGIS-1 & 2007-12-19 02:53:51 & 23.83 & 215.07005 & 53.01008 & 311.30 & VFAINT \\
1 & 9794 & AEGIS-1 & 2007-12-20 04:27:59 & 10.03 & 215.07009 & 53.01004 & 311.30 & VFAINT \\
1 & 9795 & AEGIS-1 & 2007-12-20 21:36:20 & 8.91 & 215.07008 & 53.01009 & 311.30 & VFAINT \\
1 & 9796 & AEGIS-1 & 2007-12-21 20:28:33 & 16.33 & 215.07004 & 53.01008 & 311.30 & VFAINT \\
1 & 9797 & AEGIS-1 & 2007-12-23 13:12:28 & 12.60 & 215.07007 & 53.01011 & 311.30 & VFAINT \\
1 & 9842 & AEGIS-1 & 2008-04-02 21:01:59 & 30.44 & 215.05145 & 53.00445 & 209.78 & VFAINT \\
1 & 9843 & AEGIS-1 & 2008-04-02 01:11:09 & 13.48 & 215.05143 & 53.00448 & 209.78 & VFAINT \\
1 & 9844 & AEGIS-1 & 2008-04-05 13:07:54 & 19.78 & 215.05147 & 53.00443 & 209.78 & VFAINT \\
1 & 9863 & AEGIS-1 & 2008-06-07 00:33:47 & 22.01 & 215.05733 & 52.99587 & 139.79 & VFAINT \\
1 & 9866 & AEGIS-1 & 2008-06-03 22:43:14 & 25.83 & 215.05737 & 52.99588 & 139.79 & VFAINT \\
1 & 9870 & AEGIS-1 & 2008-06-10 15:11:23 & 11.00 & 215.05736 & 52.99583 & 139.79 & VFAINT \\
1 & 9873 & AEGIS-1 & 2008-06-11 14:22:06 & 30.75 & 215.05737 & 52.99588 & 139.79 & VFAINT \\
1 & 9875 & AEGIS-1 & 2008-06-23 22:54:14 & 25.20 & 215.05968 & 52.99517 & 128.77 & VFAINT \\
1 & 9878 & AEGIS-2 & 2008-06-28 06:03:20 & 15.73 & 214.79613 & 52.80289 & 124.29 & VFAINT \\
1 & 9879 & AEGIS-2 & 2008-06-29 03:39:20 & 26.80 & 214.79612 & 52.80288 & 124.29 & VFAINT \\
1 & 9880 & AEGIS-2 & 2008-07-05 17:00:17 & 29.50 & 214.79688 & 52.80274 & 120.79 & VFAINT \\
1 & 10769 & AEGIS-3 & 2009-03-20 13:38:26 & 26.68 & 214.52497 & 52.62063 & 216.98 & VFAINT \\
1 & 10847 & AEGIS-3 & 2008-12-31 05:06:27 & 19.27 & 214.54102 & 52.62566 & 302.79 & VFAINT \\
1 & 10848 & AEGIS-3 & 2009-01-01 17:11:57 & 17.91 & 214.54109 & 52.62567 & 302.79 & VFAINT \\
1 & 10849 & AEGIS-3 & 2009-01-02 21:25:57 & 15.92 & 214.54106 & 52.62570 & 302.79 & VFAINT \\
1 & 10876 & AEGIS-3 & 2009-03-11 01:37:20 & 17.21 & 214.52626 & 52.62222 & 229.78 & VFAINT \\
1 & 10877 & AEGIS-3 & 2009-03-12 15:15:57 & 16.22 & 214.52630 & 52.62223 & 229.78 & VFAINT \\
1 & 10896 & AEGIS-3 & 2009-06-15 18:46:14 & 23.29 & 214.53123 & 52.61075 & 135.32 & VFAINT \\
1 & 10923 & AEGIS-3 & 2009-06-22 07:38:22 & 11.62 & 214.53239 & 52.61039 & 129.79 & VFAINT \\
2 & 8631 & - & 2007-11-26 00:59:04 & 8.87 & 253.14712 & 35.06573 & 10.90 & VFAINT \\
2 & 8632 & - & 2007-11-26 03:52:42 & 8.60 & 252.85635 & 35.06034 & 10.90 & VFAINT \\
2 & 8633 & - & 2007-11-26 06:30:13 & 8.60 & 253.14626 & 34.84466 & 10.90 & VFAINT \\
2 & 8634 & - & 2007-11-26 09:07:44 & 8.60 & 252.57252 & 35.05619 & 10.90 & VFAINT \\
2 & 8635 & - & 2007-11-26 11:45:15 & 8.60 & 252.29343 & 35.04576 & 10.90 & VFAINT \\
2 & 8636 & - & 2007-11-26 14:22:46 & 8.60 & 252.00739 & 35.04026 & 10.90 & VFAINT \\
2 & 8637 & - & 2007-11-26 17:00:17 & 8.60 & 251.71914 & 35.03216 & 10.90 & VFAINT \\
2 & 8638 & - & 2007-11-26 19:37:58 & 8.60 & 252.86086 & 34.84118 & 10.90 & VFAINT \\
2 & 8639 & - & 2007-11-26 22:15:40 & 8.60 & 252.57546 & 34.83892 & 10.90 & VFAINT \\
2 & 8640 & - & 2007-11-27 00:53:12 & 8.61 & 252.29954 & 34.82097 & 10.90 & VFAINT \\
2 & 8641 & - & 2007-11-28 05:53:06 & 8.92 & 252.01425 & 34.81738 & 10.90 & VFAINT \\
2 & 8642 & - & 2007-11-28 08:41:27 & 8.66 & 251.72431 & 34.81683 & 10.90 & VFAINT \\
3 & 8601 & - & 2008-08-05 04:20:00 & 9.06 & 353.25281 & 0.24568 & 242.49 & VFAINT \\
3 & 8602 & - & 2008-08-05 07:12:25 & 8.93 & 352.66172 & 0.28185 & 242.49 & VFAINT \\
3 & 8603 & - & 2008-08-05 09:48:46 & 8.84 & 353.46809 & 0.20782 & 242.49 & VFAINT \\
3 & 8604 & - & 2008-08-05 12:24:02 & 8.84 & 351.64001 & 0.25772 & 242.49 & VFAINT \\
3 & 8605 & - & 2008-08-05 14:59:49 & 8.84 & 353.37170 & 0.01529 & 242.49 & VFAINT \\
3 & 8606 & - & 2008-08-05 17:35:02 & 8.83 & 352.97330 & 0.21938 & 242.49 & VFAINT \\
3 & 8607 & - & 2008-08-05 20:09:51 & 8.83 & 351.89874 & 0.25007 & 242.49 & VFAINT \\
3 & 8608 & - & 2008-08-05 22:44:39 & 8.84 & 351.72303 & 0.01783 & 242.49 & VFAINT \\
3 & 8609 & - & 2008-08-06 01:19:39 & 8.84 & 353.09031 & -0.01102 & 242.49 & VFAINT \\
3 & 8610 & - & 2008-08-06 03:54:42 & 8.84 & 352.15938 & 0.30473 & 242.49 & VFAINT \\
3 & 8611 & - & 2008-08-06 06:29:27 & 8.83 & 352.01198 & 0.02339 & 242.49 & VFAINT \\
3 & 8612 & - & 2008-08-06 09:04:08 & 8.84 & 351.47375 & 0.02736 & 242.49 & VFAINT \\
3 & 8613 & - & 2008-08-06 11:38:49 & 8.84 & 352.25372 & 0.06485 & 242.49 & VFAINT \\
3 & 8614 & - & 2008-08-06 14:13:30 & 8.84 & 352.80328 & 0.06085 & 242.49 & VFAINT \\
3 & 8615 & - & 2008-08-06 16:48:28 & 8.84 & 351.48138 & 0.26156 & 242.49 & VFAINT \\
3 & 8616 & - & 2008-08-06 19:23:35 & 8.83 & 352.54266 & 0.05529 & 242.49 & VFAINT \\
3 & 8617 & - & 2008-08-06 21:58:23 & 8.84 & 352.42188 & 0.29140 & 242.49 & VFAINT \\
4 & 8619 & - & 2007-11-28 13:18:37 & 9.04 & 36.63964 & 0.70242 & 50.76 & VFAINT \\
4 & 8620 & - & 2007-11-28 16:09:09 & 8.66 & 36.72970 & 0.48135 & 50.76 & VFAINT \\
4 & 8621 & - & 2007-11-29 01:03:58 & 9.07 & 36.88713 & 0.70250 & 50.76 & VFAINT \\
4 & 8622 & - & 2007-11-29 03:53:59 & 8.66 & 37.14407 & 0.70447 & 50.76 & VFAINT \\
4 & 8623 & - & 2007-11-29 06:32:28 & 8.66 & 37.39158 & 0.70452 & 50.76 & VFAINT \\
4 & 8624 & - & 2007-11-29 09:10:59 & 8.66 & 37.63909 & 0.70268 & 50.76 & VFAINT \\
4 & 8625 & - & 2007-11-29 11:49:28 & 8.66 & 37.89038 & 0.76129 & 50.76 & VFAINT \\
4 & 8626 & - & 2007-12-01 11:55:28 & 8.86 & 36.97907 & 0.48142 & 50.76 & VFAINT \\
4 & 8627 & - & 2007-12-01 14:50:45 & 8.46 & 37.22656 & 0.47393 & 50.76 & VFAINT \\
4 & 8628 & - & 2007-12-01 17:25:55 & 8.47 & 37.48350 & 0.47209 & 50.76 & VFAINT \\
4 & 8629 & - & 2007-12-01 20:01:05 & 8.47 & 37.72722 & 0.47402 & 50.76 & VFAINT \\
4 & 8630 & - & 2007-12-01 22:36:15 & 8.47 & 37.88970 & 0.59744 & 50.76 & VFAINT \\
\end{longtable*}

\subsection{Data reduction} \label{sec:datared}

Basic processing was carried out using the {\it Chandra} X-ray Center
(CXC) pipeline software. In addition, further processing of the X-ray
data was carried out using the {\sc chav} (v4.3)\footnote{{\sc chav}
  is available at
  http://hea-www.harvard.edu/{\textasciitilde}alexey/CHAV/} and {\sc
  ciao} (v4.3) \footnote{{\sc ciao} is available at
  http://cxc.harvard.edu/ciao/download/} software packages combined
with custom {\sc idl} scripts. Each ACIS-I observation was analyzed
separately. Individual ACIS-I pointings were reduced from the Level-1
event file products of the standard {\it Chandra} data pipeline. We
use the {\sc ciao} tool {\tt acis\_process\_events} to remove the
standard pixel randomization, and {\sc status}=0 was used to remove
streak events, bad pixels and cosmic ray afterglow features.

\begin{figure*}[ht]
\begin{center}
\includegraphics[width=0.95\textwidth]{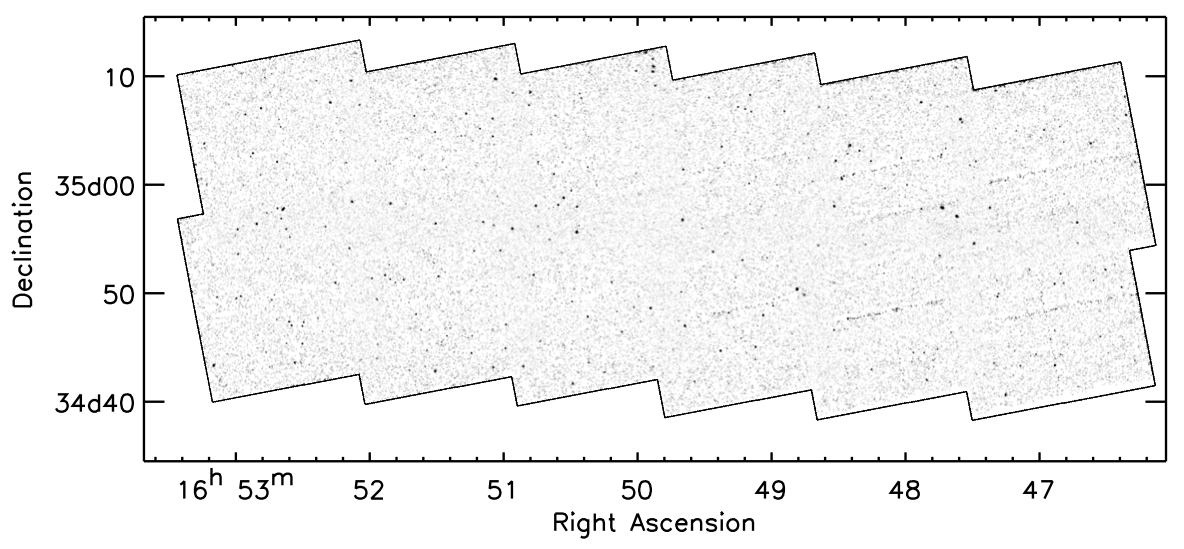}
\caption{Example of a full-band (0.5--7keV) merged raw counts image of
  an XDEEP2 field. Region shown is Field 2. Individual {\it Chandra}
  pointings have been merged using the {\sc ciao} tool {\tt
    merge\_all}. The image has been smoothed using a Gaussian kernel
  for presentation purposes only. Many sources are clearly evident
  throughout the image. Due to the presentation smoothing process,
  edge-effects (correlated streaks) can be seen along the positions of
  the chip gaps.}
\label{fig:cntsimage}
\end{center}
\end{figure*}

\begin{figure}[htb]
\centering
\includegraphics[width=0.95\linewidth]{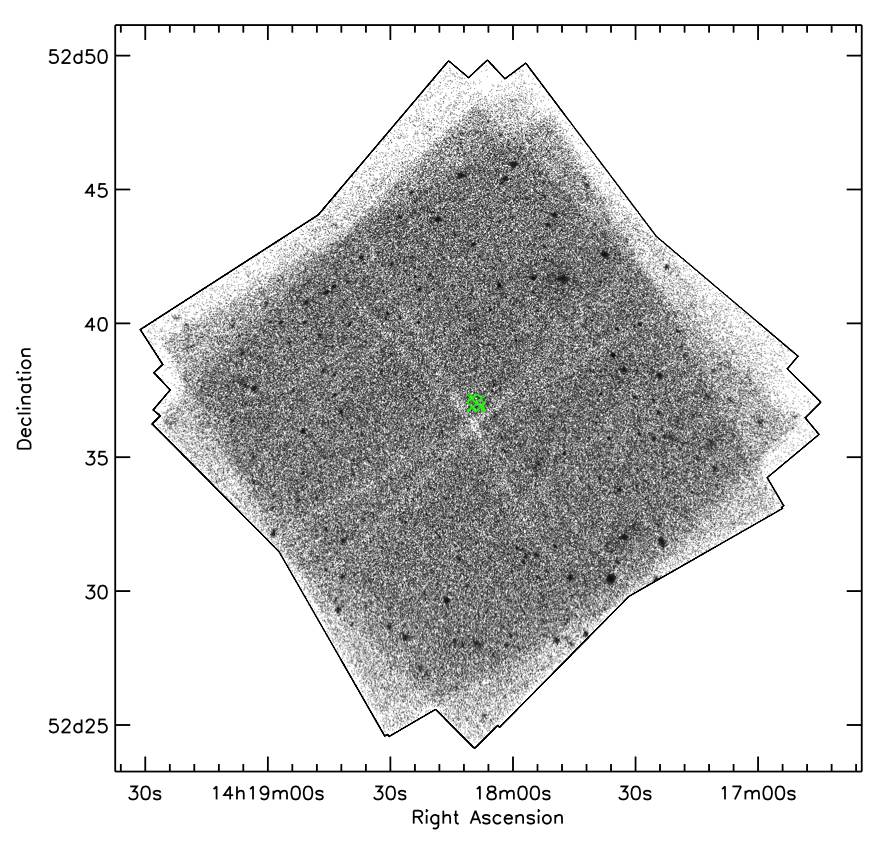}
\caption{Merged source count image of the sub-field AEGIS-3 located in
  Field~1. Aim points of the individual {\it Chandra} ObsIDs within
  the sub-field are shown with green crosses matched to the roll angle
  of the space craft. The angular separation of the aim-points is
  sufficiently small ($\sim 5$~arc-seconds) that they allow for the
  combining of the individual ObsIDs into stacked images.}
\label{fig:merge1}
\end{figure}

All observations were visually inspected for flaring and periods of
high background. The majority of the observations were found to not be
significantly contaminated. As also noted in \citet{nandra05} and
\citet{laird09}, observation 4365 does exhibit an interval ($\approx
25$~ks; $\sim 30$~\% of the observation) of elevated
background. However, unlike the previous analyses, here we
conservatively screen-out this period of high background. Final
effective exposures in good-time intervals for each observation were
generally found to be $> 90$~\% of the ``on-time'' (see
Table~\ref{tbl_obslog}).

\subsection{Creation of individual images \& exposure
  maps} \label{sec:imgexp}

\begin{figure*}[ht]
\centering
\includegraphics[width=0.95\textwidth]{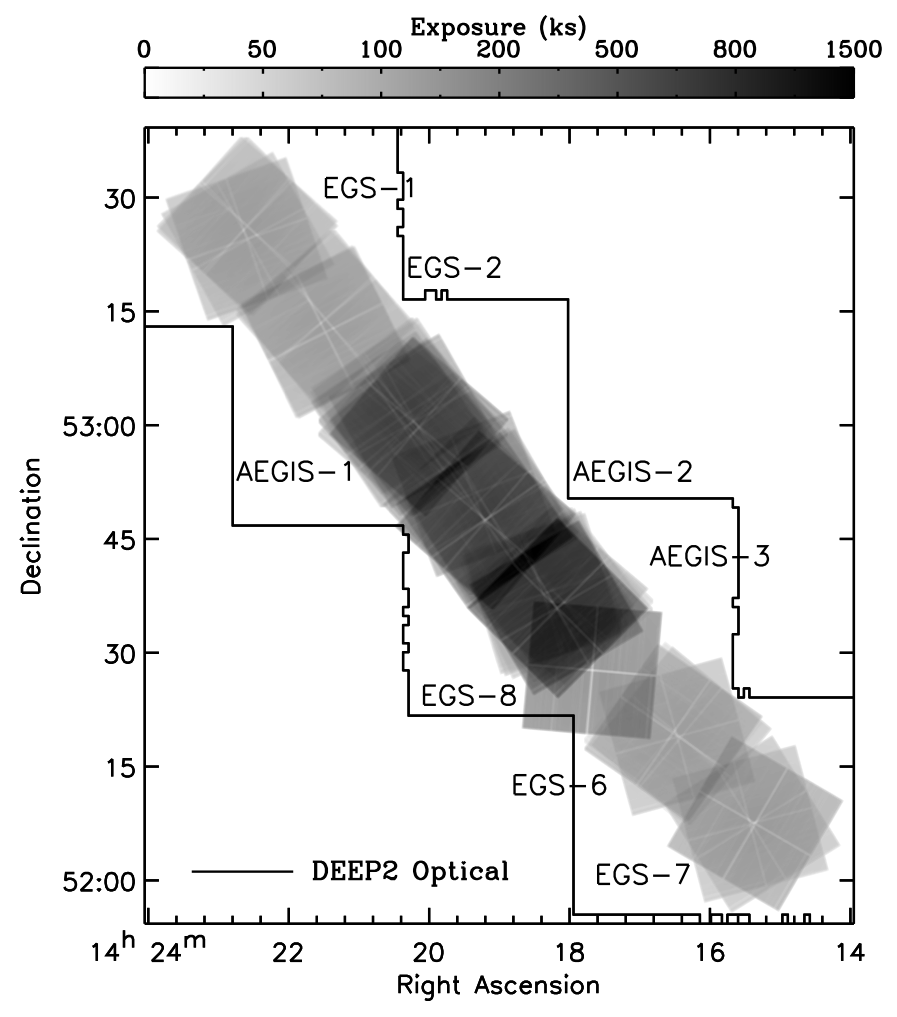}
\caption{Merged full-band (0.5-7 keV) {\it Chandra} ACIS-I exposure
  map for XDEEP2 Field 1. Reference spectra in monochromatic bands of
  $E \sim 1.0$, 4.0 and 2.5 keV and a spectral slope of $\Gamma = 1.7$
  were assumed in the creation of the exposure maps from the aspect
  histograms. The effective exposure (and hence sensitivity depth; see
  \S\ref{sec:sens}) across Field 1 is non-uniform and varies
  dramatically from $\approx 20$~ks--1.1 Ms due to the large number of
  overlapping observations. Overlaid is the nominal survey area
  covered by the DEEP2 Galaxy Redshift Survey optical observations
  (solid black line).}
\label{fig:expfld1}
\end{figure*}

\begin{figure*}[htb]
\begin{center}
\includegraphics[width=0.95\textwidth]{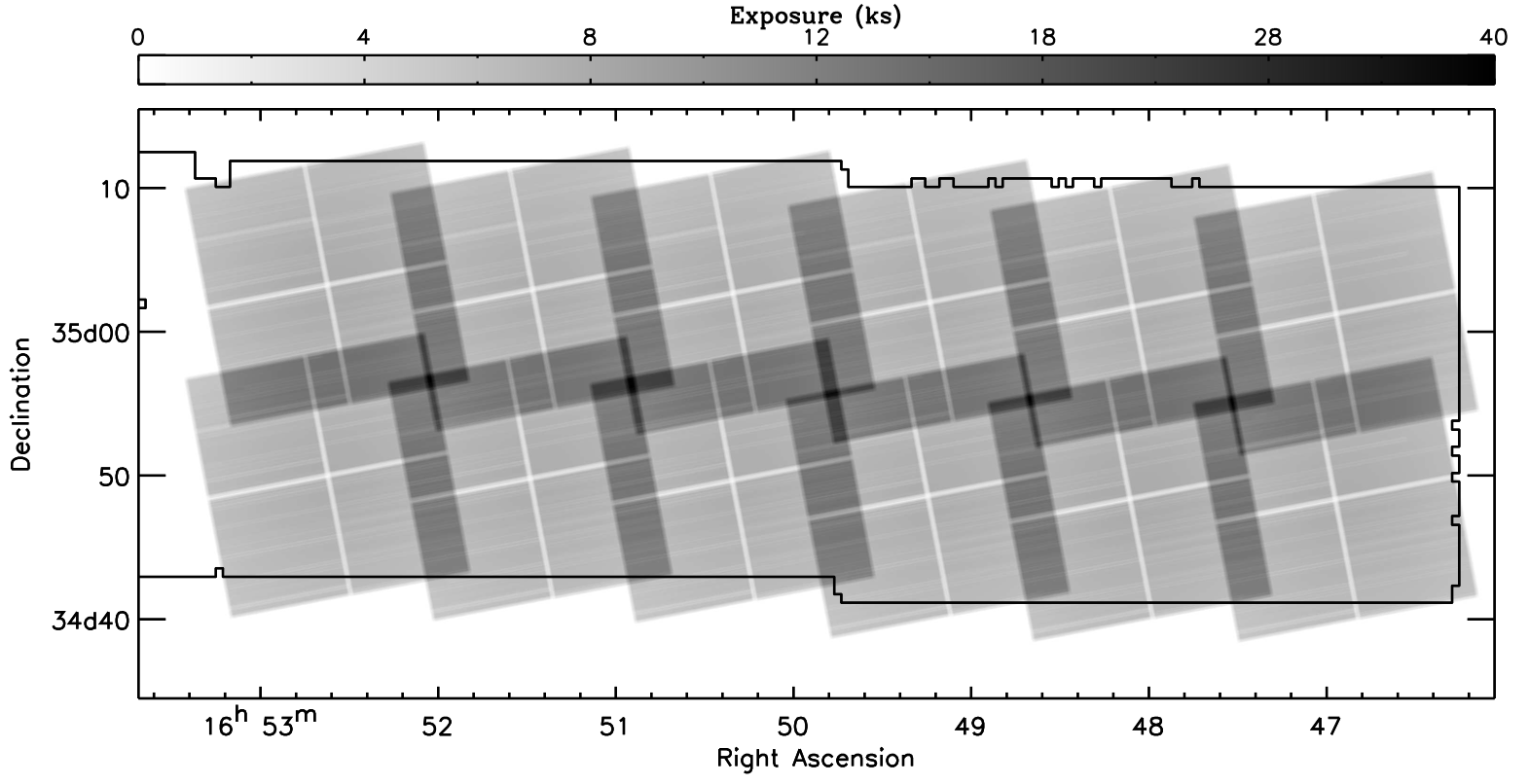}
\caption{Same as Fig.~\ref{fig:expfld1}, except field shown is XDEEP2 Field 2}
\label{fig:expfld2}
\end{center}
\end{figure*}

\begin{figure*}[htb]
\begin{center}
\includegraphics[width=0.95\textwidth]{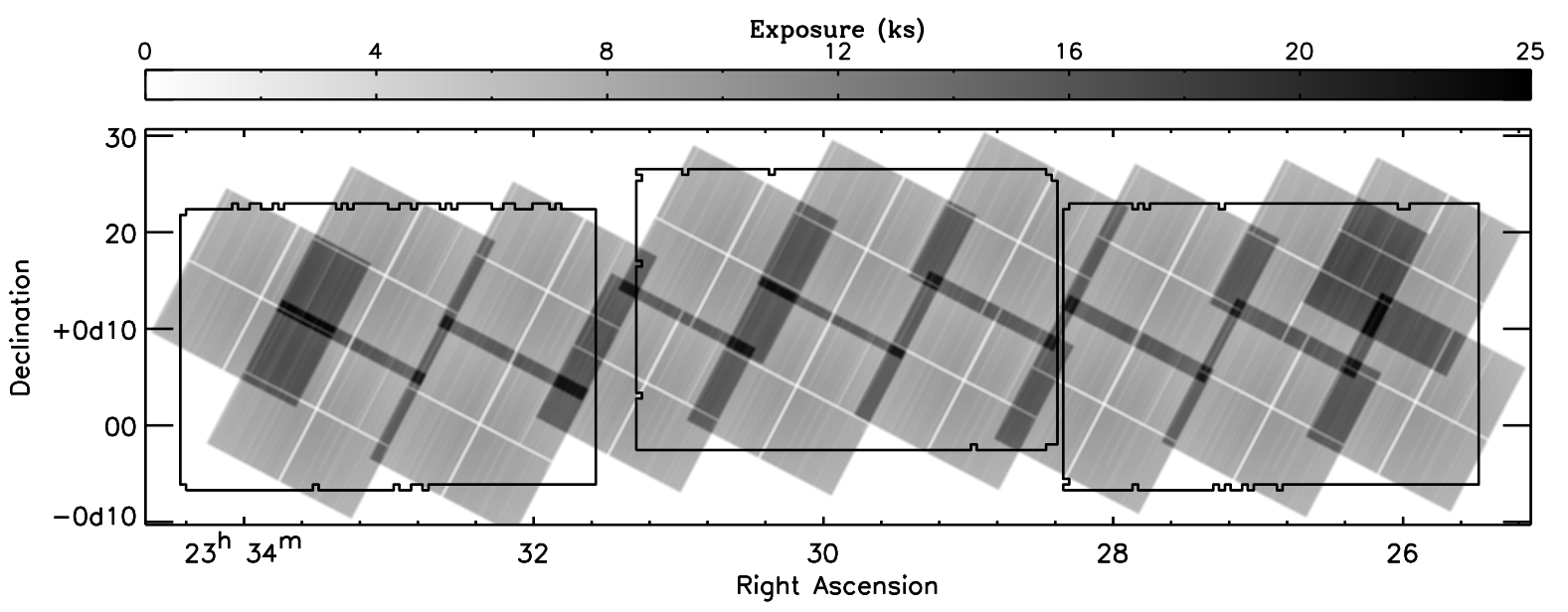}
\caption{Same as Fig.~\ref{fig:expfld1}, except field shown is XDEEP2 Field 3}
\label{fig:expfld3}
\end{center}
\end{figure*}

\begin{figure*}[htb]
\begin{center}
\includegraphics[width=0.95\textwidth]{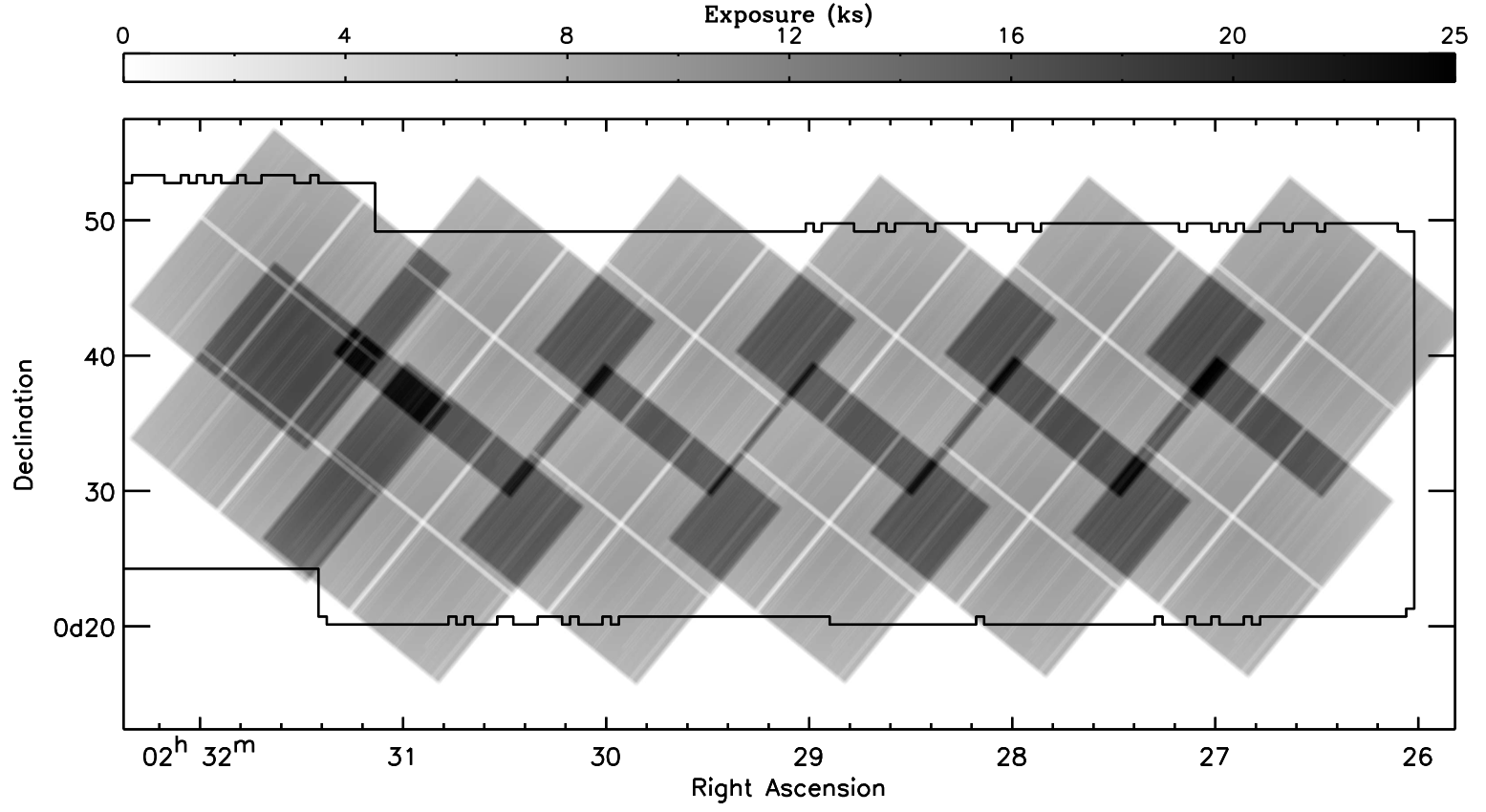}
\caption{Same as Fig.~\ref{fig:expfld1}, except field shown is XDEEP2 Field 4}
\label{fig:expfld4}
\end{center}
\end{figure*}

Events files were screened using a standard grade set ({\sc
  grade}=0,2,3,4,6) to construct images for each individual
ObsID. Images were constructed in the Full (FB; 0.5--7~keV), Soft (SB;
0.5--2~keV) and Hard (HB; 2--7~keV) bands at the full ACIS-I spatial
resolution, 0.492 arcsec/pixel. Here we limit the photon energy to
$E<7$~keV to allow a more direct comparison to sources detected in the
XBootes survey. Given the small effective area of the ACIS-I detector
at $E>7$~keV, relatively few $E>7$~keV photons are detected, and thus
this choice of energy boundary is somewhat arbitrary and will have
little effect on our conclusions. The {\sc chav} tool {\tt aspecthist}
was used to create aspect histograms in all three bands. These aspect
histograms were used to generate exposure maps by convolving them with
the standard ACIS-I chip-map ({\tt ccd\_id=0,1,2,3}) and reprojecting
to the previously created counts images. Reference spectra in
monochromatic bands of $E \sim 1.0$, 4.0 and 2.5~keV (i.e., the median
energies of the SB, HB and FB, respectively) were used in the creation
of the exposure maps.

\begin{figure*}[htb]
\includegraphics[width=0.95\textwidth]{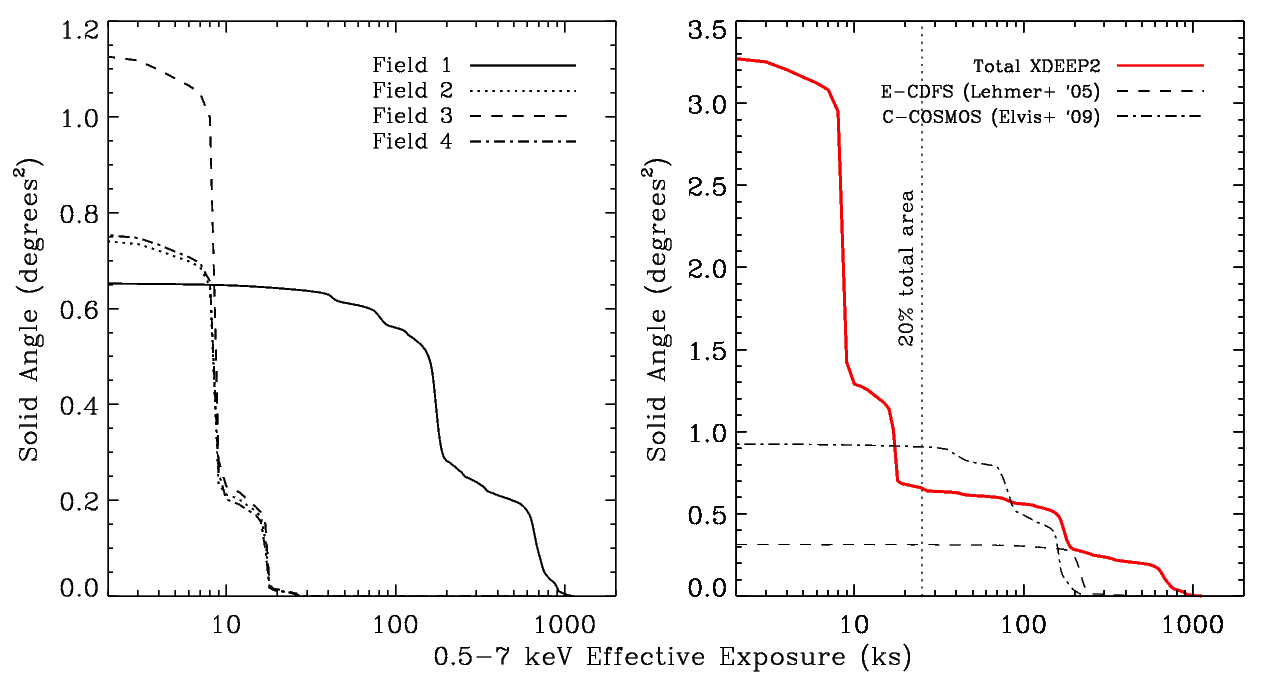}
\caption{{\bf a (left):} Cumulative survey solid angle (in degrees) as
  a function of effective exposure in the full-band (0.5--7keV) for
  the four separate XDEEP2 fields. {\bf b (right):} Total effective
  exposure across the combined XDEEP2 survey compared to the E-CDFS
  (Lehmer et al. 2005) and C-COSMOS (Elvis et al. 2009) survey
  fields. Minimum effective exposure for 20\% of the total XDEEP2 area
  is highlighted with a dotted line.}
\label{fig:expangle}
\end{figure*}

\subsection{Astrometric calibration \& observation merging} \label{sec:astrocalib}

Due to differing observing strategies and the sizes of exposure area
overlaps between individual {\it Chandra} observations within each
XDEEP2 field, X-ray observations were combined using separate methods
for Field 1 and Fields 2--4. As stated previously, Field 1 contains
eight sub-fields (see \S\ref{sec:obs}), with marginal overlap ($\sim
0.01$--0.02~deg$^2$) between one another. Each of these eight
sub-fields consists of several (3--28) individual {\it Chandra} ACIS-I
exposures with significant overlap between the observations within a
particular designated sub-field. We used the {\sc ciao} Perl script,
{\tt merge\_all} to create contiguous raw X-ray images and exposure
maps within each of the eight Field 1 sub-fields. Briefly, this script
searches for bright X-ray sources within two events tables which
spatially overlap and compares the astrometric co-ordinates of the
detected sources. By computing the average offset between the sources
within the tables, and guarding against rogue outliers, the events
table and associated aspect histograms are reprojected to the world
co-ordinate system (WCS) of the first reference observation within the
sub-field.

Given the limited area overlap (which occurs only at large off-axis
radii) between the eight sub-fields, a resultant merged events table
and images from a further use of {\tt merge\_all} to combine the
sub-fields, is likely to be highly uncertain. However, one of the
primary goals for this XDEEP2 X-ray catalog is to compare the X-ray
detected sources with the previously astrometrically-calibrated
optical DEEP2 catalog presented in Coil et~al. (2004). Hence, we may
consider the WCS astrometry of the DEEP2 optical catalog to be an
absolute reference frame. Thus, here we use the DEEP2 optical source
positions to correct the X-ray sub-fields for any systematic offsets
that may be present in the combined X-ray data. Following Brand
et~al. (2006), we use a counterpart-matching algorithm (described in
detail in \S~\ref{sec:XOPT} of this manuscript) to match X-ray sources
detected within 3 arc-minutes of the nominal observation aim-point to
optical counterparts. We calculated the median offset between the
X-ray and optical positions for the respective sources to identify any
necessary translation for the X-ray sub-fields. We present these
offsets and their associated rms uncertainties in
Figure~\ref{fig:astrometry}. Typically, $\sim 20$--30 X-ray--optical
sources were used to determine the necessary translations; the offsets
were generally found to be $< 0.25$ arc-seconds (i.e., $\sim 50$\% of
the ACIS-I pixel scale). While rotations were also allowed in the
calculation of the relative astrometries, the magnitude of the angular
rotation was always found to be negligible ($\ll 1$~degree) and
consistent with no rotation. Hence, we did not include angular
rotations and used only linear transformations for the final
corrections of the X-ray WCS to that of DEEP2. The required positional
offsets for the merged X-ray images were applied using the {\sc ciao}
tool {\tt wcsupdate}. The {\sc ciao} tool {\tt reproject\_aspect} was
used to reproject the events table and aspect solution files.

\begin{figure*}[ht]
\centering
\includegraphics[width=0.9\textwidth]{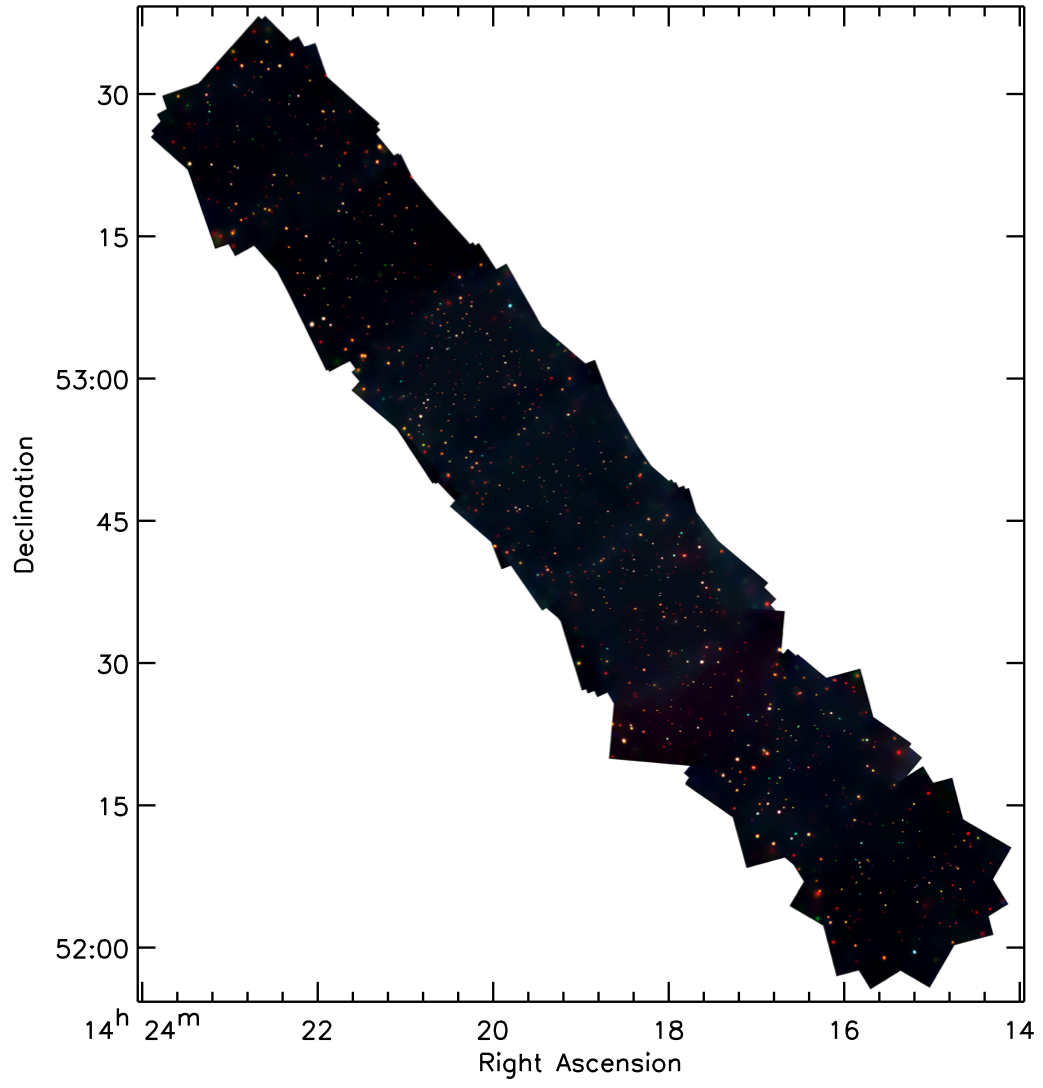}
\caption{{\it Chandra} false color image of XDEEP2 Field 1. The image
  is a merged composite of the exposure corrected 0.5--2 (red), 2--4
  (green) and 4--7~keV (blue) images within Field 1. The color-band
  images have been adaptively smoothed with varying smoothing scales
  determined from the average background counts in the stacked
  images.}
\label{fig:3color}
\end{figure*}

In Fields 2--4, the relatively shallow 9--10~ks X-ray observations
include little or no overlap area between exposures. As such, and
similar to the merged sub-fields in Field 1, {\tt merge\_all} cannot
be used to accurately co-align the relative astrometries within the
individual X-ray observations in these three fields. Hence, again we
consider the WCS reference frame of the optical DEEP2 catalog to be
absolute, and use the optical sources to align individual X-ray
observations following the same methodology described above. Given the
far shallower depth of the X-ray observations in Fields 2--4, we
include all X-ray sources with optical counterparts to a distance of
$<5$ arc-minutes from the aim-point. This larger off-axis distance
encompasses sufficient X-ray--optical source numbers (5--20 per
observation) to accurately constrain any required systematic
astrometric correction. Four of the {\it Chandra} ObsIDs (8637; 8614;
8604; 8628) included five or fewer X-ray--optical sources, and hence
we consider any astrometric corrections for these four observations to
be sufficiently uncertain that we subsequently include all detected
X-ray sources (at all off-axis distances within the observation) to
further constrain any median offset. The calculated median offsets and
associated uncertainties are also included in
Figure~\ref{fig:astrometry}. Clearly, using our adopted methodology,
we do not account for any possible field-to-field (or intra-field)
variations in the astrometric accuracy of the optical DEEP2
catalog. However, given the low number of X-ray sources within
individual {\it Chandra} observations, further investigation and/or
necessary correction to the DEEP2 catalog are beyond the scope of this
study. Overall, the required astrometric corrections (average
correction of 0.24'') for the whole of XDEEP2 are consistent, if not
slightly lower, than those found in previous wide-field X-ray surveys
(e.g., XBootes: 0.41''; Brand et al. 2006) and can be considered
sufficiently precise for our purposes.

\subsection{Merged XDEEP2 field maps}

In Figures \ref{fig:cntsimage} and \ref{fig:merge1}, we show examples
of the merged full-band (0.5--7~keV) counts images and in Figures
\ref{fig:expfld1}--\ref{fig:expfld4}, we present the merged full-band
exposure maps for the four survey fields. As shown in Figure
\ref{fig:expangle}a, the effective exposure (and hence sensitivity
depth; see \S~\ref{sec:sens}) across Field 1 is non-uniform and varies
dramatically from $\approx$20~ks--1.1~Ms. The effective exposure in
Field 1 is dependent on the number of repeat exposures, the large
number of overlapping regions and the varying space-craft roll-angles
between separate pointings. We show that at the 80th percentile, the
effective exposure in Field 1 is $\approx$140~ks. By contrast, the
effective exposures in Fields 2, 3 and 4 are relatively uniform ($\sim
9$~ks at 80\%) with constant spacecraft roll angle and only small
overlap regions between the individual ACIS-I pointings ($ <
20$\%). In Figure \ref{fig:expangle}b, we also show the effective
exposure time across the combined XDEEP2 area and compare this to the
{\it Chandra}-COSMOS (\citealt{elvis09}) and Extended-{\it Chandra}
Deep Field South fields (\citealt{lehmer05}).  It is clear that XDEEP2
complements these previous surveys: the survey depth of XDEEP2 extends
well beyond $\sim 200$~ks (the limiting effective exposure of the
E-CDF-S) to $> 600$~ks at similar survey area ($A \sim 0.2$~deg$^2$);
and XDEEP2 covers a survey area which is a factor $\approx 4$ greater
than that of {\it Chandra}-COSMOS.

The raw merged count images for each of the four XDEEP2 fields were
adaptively smoothed using custom {\sc idl} software based on the
kernel-smoothing program, {\sc asmooth} \citep{ebeling06}. Given the
wide range in exposure times across Field 1, we include a weighting
algorithm based on the average number of counts within binned
background images (see \S\ref{sec:sens}) to account for changes in
background count rate in overlapping regions. This background-weight
is applied to the calculation of the smoothing radii within our custom
version of {\sc asmooth}. The smoothing scales, which are calculated
from analysis of the merged counts images, are then applied directly
to the respective exposure maps. We use these to create false-color
exposure-corrected smoothed images in each field (see Figure
\ref{fig:3color}).

\begin{figure*}[htb]
\centering
\includegraphics[width=0.9\textwidth]{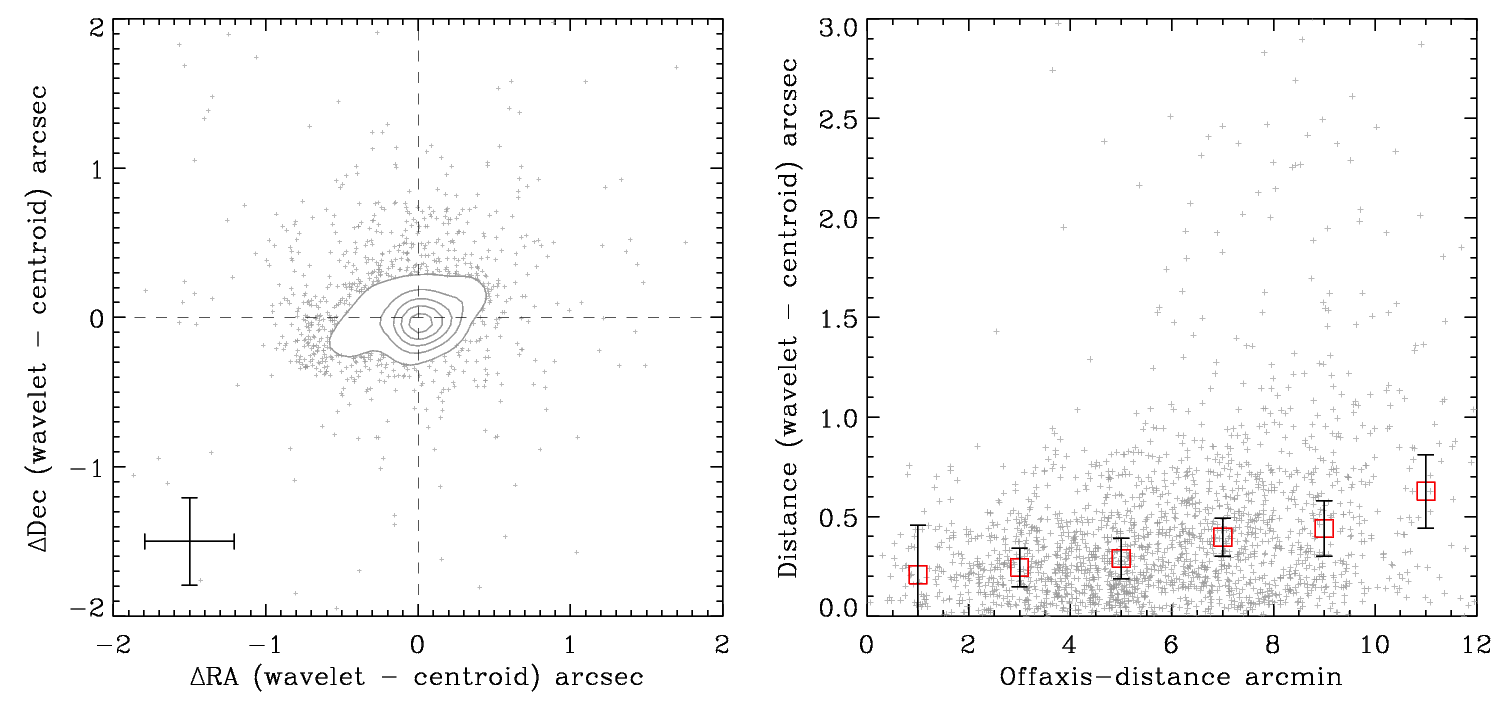}
\caption{{\bf a, left:} Positional offsets between the wavelet-centers
  produced from {\tt wvdecomp} and the events centroids. Contours
  encompass 50, 65, 80, 90 and 95\% of XDEEP2 sources. We find a small
  systematic offset of $<0.1$'' between the median source positions
  produced from the events centroid and wavelet center methods. Median
  90\% uncertainties of $\pm 0.3$'' were derived following Murray et
  al. (2005), see \S\ref{sec:src_extract}. {\bf b, right:} Angular
  separation (wavelet -- centroid) as a function of off-axis distance
  for XDEEP2. Median offset distances and the associated median RMS
  scatter are given in bins of 2' (open squares).}
\label{fig:wav_centroid}
\end{figure*}

\section{Point source detection \& spurious sources} \label{sec:srcdec}

In this section we outline the methods used to detect point-like
sources throughout the XDEEP2 fields. Following earlier analogous
methods for numerous wide-field and deep X-ray surveys, we used
wavelet decomposition software to detect sources across
XDEEP2. Indeed, previous analyses of Field 1 have used the {\sc ciao}
tool {\tt wavdetect} to detect X-ray source candidates. Here, we chose
to use {\tt wvdecomp} which is publicly available in the {\sc zhtools}
package (see \citealt{vikhlinin98}). In \S\ref{sec:laird} we perform a
comparison of the X-ray sources detected in \citet{laird09} which used
{\tt wavdetect} and additional signal-to-noise criteria to the sources
detected in this work using {\tt wvdecomp}. Briefly, we find little or
no difference between the number of sources detected in either
analyses. We find that $\sim 96$\% of the unique X-ray sources found
in the previous AEGIS-X catalog are included in our new catalog
(presented here) which now includes the more recent longer exposure
ACIS-I observations. We find that the majority of the sources which
are not included in our new catalog are relatively low significance
with few counts ($< 10$) and, in general, are detected in only one
energy-band in the Laird et al. catalog. Sources similar to these were
conservatively removed as possibly spurious detections in our new
catalog based on our extensive MARX simulations (see
\S\ref{sec:spursrc}).

\subsection{Point source detection in individual {\it Chandra} ObsIDs}

Point sources were detected in the individual (non-merged) counts
images for the SB, HB and FB energy ranges. We used a point source
detection threshold in {\tt wvdecomp} of 4.5$\sigma$ (equivalent to a
probability threshold of $1 \times 10^{-6}$). Point sources were
detected over wavelet scales of \{1, $\sqrt{2}$, 2, 4\} $\times
0.492''$. After detection of a source candidate, the event data at the
approximate wavelet position was iterated up to five times to
accurately determine the final events centroid, and hence, source
position. In Figure~\ref{fig:wav_centroid}, we present the offset
distances between the wavelet and event centroid positions. We find
that $> 90$\% of the X-ray sources have offset distances $\lesssim
0.3$'' from the wavelet position. Indeed, the vast majority of the
sources are consistent with zero offset. Furthermore, we find that the
median offset distance between the wavelet and centroid positions are
mildly correlated with the on-chip distance of the source from the
observation aim-point. Those sources at $d_{\rm OAX} < 5$' have
$\Delta({\rm wavelet} - {\rm centroid}) \sim 0.2$'', while those
sources closer to the edge of the FOV, at $d_{\rm OAX} \gtrsim 10$',
have $\Delta({\rm wavelet} - {\rm centroid}) \sim 0.65$''. These
increased offsets at large off-axis distances were most likely due to
asymmetries in the ACIS PSF shape.

Source lists, generated from the separate energy bands in the
individual observations, were cross-correlated based on their source
positions. Two-dimensional Gaussian profiles were used to represent
the sources detected in the separate energy bands with full-width half
maxima (FWHM) determined by the physical size of the 90\% encompassed
energy fraction (EEF) within the {\it Chandra} energy-band images with
the assumption of a spherically symmetric model for the ACIS-I
PSF. The centers of the Gaussian profiles were allowed to shift within
the 1$\sigma$ centroid error (see \S\ref{sec:src_extract}) of the
source positions to maximise the statistical likelihood of a source
match. A unique source was determined to exist when the summed 2-D
Gaussian profile was well-fit at the 90\% confidence level by a single
(approximately symmetrical) 2-D Gaussian profile with FWHM $< 90$\%
EEF.\footnote{We use the {\sc idl} routine {\tt mpfit2dpeak},
  available in the Markwardt software package, to fit the 2-D Gaussian
  profiles.} This methodology has the advantage that the
`matching-radius' naturally becomes a function of both the off-axis
position and the energy-band of the source detection. Hence, it
incorporates the size increase and rotation of the ACIS-I PSF radius
which, while assumed to be symmetrical about the aim-point, still
increases significantly for large off-axis distances and
simultaneously changes as a function of both azimuthal angle and
effective energy.

\subsection{Sources in overlapping observations in Field
  1} \label{sec:src_overlap}

As stated previously, sources were detected in each of the individual
ObsIDs. In Fields 2,3 and 4 there are small regions of significant
exposure ($> 10$~ks), where individual observations overlap. However,
given the large systematic uncertainties brought about by significant
differences in {\it Chandra} PSF radii, we did not attempt to combine
these observations to search for faint sources, which would be
detected in the merged deeper exposure regions. Instead, where
duplicate sources in these overlap regions appear (see previous
section), the source which is radially closest to the aim point in a
particular {\it Chandra} observation (i.e., the source which has the
smallest point spread function), is included as a unique source in the
final catalog. By contrast, given the large overlap between the {\it
  Chandra} observations in the sub-fields of Field 1, it is highly
likely that the same physical X-ray source is detected in multiple
individual exposures and that many fainter sources would be detected
in merged X-ray images. Hence, we have created merged events files of
the sub-field regions, which were defined in \citet{laird09} (see Table
1 of \citealt{laird09} and Table \ref{tbl_obslog} and Figure
\ref{fig:merge1} in this work).\footnote{We note that the analyses
  presented here now include the new 600ks observations in the
  sub-fields EGS-3, EGS-4 and EGS-5.}

When combined, the Field 1 `EGS' sub-fields show a significant
increase in the overall exposure and depth. Each of the observations
in these sub-fields have varying space-craft roll angles. However, as
shown in Table \ref{tbl_obslog}, the pointing co-ordinates are similar
($\lesssim 5$ arc-seconds; see also Figure \ref{fig:merge1}). As such,
these stacked sub-fields do not suffer from significant sensitivity
degradation due to large changes in the {\it Chandra} point spread
function (i.e., the observational setup was similar to that of the
CDF-N and CDF-S; e.g., \citealt{dma03a,xue11}). We used wavelet
decomposition to search for additional faint sources in these {\it
  merged} (stacked) sub-field images which would otherwise not be
detected in the individual observations. Candidate source lists for
Field 1, which were compiled from each of the individual ObsIDs and
those lists derived from the merged sub-field images were compared
using the same unique-source detection method outlined in the previous
section. The final unique source position and associated centroid
errors were determined by averaging and combining in quadrature the
previously calculated positions and uncertainties in the individual
and merged X-ray observations.

\subsection{Source extraction} \label{sec:src_extract}

Once the unique source locations were determined across each of the
XDEEP2 fields, we counted the number of events ($C_{\rm 50,SB/HB/FB}$,
$C_{90,SB/HB/FB}$) within the 50\% ($R_{50}$) and 90\% ($R_{90}$)
encircled energy fraction regions of the merged sub-field images
(Field 1) and the individual ACIS-I observations (Fields 2--4) for
each of the soft, hard and full bands.  Within the sub-fields of Field
1, the radii for circular extraction regions were calculated from the
off-axis radial distances in PSF simulations. We used the MARX
simulator to model a point-source, within a specific energy-band, at
varying roll angles and off-axis distances from an observation
aim-point. The modeled energy-band images were combined using the
method outlined in \S\ref{sec:astrocalib}, and the spatial extent of
the merged point-source was measured using a circular aperture to
determine accurate extraction radii for the candidate sources
identified in the Field 1 sub-fields.

\begin{figure*}[htb]
\centering
\includegraphics[width=0.97\textwidth]{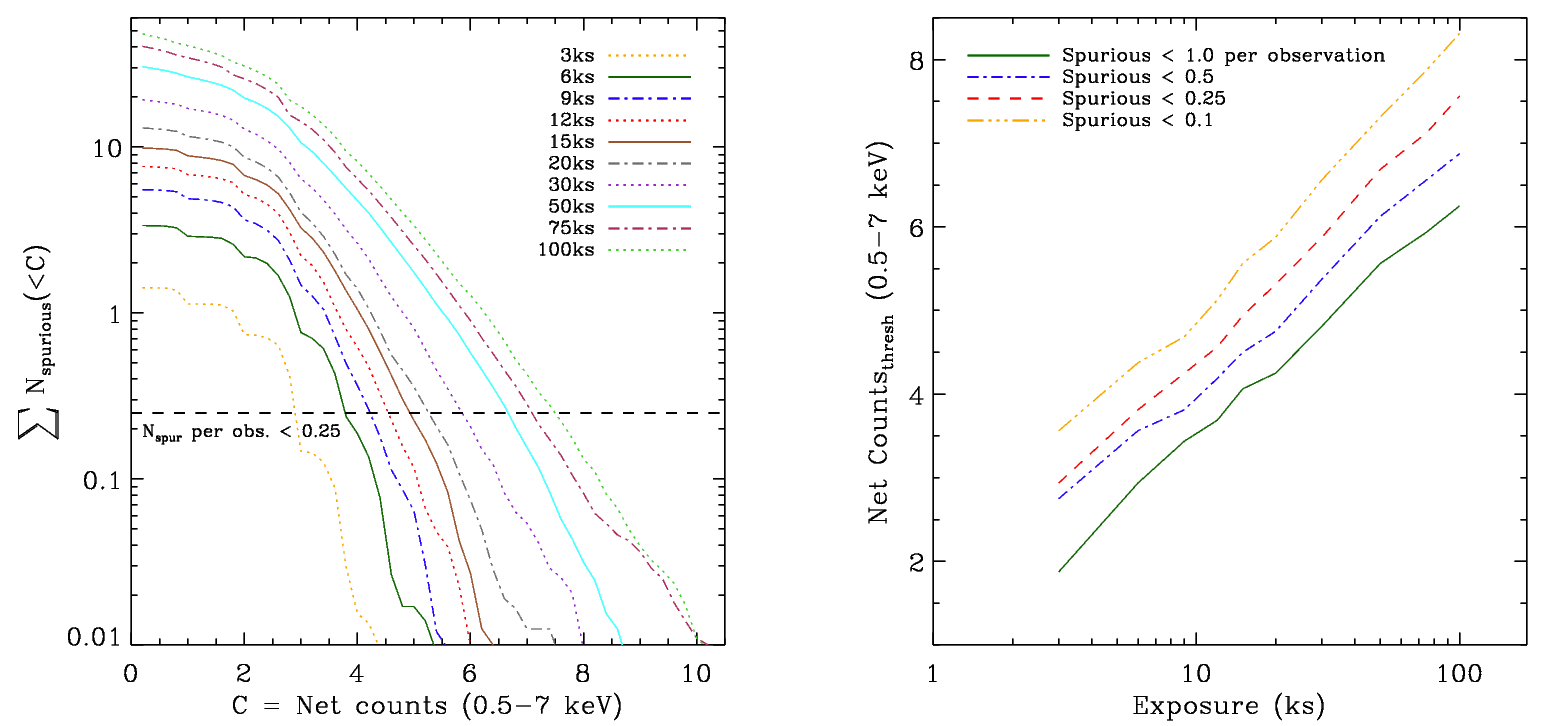}
\caption{(a; left) Average total number of spurious sources (N)
  detected in MARX simulated background images as a function of net
  photon counts in the full-band (C) measured in the spurious
  source. (b; right) Required net count threshold to ensure the
  average total number of spurious sources (derived from Monte-Carlo
  simulations) are $< 0.1$; 0.25; 0.5; 1.0 (dot-dot-dash line; dash
  line; dot-dash line; solid line, respectively) in a {\it Chandra}
  ACIS-I observation plotted as a function of exposure time in
  kiloseconds.}
\label{fig:marxsim1}
\end{figure*}

\begin{figure}[tb]
\centering
\includegraphics[width=0.9\linewidth]{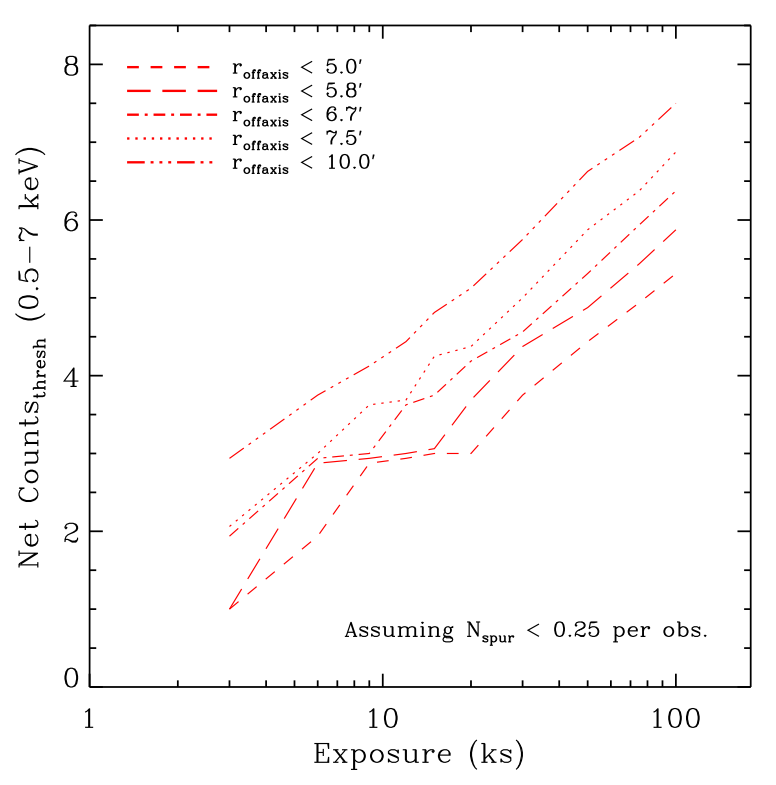}
\caption{Source count threshold cut in the 0.5--7~keV band as a
  function of exposure time in MARX simulated {\it Chandra} ACIS-I
  imaging. For a fixed total number of spurious sources of $N < 0.25$
  within simulated observations, we show the dependence of the
  threshold cut on the off-axis position of the detected spurious
  sources.}
\label{fig:marxsim2}
\end{figure}

We calculated average effective exposures for each candidate source in
the $R_{50}$ and $R_{90}$ extraction regions. Background counts were
determined for each source by extracting photon counts in annuli at
inner and outer radii $\{1.1,2.5\} \times R_{90}$, respectively, in
background images (see \S\ref{sec:sens}). Background counts were
scaled by the ratio of the areas of the EEF extraction region and the
background extraction region. Scaled background counts were subtracted
from the respective $C_{50}$ and $C_{90}$ to give final net source
counts ($C_{\rm 50, net}$; $C_{\rm 90, net}$, respectively). The 50\%
and 90\% encircled energy fraction regions were chosen to match those
used in the XBootes survey (Murray et al. 2005; Kenter et al. 2006;
Brand et al. 2006) allowing direct comparisons to be made between the
catalogs in future publications.

For a source detected in a particular energy band image, we computed
the total number of source counts in the other energy band images
using the analyses described above. We converted the net count rates
in each band (SB; HB; FB) to total fluxes ($F_{\rm SB}$; $F_{\rm HB}$;
$F_{\rm FB}$, respectively). To build a homogeneous X-ray catalog, we
assumed a single simple absorbed power-law spectrum with $\Gamma=1.7$
(i.e., the typical intrinsic slope of an AGN) for all sources and
$N_{\rm H} = \{1.24, 1.75, 3.99, 2.89\} \times 10^{20} \pcmsq$ for
those sources in Fields 1, 2, 3 and 4, respectively. Here, we use
PIMMS to calculate the count-rate--flux conversion factors assuming
the simulated ARFs from AO9 of the {\it Chandra} program. The use of
the AO9 ARFs compared to AO6 results in a $\sim 4$\% decrease in the
calculated 0.5--7~keV flux. Total galactic HI column densities were
determined using \citet{stark92}. Uncertainties on the counts and
fluxes were calculated using the formalism of \cite{gehrels86}.

Following Murray et al. (2005), we estimated the 90\% uncertainty on
the source locations as $X_{\rm err} = R_{50} / (C^{1/2}_{50}
-1)$. For those sources with $C_{50} \le 5$ counts, we set a minimum
centroid error of 0.8 arc-seconds (i.e., the 99\% positional accuracy
on the ACIS-I detector\footnote{see
  http:$/$$/$cxc.harvard.edu$/$cal$/$ASPECT$/$celmon$/$}), which takes
into account the systematic uncertainties associated with the
space-craft and detector astrometry. Random uncertainties also become
negligible for sources with large numbers of counts.

\subsection{Spurious sources} \label{sec:spursrc}

Given the widely varying exposure times, and hence varying background
levels of individual observations within XDEEP2, it is important to
apply further restrictions to the detected-source lists based on the
number of counts for a given source. For those observations with large
exposure times, the number of spurious sources with seemingly low
numbers of counts increases (see Figure \ref{fig:marxsim1}a). To limit
the number of spurious sources within our final catalog, we applied a
minimum photon count threshold of $n_{\rm counts} > n_{\rm cut}$,
where $n_{\rm cut}$ was determined through simulations of sourceless
background ACIS-I images. We used the MARX software package to
simulate 100,000 {\it Chandra} ACIS-I images of the unresolved Cosmic
X-ray background (XRB), including instrumental effects for exposure
times of 3, 6, 9, 12, 15, 20, 30, 50, 75 and 100ks. To approximate the
expected emission from the unresolved CXB, we employed a simple
absorbed power-law spectrum with $\Gamma=1.4$ (e.g.,
\citealt{hickox06}) and $f_X \sim 8.189 \times 10^{-13} \ergps$ in the
0.5--7~keV band; i.e., the XRB surface brightness measured in the
ROSAT all-sky survey in a blank-sky region of XDEEP2 Field 1, which
was then scaled to the projected area of ACIS-I. We note that this
simplification assumes the CXB emission is homogeneous across an
ACIS-I observation. We searched each of the simulated XRB ACIS-I
images for spurious sources using the same wavelet detection
thresholds defined above (see Figure \ref{fig:marxsim1}a). To build
source lists which were both relatively complete while limiting the
number of spurious sources, we cut the source-lists where the expected
total number of spurious sources $\Sigma n$ for a given exposure $i$
was $\langle \Sigma n_i \rangle < 0.25$ (see Figure
\ref{fig:marxsim1}b). By adopting a threshold of $\langle \Sigma n_i
\rangle < 0.25$, we expect a spurious source detection rate of $< 1$\%
in the final catalog.

As we show in Figure \ref{fig:marxsim2}, we find that the spurious
net count threshold is both a function of exposure time ($t_{\rm exp}$)
and off-axis position ($x_{\rm OAX[']}$) of the source within an ACIS-I
observation. This count threshold can be approximated by the empirical
formula,
\begin{equation}
n_{thresh} = - \frac{5}{3} + \frac{3}{10}x_{\rm OAX}  +  \frac{{ln}[60x_{\rm OAX}-30]}{2}{\rm log } t_{exp}
\end{equation}
and we use this to derive $n_{thresh}$ for a given fixed off-axis
position and exposure. To verify that this parametrization of the
count threshold can be extrapolated to larger exposure times (i.e.,
for the merged AEGIS-1, 2 and 3 sub-fields in XDEEP2 Field 1), we
simulated 100 1~Ms ACIS-I exposures using MARX. On average, we
detected $< 1$ spurious source in each 1~Ms simulation by using
$n_{thresh} > 20.3$. Hence, within the Poisson error, we detected the
same number of spurious sources expected when extrapolating the above
equation to $t_{exp} = 1$~Ms. By conservatively adopting a threshold
of $\langle \Sigma n_i \rangle < 0.25$ across the 126 XDEEP2 pointings
we expect $\lesssim 30$ spurious sources in the final XDEEP2 catalog.

\section{The XDEEP2 catalog} \label{numcnts}

\begin{figure}[tb]
\centering
\includegraphics[width=0.97\linewidth]{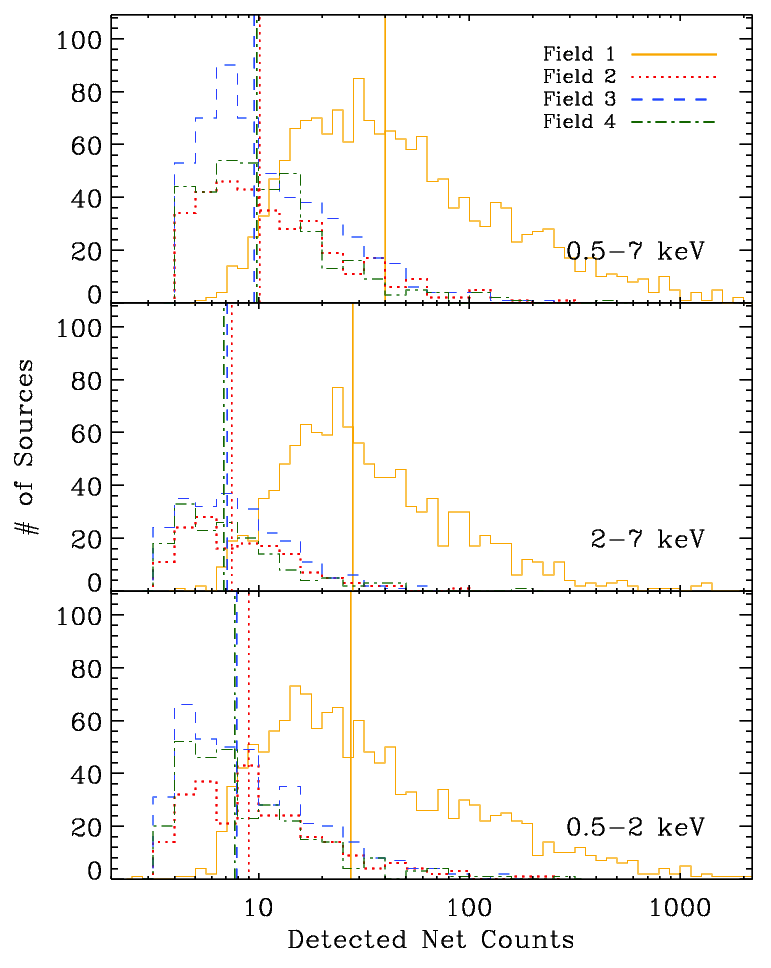}
\caption{Distribution of X-ray counts for sources detected in each of
  the four XDEEP2 fields in the full-band (0.5--7keV; top panel),
  hard-band (2--7keV; middle panel), and soft-band (0.5--2keV; bottom
  panel). Median source counts for each energy band in the associated field are
  shown with vertical lines.}
\label{fig:cntshisto}
\end{figure}

The XDEEP2 point source catalog contains 2976 unique sources, with
1720, 342, 528 and 386 sources in Fields 1, 2, 3 and 4,
respectively. For the purposes of our point source catalog, we do not
discuss those sources which are extended (e.g., the galaxy clusters)
as these will be the subject of a future publication. In Table
\ref{tbl_main_src} we show a short extract from the main source table,
which is available electronically. In Figure \ref{fig:cntshisto}, we
show the distribution of source counts across the four XDEEP2 counts
in the soft, hard and full bands. It is clear that both the
wide-spatial area of XDEEP2 combined with the smaller regions of
sensitive long-exposures, are extremely complementary to one
another. A significant cut-off is observed for sources with $C_{\rm
  90, SB} \lesssim 9$ in Field 1 since relatively few sources ($\sim
100$) are detected with 5--10 counts within $R_{90}$ due to the long
integrated exposures, even in the soft-band. However, many more
sources, down to $C_{\rm 90, FB} \sim 5$ are detected when the other
three XDEEP2 fields are included. Hence, within the point source
catalog we detect sources down to $\sim 4$ net counts in the SB, with
a completeness to 20 net counts in the HB and 15 counts in the
FB. Furthermore, we detect 70 rare bright sources with $C_{\rm 90, FB}
> 500$, which is due to the advantage of the wide-area across the
XDEEP2 survey.

\begin{table}
\begin{center}
\caption{Sources detected in separate energy bands \label{tbl_breakdown}}
\begin{tabular}{ccccc}
\tableline\tableline
\multicolumn{1}{c}{Detection band \tablenotemark{a}} &
\multicolumn{1}{c}{\#} &
\multicolumn{3}{c}{Non-detection band \tablenotemark{b}} \\
\multicolumn{1}{c}{(keV)} &
\multicolumn{1}{c}{} &
\multicolumn{1}{c}{Full} &
\multicolumn{1}{c}{Soft} &
\multicolumn{1}{c}{Hard} \\
\tableline
Full & 2849 & -   & 661 & 1196 \\
Soft & 2301 & 111 & -   & 1006 \\
Hard & 1663 & 12  & 372 & -    \\

\tableline
\end{tabular}
\end{center}
{\sc Notes}-- \\
$^{a}$Energy band which a source has been detected in \\
$^{b}$Number of sources where there is a non-detection in
  a particular energy-band when it has been detected in a different
  band. \\
\end{table}

In Table \ref{tbl_breakdown} we show the breakdown of the numbers of
sources detected and formally undetected in individual energy bands
within the main XDEEP2 source catalog. Those X-ray sources which are
not formally detected in a particular energy band are denoted by
``-1'' in the relevant net count and flux error columns of
Table~\ref{tbl_main_src} (e.g., {\tt NET COUNTS ERROR SB/HB/FB} and
{\tt FLUX ERROR SB/HB/FB}). For these `non-detections', we use the
formalism of \citet{gehrels86} to dervie $3 \sigma$ upper-limits from
the number of counts observed in the background images (see
\S~\ref{sec:sens}) within the source region. There upper-limits are
given in the appropriate {\tt NET COUNT} and {\tt FLUX} energy-band
columns.

\subsection{Background \& sensitivity analysis} \label{sec:sens}

As is clearly evident from the merged exposure maps, many of the
XDEEP2 ObsIDs spatially overlap with one another; however, a subset of
these observations, specifically in Field 1, were performed up to
seven years apart. Hence, care was taken to analyze changes between
the overlapping images as a result of the physical changes in the
detector and varying background levels. Background images were
constructed separately for each ObsID in the SB, HB and FB
energies. Source counts for candidates which were identified as being
significant in a particular energy-band using {\tt wvdecomp} were
masked. Background annuli, with inner radii $1.1 \times R_{90}$ and
outer radii $5 \times R_{90}$ centered at the source position, were
used to calculate the mean local background surrounding the candidate
source. The masked source region was re-populated with Poisson noise
with a mean distribution equal to that of the local background. The
same procedure was additionally used to create background maps of the
merged sub-fields in Field 1. While this procedure will remove the
count contributions from all point-sources, it will not remove
extended emission from sources such as clusters (e.g.,
\citealt{bauer02}). Hence, the background count levels derived from
this method are somewhat conservative, as they will be slightly
over-estimated.

\begin{table*}
\begin{center}
\caption{Background analysis of XDEEP2 fields\label{tbl_bkg}}
\begin{tabular}{ccccccc}

\tableline\tableline
\multicolumn{1}{c}{\textbf{Field \#}\tablenotemark{a}} &
\multicolumn{1}{c}{\textbf{Sub-field}\tablenotemark{b}} &
\multicolumn{1}{c}{\textbf{Energy band}\tablenotemark{c}} &

\multicolumn{1}{c}{Mean background \tablenotemark{d}} &
\multicolumn{1}{c}{Background $\sigma$ \tablenotemark{e}} &
\multicolumn{1}{c}{Total Background \tablenotemark{f}} \\
\multicolumn{1}{c}{} &
\multicolumn{1}{c}{} &
\multicolumn{1}{c}{} &
\multicolumn{1}{c}{(counts pixel$^{-1}$)} &
\multicolumn{1}{c}{(counts pixel$^{-1}$)} &
\multicolumn{1}{c}{($10^4$ counts)} \\
\tableline
1 &   AEGIS 1  &    Full   &      0.0841  &     0.2898  &      52.5 \\
1 &   AEGIS 1  &    Soft   &      0.0242  &     0.1539  &      15.1 \\
1 &   AEGIS 1  &    Hard   &      0.0599  &     0.2425  &      37.4 \\
1 &   AEGIS 2  &    Full   &      0.0842  &     0.2900  &      51.6 \\
1 &   AEGIS 2  &    Soft   &      0.0236  &     0.1524  &      14.5 \\
1 &   AEGIS 2  &    Hard   &      0.0605  &     0.2448  &      37.1 \\
1 &   AEGIS 3  &    Full   &      0.0991  &     0.3033  &      61.3 \\
1 &   AEGIS 3  &    Soft   &      0.0284  &     0.1609  &      17.6 \\
1 &   AEGIS 3  &    Hard   &      0.0706  &     0.2548  &      43.7 \\
1 &   EGS 1    &    Full   &      0.0243  &     0.1438  &      13.1 \\
1 &   EGS 1    &    Soft   &      0.0070  &     0.0773  &       3.8 \\
1 &   EGS 1    &    Hard   &      0.0165  &     0.1186  &       8.9 \\
1 &   EGS 2    &    Full   &      0.0235  &     0.1419  &      12.0 \\
1 &   EGS 2    &    Soft   &      0.0068  &     0.0765  &       3.5 \\
1 &   EGS 2    &    Hard   &      0.0159  &     0.1167  &       8.1 \\
1 &   EGS 6    &    Full   &      0.0271  &     0.1531  &      14.8 \\
1 &   EGS 6    &    Soft   &      0.0077  &     0.0815  &       4.2 \\
1 &   EGS 6    &    Hard   &      0.0184  &     0.1257  &      10.0 \\
1 &   EGS 7    &    Full   &      0.0241  &     0.1429  &      13.3 \\
1 &   EGS 7    &    Soft   &      0.0070  &     0.0769  &       3.9 \\
1 &   EGS 7    &    Hard   &      0.0163  &     0.1176  &       9.0 \\
1 &   EGS 8    &    Full   &      0.0332  &     0.1684  &      14.7 \\
1 &   EGS 8    &    Soft   &      0.0111  &     0.0980  &       4.9 \\
1 &   EGS 8    &    Hard   &      0.0202  &     0.1310  &       8.9 \\
& & & & & \\
\tableline
& & & & & \\
2 &   -        &    Full   &      0.0018  &     0.0428  &      11.1 \\
2 &   -        &    Soft   &      0.0005  &     0.0231  &       3.2 \\
2 &   -        &    Hard   &      0.0013  &     0.0361  &       7.9 \\
& & & & & \\
\tableline
& & & & & \\
3 &   -        &    Full   &      0.0018  &     0.0432  &       7.4 \\
3 &   -        &    Soft   &      0.0005  &     0.0234  &       2.1 \\
3 &   -        &    Hard   &      0.0013  &     0.0363  &       5.2 \\
& & & & & \\
\tableline
& & & & & \\
4 &   -        &    Full   &      0.0018  &     0.0431  &       7.6 \\
4 &   -        &    Soft   &      0.0005  &     0.0231  &       2.2 \\
4 &   -        &    Hard   &      0.0013  &     0.0364  &       5.4 \\
& & & & & \\
\tableline
\end{tabular}
\end{center}
\footnotesize
{\sc Notes}-- \\
$^{a}$XDEEP2 field number \\
$^{b}$XDEEP2 sub-field name\\
$^{c}$X-ray energy band of background image: Full 0.5--7keV; Soft 0.5--2keV; Hard 2--7keV\\
$^{d}$Mean number of background counts per pixel within the non-zero exposure area of the merged images.\\
$^{e}$Standard deviation of the background counts within the merged images.\\
$^{f}$Total number of background counts within the merged images.\\
\end{table*}
\normalsize

The mean background counts, their associated standard deviation and
total number of background counts for each field (and sub-field) are
shown in Table \ref{tbl_bkg}. As expected, the average background
counts are a factor of $\approx 15$--50 greater in Field 1 than those
in Fields 2--4, owing to the much longer exposure times in Field 1. We
find that the average backgrounds appear to be relatively stable
across the deep sub-fields AEGIS-1 and AEGIS-2, with a slightly higher
($\approx 15$\%) average background count in AEGIS-3. However, we note
that the observations in AEGIS-3 occurred 6--12 months after those
observations in AEGIS-1 and AEGIS-2. The background levels in XDEEP2
Fields 2, 3 and 4 are almost identical for each of the three
energy-bands.

For the purposes of comparing the X-ray point sources detected within
each of the XDEEP2 fields, as well as comparing with previous X-ray
surveys, it is important to understand the flux sensitivity
limitations of a particular X-ray field. The faintest sources detected
in the XDEEP2 fields have $f_{\rm X,SB} \sim 3.1 \times 10^{-17}
\ergpcmsqps$ and $f_{\rm X,HB} \sim 1.2 \times 10^{-16} \ergpcmsqps$.
While these fluxes are good indicators of the ultimate sensitivity of
the survey, sources similar to these may only be detected in stacked
images close to the center of several ObsID aim-points where exposure
levels are sufficiently high ($\sim 800$~ks) and the combined PSF is
relatively small. Hence, given an observing strategy with varying
levels of exposure across the fields, X-ray sources at these low flux
levels cannot be uniformly detected across the whole of each field. To
quantify the expected number of sources as a function of survey area,
we have constructed flux sensitivity maps for each merged field in the
0.5--2~keV, 2--7~keV and 0.5--7~keV bands.

\begin{figure*}[htb]
\centering
\includegraphics[width=0.95\textwidth]{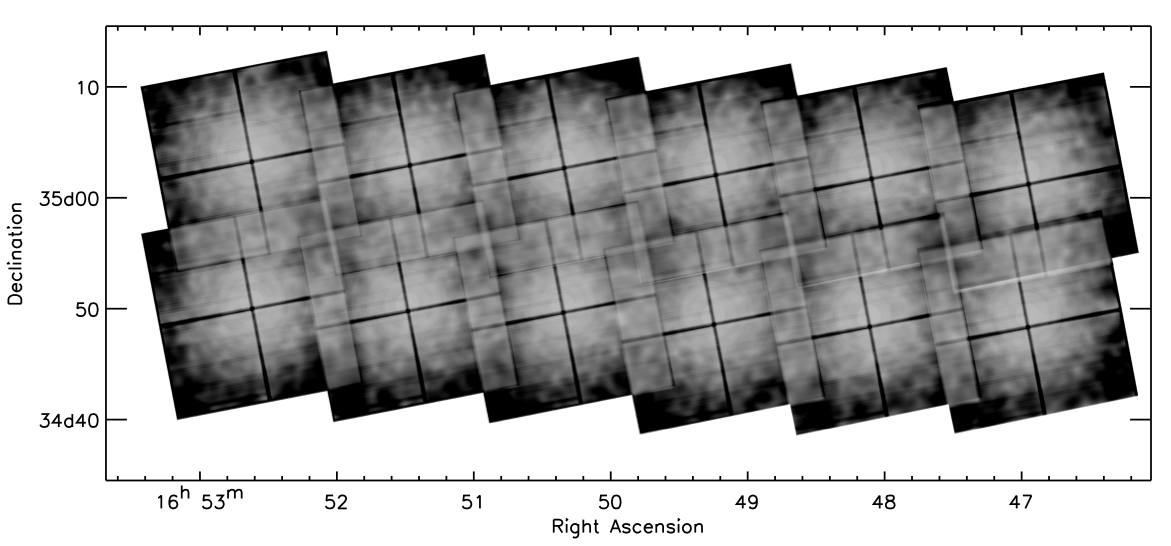}
\caption{Example of an exposure-corrected full-band flux sensitivity
  map for an XDEEP2 field. The sensitivity map has been created as
  described in section \ref{sec:sens}. Areas with lightest (darkest) colors
  correspond to those regions of the map with the greatest (poorest)
  sensitivity.}
\label{fig:sensmap}
\end{figure*}

\begin{figure*}[htb]
\centering
\includegraphics[width=0.97\textwidth]{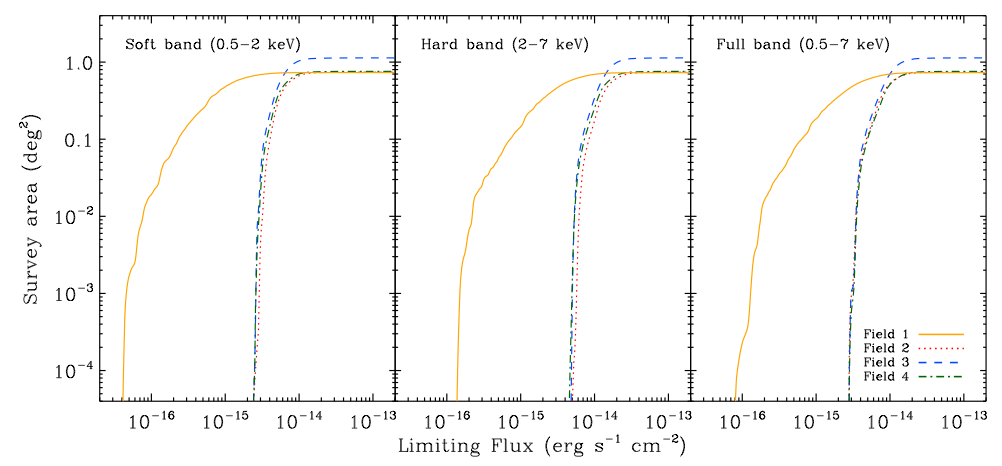}
\caption{Survey solid angle as a function of the limiting flux in the
  soft-band, hard-band and full-band (left, center and right panels,
  respectively) for each XDEEP2 field. Limiting fluxes in the
  full-band where at least 10\% of the survey field area are sensitive
  are $f_{X,1} > 2.8 \times 10^{-16} \ergpcmsqps$, $f_{X,2} > 4.5
  \times 10^{-15} \ergpcmsqps$, $f_{X,3} > 4.6 \times 10^{-15}
  \ergpcmsqps$ and $f_{X,4} > 4.6 \times 10^{-15} \ergpcmsqps$.}
\label{fig:limflux}
\end{figure*}

Maps of the {\it Chandra} point spread functions for an enclosed
energy fraction of 90\% were simulated at $E \sim 1.0$, 4.0 and
2.5~keV (mean SB, HB and FB energies, respectively) for each ObsID
using the {\sc ciao} tool {\tt mkpsfmap}. These maps were then merged
for all overlapping fields to calculate the mean $R_{90}$ in each
image pixel for a merged counts image in the soft, hard and
full-bands. We used the formalism of Lehmer et al. (2005) and employed
a Poisson model to calculate the average number of counts ($N$)
required to detect a source in a given image pixel for the background
counts ($b$) enclosed within the mean $R_{90}$ calculated in the
merged PSF model,

\begin{equation}
{\rm log} (N) = \alpha + \beta {\rm log} b + \gamma({\rm log} b)^2 +
\delta({\rm log} b)^3
\label{eqn:lehmer}
\end{equation}

where $\alpha = 0.967$, $\beta = 0.414$, $\gamma = 0.0822$ and $\delta
=0.0051$ (\citealt{lehmer05}). Using equation \ref{eqn:lehmer} we
convolve the merged PSF and background images at each image pixel and
normalize to the appropriate merged exposure maps to create final
fluxed sensitivity images in each energy band (three per field; an
example sensitivity image is shown in Figure \ref{fig:sensmap}).

We calculate empirical sensitivity curves in the SB, HB and FB for
each of the four XDEEP2 fields using the sensitivity images derived
above (see Figure \ref{fig:limflux}). Due to the small overlapping
regions in Fields 2--4, the sensitivity curves are found to be
relatively smooth over the entire survey region with relatively sharp
cut-offs at $f_{X,SB} \sim 4 \times 10^{-15} \ergpcmsqps$, $f_{X,HB}
\sim 9 \times 10^{-15} \ergpcmsqps$ and $f_{X,FB} \sim 7 \times
10^{-15} \ergpcmsqps$. Hence, the sensitivity limit is approximately
uniform across the majority of the survey area in Fields 2, 3 and
4. By contrast, the wedding-cake style observational setup of Field 1
combined with changing roll angles produces small ($\sim 0.1$~deg$^2$)
regions of high sensitivity, which combine over the field to produce a
much more shallow sensitivity curve (i.e., the sensitivity is
non-uniform). However, as the average exposure across Field 1 is
$\approx 20$--100 times greater than Fields 2--4, the mean sensitivity
to the {\it detection} of faint sources is vastly improved in Field
1. We find that the limiting flux in the 0.5--7~keV band for source
detection, which includes at least 10\% of the survey area, is a
factor $\approx 16$ lower in Field 1 ($f_{X,FB} > 2.8 \times 10^{-16}
\ergpcmsqps$) than in Fields 2--4 ($f_{X,FB} > 4.5 \times 10^{-15}
\ergpcmsqps$, $f_{X,FB} > 4.6 \times 10^{-15} \ergpcmsqps$ and
$f_{X,FB} > 4.6 \times 10^{-15} \ergpcmsqps$, respectively).

\subsection{Comparison of X-ray sources in Field 1 to Laird
  et~al. (2009)} \label{sec:laird}

\begin{turnpage}
\begin{table*}
\begin{center}
\tiny
\setlength{\tabcolsep}{0.25mm}
\caption{XDEEP2 source catalog \label{tbl_main_src}}
\begin{tabular}{ccccccccccccccccccccccccccccccccc}

\tableline\tableline
\multicolumn{3}{c}{Nomenclature} &
\multicolumn{3}{c}{Positional\tablenotemark{d}} &
\multicolumn{2}{c}{Radii\tablenotemark{e}} &
\multicolumn{1}{c}{} &
\multicolumn{4}{c}{Net Counts Soft\tablenotemark{g}} &
\multicolumn{4}{c}{Net Counts Hard\tablenotemark{h}} &
\multicolumn{4}{c}{Net Counts Full\tablenotemark{i}} &
\multicolumn{6}{c}{Flux\tablenotemark{j}} &
\multicolumn{3}{c}{Hardness Ratio\tablenotemark{k}} &
\multicolumn{3}{c}{Flux Ratio\tablenotemark{l}} \\
\multicolumn{1}{c}{DEEP2\tablenotemark{a}} &
\multicolumn{1}{c}{DEEP2\tablenotemark{b}} &
\multicolumn{1}{c}{CSC\tablenotemark{c}} &
\multicolumn{1}{c}{$\alpha_{\rm J2000}$ } &
\multicolumn{1}{c}{$\delta_{\rm J2000}$} &
\multicolumn{1}{c}{Err} &
\multicolumn{1}{c}{$R_{50}$} &
\multicolumn{1}{c}{$R_{90}$} &
\multicolumn{1}{c}{$D_{\rm OA}$\tablenotemark{f}} &
\multicolumn{1}{c}{${50}$ } &
\multicolumn{1}{c}{${50_{\rm er}}$} &
\multicolumn{1}{c}{${90}$} &
\multicolumn{1}{c}{${90_{\rm er}}$} &
\multicolumn{1}{c}{${50}$} &
\multicolumn{1}{c}{${50_{\rm er}}$} &
\multicolumn{1}{c}{${90}$} &
\multicolumn{1}{c}{${90_{\rm er}}$} &
\multicolumn{1}{c}{${50}$} &
\multicolumn{1}{c}{${50_{\rm er}}$} &
\multicolumn{1}{c}{${90}$} &
\multicolumn{1}{c}{${90_{\rm er}}$} &
\multicolumn{1}{c}{$S$} &
\multicolumn{1}{c}{$S_{\rm er}$} &
\multicolumn{1}{c}{$H$} &
\multicolumn{1}{c}{$H_{\rm er}$} &
\multicolumn{1}{c}{$F$} &
\multicolumn{1}{c}{$F_{\rm er}$} &
\multicolumn{1}{c}{$HR$} &
\multicolumn{1}{c}{$er_{\rm lo}$} &
\multicolumn{1}{c}{$er_{\rm up}$} &
\multicolumn{1}{c}{$FR$} &
\multicolumn{1}{c}{$er_{\rm lo}$} &
\multicolumn{1}{c}{$er_{\rm up}$} \\
\multicolumn{1}{c}{Name} &
\multicolumn{1}{c}{Field} &
\multicolumn{1}{c}{Name} &
\multicolumn{1}{c}{($^o$)} &
\multicolumn{1}{c}{($^o$)} &
\multicolumn{1}{c}{('')} &
\multicolumn{1}{c}{('')} &
\multicolumn{1}{c}{('')} &
\multicolumn{1}{c}{('')} &
\multicolumn{4}{c}{(counts)} &
\multicolumn{4}{c}{(counts)} &
\multicolumn{4}{c}{(counts)} &
\multicolumn{6}{c}{($10^{-16} \ergpcmsqps$)} &
\multicolumn{6}{c}{} \\
\tableline

aeg1\_001 & F1\_AEG1 & CXOJ141907.7+525946 & 214.78246 & 52.99710 & 0.84 & 4.48 & 10.36 & 632.0 &  16.62 &  5.77 &  31.17 &  8.86 &  23.17 &  7.15 &  48.97 & 12.25 &  39.81 &  8.74 &  80.26 & 14.70 & 4.80 & 1.44 & 20.6 & 5.62 & 19.2 & 3.78 &  0.20 &  0.04 &  0.43 &    4.75 &  3.18 & 7.66 \\
aeg1\_002 & F1\_AEG1 & CXOJ141907.8+530025 & 214.78334 & 53.00712 & 0.32 & 4.45 & 10.30 & 628.4 &  162.4 & 13.99 &  311.9 & 19.55 &  57.06 &  9.42 &  95.73 & 13.98 &  219.1 & 16.46 &  406.0 & 23.62 & 50.6 & 3.21 & 40.6 & 6.48 & 102. & 6.14 & -0.53 & -0.59 & -0.47 &    0.88 &  0.69 & 0.97 \\
aeg1\_003 & F1\_AEG1 & -                   & 214.79521 & 52.98033 & 1.18 & 6.91 & 12.42 & 611.5 &  14.31 &    -1 &  22.97 &    -1 &  35.58 &  9.37 &  54.85 & 13.96 &  46.79 & 10.76 &  71.22 & 16.18 & 3.66 &   -1 & 25.1 & 6.30 & 18.3 & 4.09 &  0.53 &  0.30 &  0.84 &   12.37 &  7.12 &    -1 \\
aeg1\_004 & F1\_AEG1 & CXOJ141911.2+530320 & 214.79699 & 53.05600 & 1.24 & 4.48 & 10.33 & 623.2 &  12.51 &  5.10 &  20.42 &  7.36 &  12.09 &    -1 &  23.29 &    -1 &  21.19 &  6.82 &  27.06 & 10.65 & 474. & 91.2 & 824. &   -1 & 1710 & 208. & -0.55 & -1.00 & -0.40 &    1.17 &  0.18 & 3.78 \\
aeg1\_005 & F1\_AEG1 & CXOJ141919.9+530254 & 214.83506 & 53.04790 & 0.84 & 3.42 &  7.95 & 536.8 &  10.13 &  4.84 &  43.10 &  9.10 &  15.51 &  6.08 &  46.83 & 11.03 &  25.57 &  7.31 &  89.53 & 13.87 & 5.98 & 1.31 & 17.7 & 4.55 & 19.3 & 3.18 &  0.02 & -0.11 &  0.20 &    3.26 &  2.25 & 4.34 \\
aeg1\_006 & F1\_AEG1 & CXOJ141920.6+530028 & 214.83600 & 53.00792 & 0.29 & 3.03 &  7.11 & 514.7 &  117.3 & 12.04 &  226.4 & 16.77 &  12.68 &    -1 &  28.79 & 10.21 &  127.6 & 12.91 &  255.2 & 19.24 & 23.7 & 1.76 & 8.15 & 3.07 & 42.3 & 3.22 & -0.79 & -0.85 & -0.70 &    0.41 &  0.25 & 0.56 \\
aeg1\_007 & F1\_AEG1 & CXOJ141922.8+530132 & 214.84506 & 53.02555 & 0.21 & 2.86 &  6.75 & 498.7 &  100.1 & 11.24 &  161.1 & 14.59 &  121.0 & 12.44 &  222.9 & 17.61 &  221.2 & 16.32 &  384.7 & 22.44 & 18.4 & 1.66 & 73.1 & 5.75 & 70.0 & 4.07 &  0.16 &  0.10 &  0.21 &    4.05 &  3.54 & 4.44 \\
aeg1\_008 & F1\_AEG1 & -                   & 214.85376 & 52.99871 & 2.53 & 3.14 &  5.65 & 477.5 &   5.61 &  4.29 &   7.76 &  5.82 &   4.46 &    -1 &   7.09 &    -1 &   5.03 &    -1 &   8.13 &    -1 & 0.76 & 0.59 & 0.62 &   -1 & 0.78 &   -1 & -0.54 & -1.00 & -0.45 &   2865. &  0.06 & -1 \\
aeg1\_009 & F1\_AEG1 & -                   & 214.85694 & 53.00549 & 0.43 & 2.53 &  6.01 & 469.5 &  20.15 &  5.88 &  42.87 &  8.77 &  27.37 &  6.90 &  41.56 & 10.01 &  47.39 &  8.60 &  83.71 & 12.87 & 4.32 & 0.88 & 12.0 & 2.90 & 13.4 & 2.07 & -0.03 & -0.17 &  0.13 &    2.88 &  2.14 & 3.96 \\
aeg1\_010 & F1\_AEG1 & -                   & 214.85765 & 53.01971 & 0.96 & 2.53 &  6.06 & 469.6 &   9.02 &    -1 &  16.26 &    -1 &  12.63 &  5.56 &  34.74 &  9.91 &  13.20 &    -1 &  36.99 & 11.10 & 1.63 &   -1 & 9.82 & 2.86 & 5.75 & 1.78 &  0.80 &  0.74 &  1.00 &   9117. & 18.87 &    -1 \\
aeg1\_011 & F1\_AEG1 & -                   & 214.86239 & 53.03122 & 0.54 & 2.56 &  5.98 & 464.7 &  22.14 &  6.08 &  44.39 &  8.83 &  11.39 &    -1 &  24.43 &  9.07 &  33.24 &  7.62 &  69.73 & 12.22 & 4.56 & 0.91 & 7.06 & 2.70 & 11.3 & 2.02 & -0.34 & -0.51 & -0.10 &    1.92 &  1.11 & 3.12 \\
aeg1\_012 & F1\_AEG1 & -                   & 214.86615 & 53.02515 & 0.56 & 2.44 &  5.75 & 453.7 &  10.17 &  4.71 &  29.25 &  7.81 &  18.13 &  6.08 &  27.73 &  9.26 &  28.58 &  7.23 &  58.53 & 11.67 & 2.91 & 0.78 & 7.87 & 2.67 & 9.24 & 1.86 & -0.06 & -0.23 &  0.20 &    3.14 &  1.96 & 4.95 \\
aeg1\_013 & F1\_AEG1 & CXOJ141928.0+525840 & 214.86670 & 52.97822 & 0.25 & 2.42 &  5.70 & 461.1 &  76.61 &  9.93 &  149.7 & 13.82 &  37.22 &  7.62 &  58.93 & 10.64 &  113.9 & 12.08 &  208.9 & 17.02 & 15.9 & 1.47 & 18.0 & 3.26 & 35.5 & 2.89 & -0.43 & -0.51 & -0.35 &    1.18 &  0.97 & 1.44 \\
aeg1\_014 & F1\_AEG1 & -                   & 214.87337 & 53.03977 & 0.29 & 2.37 &  5.63 & 448.2 &  52.24 &  8.47 &  87.44 & 11.22 &  29.69 &  7.07 &  59.41 & 10.85 &  81.87 & 10.58 &  146.5 & 15.17 & 8.73 & 1.13 & 16.8 & 3.15 & 23.2 & 2.44 & -0.19 & -0.29 & -0.09 &    2.12 &  1.62 & 2.50 \\
aeg1\_015 & F1\_AEG1 & -                   & 214.87634 & 53.04383 & 1.07 & 3.21 &  5.78 & 446.2 &  13.05 &  5.34 &  22.20 &  7.27 &   4.52 &    -1 &  12.87 &  8.22 &  15.98 &  6.83 &  33.72 & 10.53 & 2.26 & 0.75 & 0.41 &   -1 & 5.35 & 1.74 & -0.36 & -0.70 &  0.05 &    0.24 &  0.09 & 0.64 \\
aeg1\_016 & F1\_AEG1 & -                   & 214.87842 & 53.00748 & 1.44 & 3.34 &  6.01 & 422.9 &  12.95 &  5.34 &  21.81 &  7.27 &   4.74 &    -1 &   7.56 &    -1 &  11.05 &  6.47 &  17.48 &  9.83 & 2.11 & 0.71 & 0.64 &   -1 & 2.67 & 1.53 & -0.81 & -1.00 & -0.78 &    0.27 &  0.02 & 1.31 \\
aeg1\_017 & F1\_AEG1 & CXOJ141930.8+525915 & 214.87886 & 52.98781 & 0.42 & 2.11 &  5.00 & 428.3 &  30.90 &  6.81 &  62.29 &  9.67 &   9.90 &    -1 &  18.33 &    -1 &  36.53 &  7.62 &  79.83 & 11.87 & 6.30 & 0.98 & 5.35 &   -1 & 13.0 & 1.91 & -0.61 & -0.75 & -0.43 &    0.86 &  0.53 & 1.58 \\
aeg1\_018 & F1\_AEG1 & -                   & 214.88704 & 53.04167 & 1.12 & 2.96 &  5.33 & 421.8 &  10.14 &  4.85 &  14.38 &  6.31 &   4.26 &    -1 &   6.75 &    -1 &  13.21 &  6.28 &  20.96 &  9.27 & 1.51 & 0.65 & 0.61 &   -1 & 3.62 & 1.53 & -0.43 & -1.00 & -0.23 &    0.78 &  0.16 & 4.05 \\
aeg1\_019 & F1\_AEG1 & -                   & 214.88706 & 52.99963 & 0.82 & 1.86 &  4.58 & 405.0 &   7.57 &    -1 &  12.61 &    -1 &  10.46 &  4.84 &  17.61 &  7.37 &  10.76 &  5.10 &  22.31 &  8.48 & 1.23 &   -1 & 4.92 & 2.08 & 3.44 & 1.32 &  0.52 &  0.39 &  1.00 &   7412. &  6.61 & -1 \\
aeg1\_020 & F1\_AEG1 & -                   & 214.88917 & 53.09005 & 1.37 & 4.92 &  8.88 & 496.5 &  20.36 &  6.37 &  24.09 &  8.12 &  16.40 &    -1 &  26.88 &    -1 &  21.16 &  8.00 &  31.18 &    -1 & 3.20 & 1.29 & 12.2 &   -1 & 7.86 &   -1 & -0.71 & -1.00 & -0.64 &    2.09 &  0.16 &       6.51 \\

\tableline

\end{tabular}
\end{center}
\footnotesize
{\sc Notes}-- \\
$^{a}$Unique source identifier \\
$^{b}$XDEEP2  ObsID/sub-field name \\
$^{c}$Unique source identifier for matched XDEEP2 sources present in the
{\it Chandra} X-ray Source Catalog (CSC) \\
$^{d}$X-ray position in J2000 co-ordinates (degrees) and associated centroid
positional error (arc-seconds) \\
$^{e}$Aperture radius in arc-seconds at  50\% and 90\% the effective area of ACIS-I given the off-axis distance of the X-ray source \\
$^{f}$Off-axis distance in arc-seconds of X-ray source from aim-point of observation \\
$^{g}$Soft-band (0.5--2~keV) net counts and associated errors in the
$R_{50}$ and $R_{90}$ apertures \\
$^{h}$Hard-band (2--7~keV) net counts and associated errors in the
$R_{50}$ and $R_{90}$ apertures \\
$^{i}$Full-band (0.5--7~keV) net counts and associated errors in the
$R_{50}$ and $R_{90}$ apertures \\
$^{j}$Total soft-band (S), hard-band (H) and full-band (F) fluxes and
associated errors in units  of $10^{-16} \ergpcmsqps$ \\
$^{k}$Classical hardness ratios ($HR = (C_{\rm H} - C_{\rm S}) /
(C_{\rm H} + C_{\rm S})$) and associated 1$\sigma$ upper and lower
limits calculated using the BEHR method \\
$^{l}$Flux ratios ($FR = F_{\rm HB} / F_{\rm SB}$) and associated 1$\sigma$ upper and lower
limits calculated using the BEHR method  \\
\end{table*}
\end{turnpage}
\normalsize

As stated previously, while the analyses presented here include the
recent 600ks observations of AEGIS 1--3, the 200ks X-ray source
catalog for Field 1 ({\it AEGIS-X}) has been previously presented in
Laird et al. (2009). Furthermore, the new 600ks observations will also
be presented in a forthcoming paper (Nandra et al. in prep.) using
similar detection and Bayesian-style sensitivity analyses to that used
for the previous {\it AEGIS-X} catalog. Since the source detection and
extraction analyses differ significantly between {\it AEGIS-X} and the
XDEEP2 catalog presented here, we now compare the detection methods
and results.

\begin{figure*}[htb]
\includegraphics[width=0.97\textwidth]{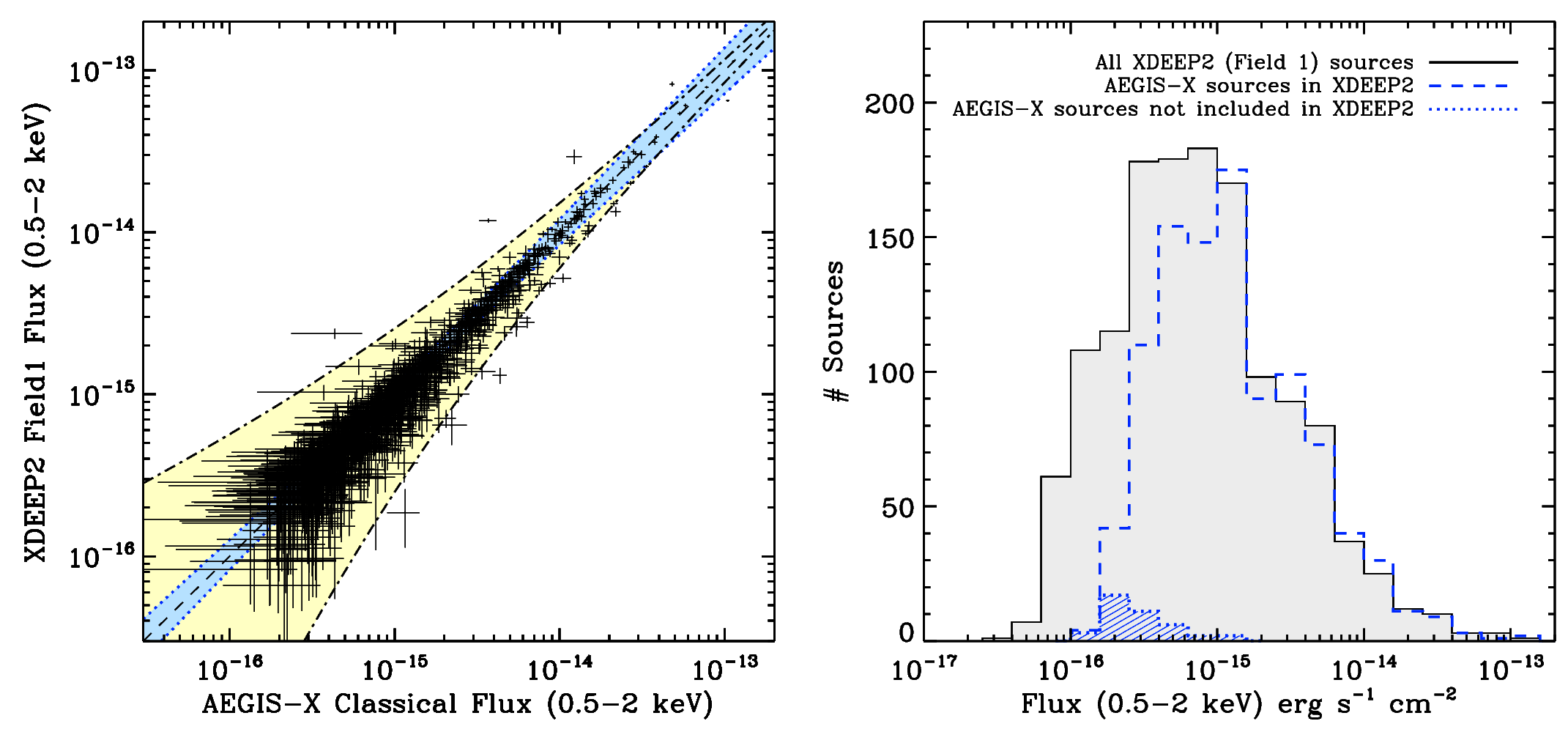}
\caption{{\bf a (left):} Comparison of X-ray source fluxes in the
  0.5--2~keV energy band for the 1260 sources in common between the
  previous AEGIS-X source catalog and the XDEEP2 catalog presented
  here. The AEGIS-X sources are corrected for galactic absorption. The
  XDEEP2 fluxes are converted to $\Gamma = 1.4$ to match the AEGIS-X
  catalog. The dashed-line is the best-fit linear-bisector to the
  logarithms of the source fluxes. The blue-shaded region
  (dotted-lines) represents the 3$\sigma$ uncertainty on the linear
  correlation calculated using 2-dimensional linear regression
  analyses. The yellow-shaded region (dash-dot-lines) represents the
  $3\sigma$ Poissonian error on the source fluxes due to
  photon-counting statistics, derived using the formalism of
  \citet{gehrels86}. We find a very strong agreement at the 99.99\%
  level between XDEEP2 and AEGIS-X source fluxes. The $\lesssim 0.1$\%
  of outliers are significantly extended in the ACIS-I images and/or
  have $\gg 100$ counts, suggesting that these sources are strong
  candidates for galaxy clusters and/or moderately variable QSOs. {\bf
    b (right):} Soft-band source flux distributions (0.2 dex
  bin-width) for all X-ray sources detected in XDEEP2 (gray-shaded
  region) and AEGIS-X (blue-dash). The flux distribution of the 59
  AEGIS-X source candidates which lack secure matches in XDEEP2 are
  highlighted with blue-hashed shading.}
\label{fig:xd2vslaird}
\end{figure*}

Briefly, detection of sources in {\it AEGIS-X} was carried out using a
custom implementation of the CIAO {\tt wavdetect} tool. Laird et
al. (2009) perform several runs of the detection software using
different probability thresholds to build seed catalogs and to derive
multiple estimates of the X-ray background in the observation. The
final probability threshold for which a particular candidate is
determined to be false in {\it AEGIS-X} is comparable to that used in
our analyses. Laird et al. detected source candidates separately in
the full, soft, hard and ultra-hard band images.\footnote{In {\it
    AEGIS-X} the ultra-hard band is defined in the energy range
  4--7~keV.} These source candidates were then combined into an
individual source catalog using Bayesian techniques to statistically
associate the source candidates and calculate the fluxes in the
respective energy bands. The {\it AEGIS-X} catalog contains 1325
sources. Two sources (EGS4\_258; EGS7\_204, nomenclature adopted from
Laird et al. 2009) in the {\it AEGIS-X} catalog were only detected in
the ultra-hard band, i.e., an energy-band which we do not use due to
the relatively small effective area of the telescope at these higher
energies. Furthermore, four {\it AEGIS-X} sources (EGS4\_240;
EGS6\_185; EGS7\_194; EGS8\_127) have 90\% effective-area extraction
regions which significantly ($> 50$\%) overlap with those extraction
radii of other sources in the {\it AEGIS-X} catalog; from visual
inspection we find that these four {\it AEGIS-X} sources (and their
neighbors) are consistent with being single point sources. Hence, we
remove these six sources from further comparison between the XDEEP2
and {\it AEGIS-X} catalogs.

We compared the 1319 unique source candidates identified in {\it
  AEGIS-X} to the 1720 source candidates identified in Field 1 of our
XDEEP2 catalog solely on the basis of source position using the same
varying matching radius method described in \S\ref{sec:XOPT}. We find
that 1260 ($\approx 96$\%) of the source candidates identified in {\it
  AEGIS-X} are included in our new catalog. We have visually inspected
each of the 59 {\it AEGIS-X} sources which were not identified in the
XDEEP2 catalog. We find that the majority (44/59; $\sim 63$\%) of the
{\it AEGIS-X} sources, which are not included as part of the XDEEP2
catalog, were detected by {\tt wvdecomp} as source candidates in one
energy band. However, on the basis of our MARX simulations, these 44
non-matched {\it AEGIS-X} sources did not meet our ultimate and more
conservative count detection threshold and were removed as possibly
spurious based on their low net counts ($C_{\rm 90,net} \sim
5$--10). A further seven of the 59 non-matched sources were flagged as
`non-standard' and possibly spurious; we discuss these seven sources
below. Finally, eight of the 59 non-matched {\it AEGIS-X} source
candidates are not detected using {\tt wvdecomp} after the inclusion
of the more recent 600ks data.

We now briefly discuss the seven of the 59 non-matched {\it AEGIS-X}
source candidates (EGS2\_052; EGS5\_105; EGS6\_073; EGS6\_093;
EGS7\_180; EGS8\_093; EGS8\_134) that were initially detected by {\tt
  wvdecomp} and then highlighted by our routine as `possibly
spurious'. Visual inspection shows that three of these seven source
candidates, EGS6\_093, EGS5\_105 and EGS7\_180) have their expected
source PSFs partially blended with secondary brighter sources. Indeed,
EGS6\_093, is located between ($<2.5$~arc-seconds) two significantly
brighter X-ray sources (EGS6\_164 and EGS6\_165; both these sources
are included in the {\it AEGIS-X} and XDEEP2 catalogs) causing
sufficient detection ambiguity and EGS7\_180 has an X-ray morphology
consistent with that of a jet. These three sources, while initially
detected in XDEEP2, are not included in our final catalog due to our
inability to accurately separate the flux contribution from the
neighboring bright source. Furthermore, EGS6\_073 falls on a chip-gap;
EGS2\_052 has $C_{\rm FB,net} < 6$; and EGS8\_134 has $\sim 50$\% of
its low source counts ($C_{\rm FB,net} \sim 7$) in one ACIS-I pixel,
and is conservatively removed based on our MARX simulation
analyses. Finally, EGS8\_093 is possibly part of an extended source
which appears extremely diffuse and only has a marginal detection
($P_{\rm SB} \sim 1.4 \times 10^{-6}$) in the {\it AEGIS-X} catalog.

As above, eight of the 59 non-matched {\it AEGIS-X} source candidates
are not detected using {\tt wvdecomp} after the inclusion of the 600ks
data; while these sources were detected in the {\it AEGIS-X} analyses
with $P_{\rm band} > 10^{-6}$, we note here that these sources may
still be real, but are no longer detected due to intrinsic variability
of the source. Similarly, Nandra et al. (in prep.) find that from a
re-analysis of {\it AEGIS-X}, 17 of the {\it AEGIS-X} sources are no
longer detected in the three sub-fields which include the new 600ks
data. Assuming a similar number of non-detected source candidates
across all of Field 1, this would suggest a false source contamination
rate of $\approx 45$ sources ($\approx 3.5$\%) in {\it AEGIS-X}.

In Figure \ref{fig:xd2vslaird} we show a comparison between the
soft-band fluxes for isolated and formally-detected point-sources in
the {\it AEGIS-X} (classically derived flux) and XDEEP2 catalogs. We
have converted the fluxes we derived using $\Gamma = 1.7$ in XDEEP2 to
$\Gamma = 1.4$, as used in {\it AEGIS-X}, using a conversion of 1.031
and we have corrected the {\it AEGIS-X} fluxes for galactic absorption
(a factor of 1.042; Laird et al. 2009). We find excellent agreement
between the fluxes derived in the XDEEP2 and {\it AEGIS-X} catalogs
with a Spearman's rank coefficient of $r \sim 0.963$ which is
significant at $P > 99.99$\% level. Additionally, we have used
two-dimensional linear-regression analyses to calculate the $3 \sigma$
uncertainty on the derived correlation between the XDEEP2 and AEGIS-X
source fluxes (dotted-lines in Figure \ref{fig:xd2vslaird}), and the
$3 \sigma$ error region on the photon counts used to derive the source
fluxes (dash-dot-lines in Figure \ref{fig:xd2vslaird}). As expected,
the Poissonian error due to low source counts significantly dominate the
uncertainty towards low fluxes. We find that 12 ($\lesssim 0.1$\%) of
the matched XDEEP2--{\it AEGIS-X} sources lie substantially outside
the $3\sigma$ error region. These outlying sources have large numbers
of counts ($\gg 100$) and/or are significantly extended beyond the
expected 90\% EEF angular size in the merged ACIS-I images. This
suggests that these outlying sources are strong candidates for galaxy
clusters and/or moderately variable quasi-stellar objects
(QSOs).\footnote{As noted previously, the cluster candidates and their
  properties will be discussed in detail in a forthcoming
  publication.} Furthermore, variations in extraction radii at large
offaxis distances, due to the introduction of the more recent ACIS-I
observations (which were performed with substantially different
roll-angles) may potentially cause significant differences in measured
counts/flux for bright X-ray sources with non-point-like profiles,
such as galaxy clusters. Indeed, we find that when considering only
the previous 200ks observations studied in Laird et al., with matched
extraction apertures, the fluxes for all of the matched XDEEP2--{\it
  AEGIS-X} sources are consistent to within 1$\sigma$.

In Figure \ref{fig:xd2vslaird} we show the source flux distributions
in {\it AEGIS-X} and XDEEP2 including the 460 new XDEEP2 sources which
are detected in the new deeper 600~ks observations. As expected, the
majority of these 460 new XDEEP2 sources have $f_{\rm 0.5-2keV} \sim
(8$--$80) \times 10^{-16} \ergpcmsqps$, extending the distribution of
the previous catalog to lower source fluxes. We additionally highlight
the fluxes of the 59 {\it AEGIS-X} source candidates, which we
conservatively do not include in XDEEP2. Each of these non-matched
sources have $f_{\rm 0.5-2keV} < 1.3 \times 10^{-15} \ergpcmsqps$,
with the vast majority at the extreme low-flux end of the main {\it
  AEGIS-X} source-flux distribution ($f_{\rm 0.5-2keV} \sim (1$--$4)
\times 10^{-16} \ergpcmsqps$). Using a Bayesian counterpart matching
algorithm, which we present in \S~\ref{sec:XOPT}, we have attempted to
assign DEEP2 optical counterparts to the 59 {\it AEGIS-X} source
candidates. We find that the majority (35/59; $\sim 60$\%) of these
{\it AEGIS-X} source candidates lack secure optical counterparts; this
is a factor $\sim 2$ larger than the fraction of X-ray sources which
lack counterparts across the entire XDEEP2 sample ($\sim
29$\%). However, based on simulations of a purely random set of 59
source positions, we would expect only $\sim 3$--7 spurious
counterpart matches using our Bayesian matching algorithm. Hence, the
24 {\it AEGIS-X} sources found to coincide with an optical counterpart
is a factor $\sim 3$--8 larger than the random expectation of spurious
counterparts, suggesting that some of these X-ray sources may be real.

Based on our rigorous comparison of the {\it AEGIS-X} catalog and our
XDEEP2 catalog, we suggest that the two catalogs appear to be in
excellent agreement, despite the use of different detection algorithms
({\tt wavdetect} versus {\tt wvdecomp}). In general, the small
($\approx 4$\%) discrepancy between the catalogs can be attributed to
the removal of low significance sources in the XDEEP2 catalog based on
our MARX simulations. Additionally, we stress that since 51 of the 59
low significance sources are initially identified by both {\tt
  wavdetect} and {\tt wvdecomp}, we cannot rule out that they are real
sources, although they ultimately did not meet our more conservative
detection criteria.

\subsection{Comparison of X-ray sources in Fields 2--4 to the {\it
    Chandra} Source Catalog} \label{sec:CSC}

\begin{figure*}[htb]
\centering
\includegraphics[width=0.97\textwidth]{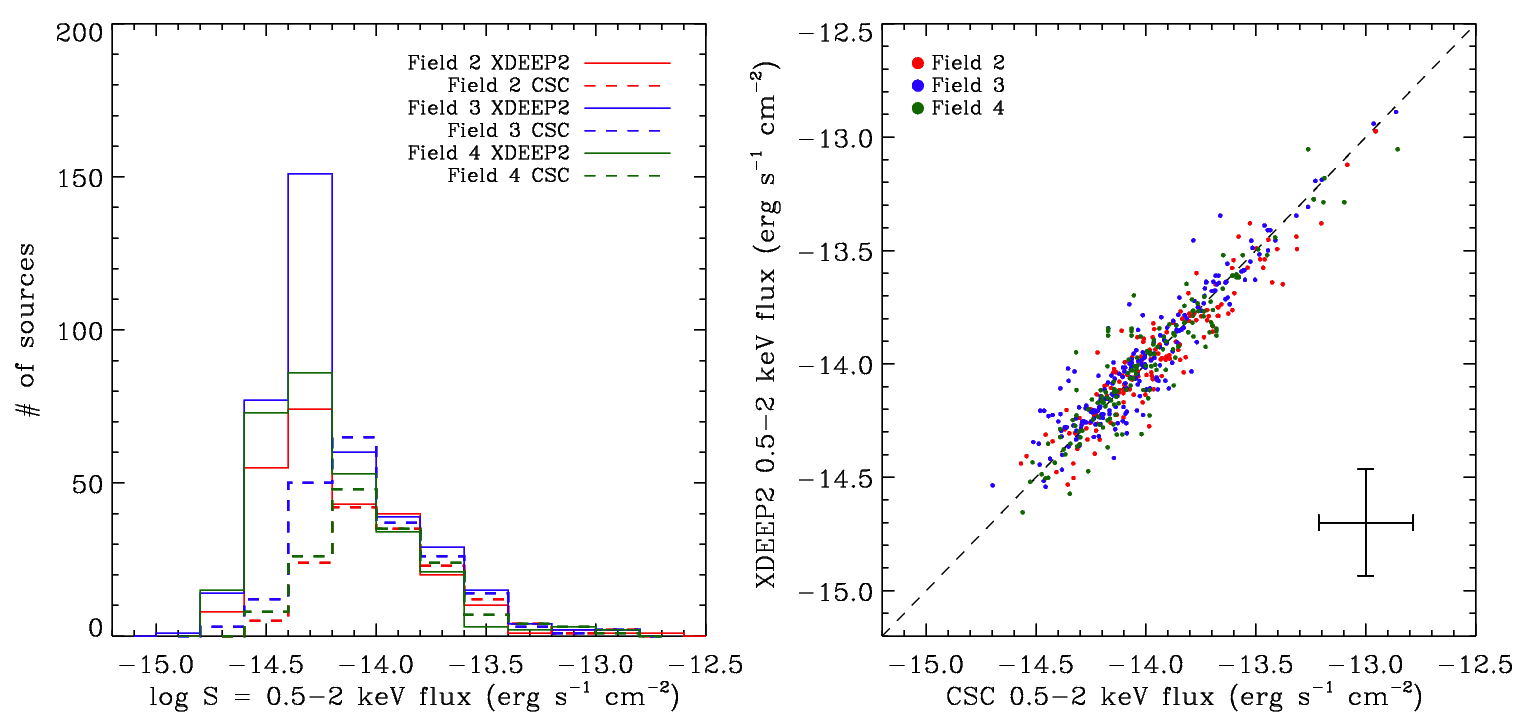}
\caption{{\bf a (left):} Soft-band (0.5--2~keV) flux distributions for
  all X-ray sources identified in the {\it Chandra} ACIS-I
  observations of XDEEP2 Fields 2 (red), 3 (blue) and 4 (green) within
  the XDEEP2 (solid lines) and CSC (dashed-lines) catalogs. {\bf b
    (right):} Comparison of 0.5--2~keV fluxes for all XDEEP2 sources
  associated with CSC sources in Fields 2 (red dots), 3 (blue dots) and
  4 (green dots). Error bar represents the median uncertainty in the
  flux estimates for the CSC and XDEEP2 sources.}
\label{fig:csc_comparison}
\end{figure*}

The Chandra Source Catalog (CSC) is a compilation of all relatively
bright X-ray sources detected in single ACIS and HRC imaging
observations by the Chandra X-ray Observatory in the first eight years
of the mission (\citealt{evans10}). In principle, the X-ray sources
detected in XDEEP2 Fields 2--4 and by the CSC are likely to be
equivalent. Similar to the CSC, we have not attempted to merge events
in overlapping regions of Fields 2--4 as, in general, these regions
occur at large off-axis distances where the {\it Chandra} PSF is
poor. In this section, we compare the XDEEP2 source properties to
those detected in the CSC release 1.1. The current CSC data release
contains X-ray data products and information (positions; spatial;
temporal multi-band count rates; fluxes) for distinct point sources
and compact sources, with observed spatial extents $\lesssim 30$''
observed in publicly released data to the end of 2009. Highly extended
sources, and sources located in selected fields containing bright,
highly extended sources are excluded in the CSC. See
\url{http://cxc.cfa.harvard.edu/csc/index.html} for further
information.

We have used the publicly available java-applet, CSCview to associate
the XDEEP2 X-ray sources in Fields 2--4 to the CSC Master
Catalog. Although we do not attempt to merge the individual ACIS-I
observations in Fields 2--4, we find that only $\sim 41.9 \pm 2.2$\%
(i.e., 150/342; 218/528; 158/386 sources in Fields 2, 3 and 4,
respectively) of the XDEEP2 sources are identified in the CSC. In
Figure~\ref{fig:csc_comparison}a we show a comparison of the flux
distributions for all XDEEP2 sources and CSC sources within the area
covered by Fields 2, 3 and 4 of XDEEP2. The 90\% EEF aperture fluxes
produced by the CSC are derived using a simple absorbed powerlaw with
$\Gamma = 1.7$ and $N_{\rm H} = 3 \times 10^{20} \pcmsq$. Hence, for
the purposes of comparison we convert the field-specific $N_{\rm H}$
used to derive the XDEEP2 fluxes to match the CSC fluxes.

We find that while all CSC sources with $f_{\rm 0.5-2} \gtrsim 6
\times 10^{-15} \ergpcmsqps$ are identified in the XDEEP2 catalog, the
vast majority of the lower flux XDEEP2 sources are not included in the
CSC. By design, the detected CSC X-ray sources have $C_{\rm net}
\gtrsim 10$ counts for an on-axis source ($\gtrsim 20$--30 counts for
an off-axis source), i.e., the CSC catalog only includes sources whose
flux estimates are greater than three times their estimated 1$\sigma$
uncertainties. However, as we have shown in Figures~\ref{fig:marxsim1}
and \ref{fig:marxsim2}, and has been shown conclusively by many other
deep and wide-field X-ray surveys (e.g., CDF-N; CDF-S; C-COSMOS;
AEGIS-X; XBootes), many X-ray sources can be significantly identified
with only $\sim 3$--5 net counts, although the source flux will remain
relatively unconstrained due to Poisson uncertainties. Indeed,
$\gtrsim 98$\% of the XDEEP2 sources not identified in the CSC catalog
have $C_{\rm net} < 20$ counts. Furthermore, to within $1 \sigma$, we
find excellent agreement for the X-ray fluxes of the sources in common
between XDEEP2 and the CSC (see Figure~\ref{fig:csc_comparison}b).

Given that all of the CSC sources within the survey area are
identified in the XDEEP2 catalog and the non-matched sources have
lower counts/flux which lie above the thresholds derived from our
extensive simulation analyses, we find that the CSC provides a more
conservative identification of X-ray sources within the XDEEP2
fields. For completeness, we have also associated the X-ray sources
identified in Field 1 to the CSC catalog, and find there are 689
distinct X-ray sources in common between the catalogs. The faintest
CSC sources in Field 1 have $f_{\rm 0.5-2} \gtrsim 5 \times 10^{-16}
\ergpcmsqps$, but with the majority at $f_{\rm 0.5-2} \gtrsim 2 \times
10^{-15} \ergpcmsqps$ (i.e., an average factor $\sim 3$ more sensitive
per individual observation than Fields 2--4). For ease of comparison
with future surveys, we include the CSC source identifiers as part of
the XDEEP2 catalog, for all XDEEP2 sources with CSC counterparts.

\subsection{Source spectral properties: hardness ratios} \label{sec:HR}

Using the Bayesian Estimator of Hardness Ratio (BEHR) method
(\citealt{park06}), hardness count ratios (HR), defined as ${\rm HR} =
(C_{\rm HB}-C_{\rm SB})/(C_{\rm HB}+C_{\rm SB})$, where $C_{\rm SB}$
and $C_{\rm HB}$ are the counts in the soft and hard bands
respectively, as well as the hardness flux ratios (FR), defined as
${\rm FR} = F_{\rm HB}/F_{\rm SB}$, were calculated for all detected
sources in the XDEEP2 catalog. FR and HR and their associated
uncertainties calculated using BEHR are available in the main XDEEP2
source table. Briefly, BEHR treats the detected source and background
X-ray photons as independent Poisson random variables, and uses a
Monte Carlo based Gibbs sampler to select samples from posterior
probability count distributions to correctly propagate the
non-Gaussian uncertainties, which derive from the calculation of
hardness ratios. BEHR is particularly powerful in the low-count
Poisson regime, and computes a realistic uncertainty for the HR and
FR, regardless of whether the X-ray source is detected in both energy
bands. In Table~\ref{tbl_main_src}, we include the FR and HR ratios
with the associated $1\sigma$ upper and lower limits for all XDEEP2
sources. Sources with unconstrained upper or lower limits due to
non-detections are denoted by ``-1'' in the appropriate uncertainty
column.

\begin{figure}[htb]
\includegraphics[width=0.97\linewidth]{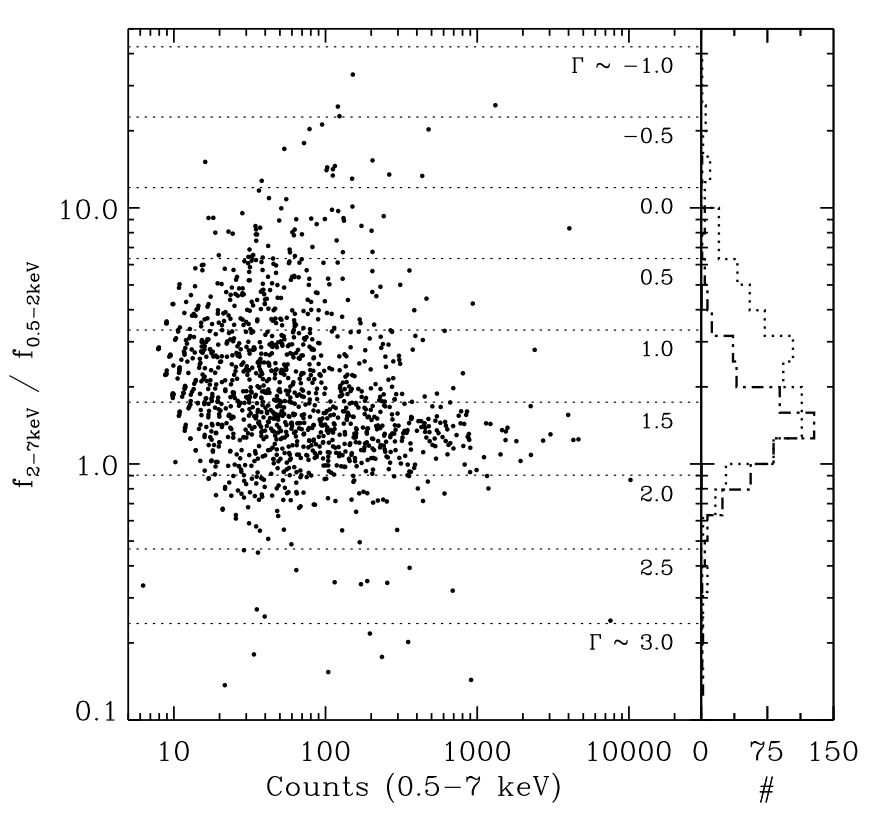}
\caption{{\bf Main panel:} Flux band ratio defined as ${\rm FR} =
  f_{\rm 2-7keV}/f_{\rm 0.5-2keV}$ as a function of full-band counts
  ($C_{0.5-7keV}$) for all XDEEP2 sources detected in the soft and
  hard energy bands. Average spectral slopes for fixed values of FR
  established from X-ray spectral simulations using {\tt Sherpa} are
  highlighted with horizontal dotted lines. {\bf Right panel:} FR
  distributions for all detected sources within XDEEP2 with $f_{\rm
    0.5-7keV}\gtrsim 4 \times 10^{-15} \ergpcmsqps$ (dot-dashed) and
  $f_{\rm 0.5-7keV}\lesssim 4 \times 10^{-15} \ergpcmsqps$ (dotted). }
\label{fig:HR}
\end{figure}

In Figure~\ref{fig:HR} we show the FR distribution for the XDEEP2
sources. Typically, the XDEEP2 sources which are detected in both the
hard and soft-bands have FR in the range $\sim 0.7$--7, with
distribution tails at low and high values of FR. Following previous
studies (e.g., \citealt{bauer02,dma03a,Luo08}), we divide the X-ray
sources with low and high-flux at $f_{\rm FB} \sim 4 \times 10^{-15}
\ergpcmsqps$ (i.e., the 10\% flux limit of the shallow exposure XDEEP2
fields). While the choice of cut is somewhat arbitrary, clearly we
find the same general trend towards higher values of FR for X-ray
sources with low-fluxes as has been observed previously
(e.g.,\citealt{hasinger93,vikhlinin95,giacconi02,tozzi06}). We find
that the distribution of FR values is moderately peaked at ${\rm FR}
\sim 1.3$ sources with high flux ($f_{\rm FB} \gtrsim 4 \times
10^{-15} \ergpcmsqps$). By contrast, lower flux sources have a more
extended distribution, with a median value of ${\rm FR} \sim 2.1$ and
tailing to higher values of FR. Using the {\sc ciao} spectral analysis
package, {\tt sherpa}, we have simulated X-ray spectra for AGN
populations at $0 < z < 6$ in order to quantify the evolution of X-ray
spectral slopes due to the k-correction of the observed AGN spectra
towards high-z. Based on these simulations, we find that the two peaks
observed in the FR distributions are co-incident with the spectral
slopes expected for two separate AGN populations with $\Gamma \sim
1.2$--1.4 and $\Gamma \sim 1.7$--1.8. Further, we find that the
majority of the 460 low-flux sources in Field 1, which were not
previously identified in AEGIS-X due to insufficient survey depth (see
\S~\ref{sec:laird}), have a similarly wide FR distribution ($\sim
0.8$--10) to the AEGIS-X source candidates and the sources identified
in Fields 2--4. However, the median FR for the new faint Field 1
sources is shifted slightly higher with FR$\sim$3 (i.e., harder
spectral indices), suggesting that these new sources have flatter
X-ray spectral slopes, and are likely to be more heavily
obscured. Hence, their previous non-detection in the 200ks data is due
to the combined result of AGN luminosity, distance and intrinsic
obscuration.

\begin{figure*}[htb]
\begin{center}
\includegraphics[width=0.9\textwidth]{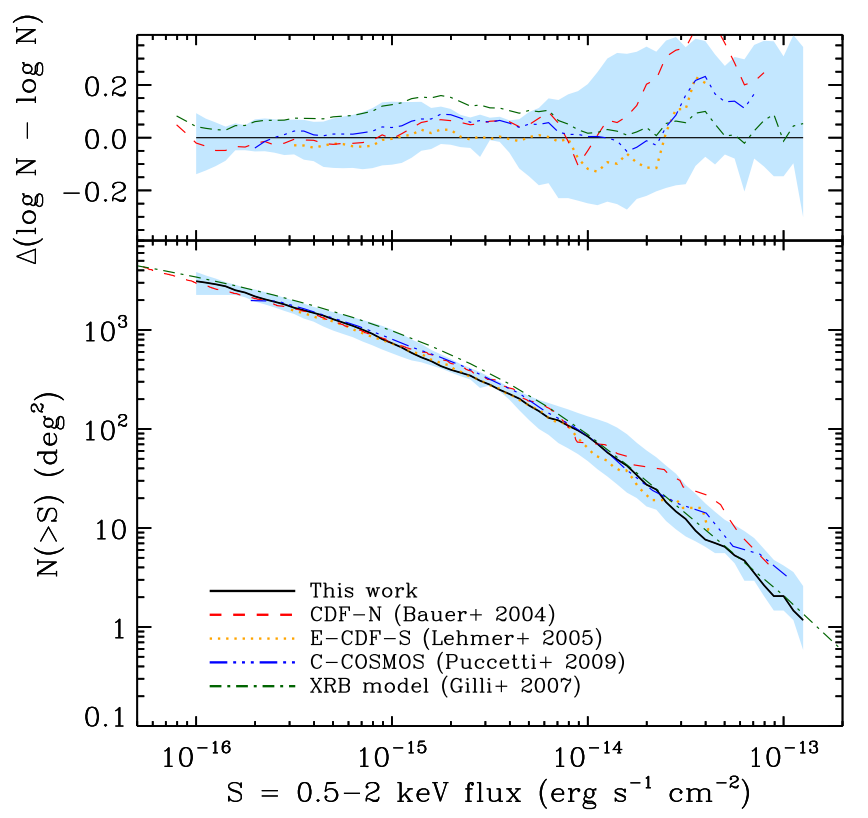}
\caption{{\bf Main panel:} Logarithm of the number of detected sources
  within the XDEEP2 catalog brighter than a given soft-band flux
  ($N(>S)$; black solid line) versus the logarithm of soft-band flux
  (i.e., the log$N$--log$S$ distribution). For comparison to previous
  surveys, source fluxes were converted using $\Gamma = 1.4$. We use a
  Monte-Carlo simulation to assess the 90\% uncertainty on the XDEEP2
  distribution due to flux errors and Poisson counting statistics
  (shaded region). We compare this $N(>S)$ distribution to previous
  surveys fields [CDF-N (Bauer et al. 2004); Extended-CDF-S (Lehmer et
  al. 2005); {\it Chandra}-COSMOS (Puccetti et al. 2009)] and to an
  X-ray background synthesis model (Gilli et al. 2007). {\bf Top
    panel:} Logarithm of the residuals between $N(>S)$ XDEEP2 and the
  comparison $N(>S)$ curves. The logarithm of the uncertainty for the
  log$N$--log$S$ is shown by the shaded region. We find good agreement
  ($< 10$\% deviation) between XDEEP2 and all previous observation
  surveys in the regime $f_X < 1 \times 10^{-14}
  \ergpcmsqps$. However, we show a mild systematic offset towards
  lower $N$ for sources with $f_X > 2 \times 10^{-14} \ergpcmsqps$, in
  closer agreement with XRB models.}
\label{fig:lognlogs}
\end{center}
\end{figure*}

\subsection{XDEEP2 source number counts} \label{sec:lognlogs}

We have calculated the cumulative number of sources in the XDEEP2
catalog ($N(>S)$) detected per square degree that are brighter than a
given flux in the soft (0.5--2 keV) band, i.e., the ${\rm log} N -
{\rm log} S$ distribution (see Figure~\ref{fig:lognlogs}). This
provides a good check that the merging of the datasets and the
extensive calibrations were performed correctly, as well as an
excellent comparison to previous X-ray surveys. We choose to compare
in the soft-band as this is the most sensitive energy and the specific
energy range definition of the soft-band (0.5--2~keV) is consistent
across previous surveys. As a consequence of (1) the changing slope of
the ${\rm log} N - {\rm log} S$ distribution towards fainter fluxes,
and (2) observationally fainter sources possibly being more obscured
and/or lower accretion rate AGN than brighter sources, the so-called
`Eddington bias' introduces many statistically low-significance
sources at the sensitivity limit of the X-ray survey. Hence, we have
empirically restricted our analyses presented in this section to only
those sources detected with $f_{\rm SB} > 4.5\sigma_{\rm bkg,field}$,
i.e., on-axis 0.5--2~keV fluxes of $f_X \gtrsim 9 \times 10^{-17}
\ergpcmsqps$ in Field 1 and $f_X \gtrsim 4 \times 10^{-15}
\ergpcmsqps$ in Fields 2--4 (equivalent to $C_{\rm SB,net} > 10$ and
$C_{\rm SB,net} > 6$, respectively). For the purpose of comparison, we
have converted all source and field fluxes to $\Gamma = 1.4$ and use
the combined flux limits (see \S\ref{sec:sens}) to construct the ${\rm
  log} N - {\rm log} S$ distribution.

To quantify the uncertainties on the derived ${\rm log} N - {\rm log}
S$, we have used a Monte-Carlo (MC) style simulation. Using the formal
$1\sigma$ error on the source flux, we built symmetrical probability
flux distributions ($P(f_X)$) for each source to be input to 10,000
realizations of our simulation. Within the MC simulation, we randomly
assign fluxes to each source within the sample based on the individual
$P(f_X)$, and recompute the ${\rm log} N - {\rm log} S$
distribution. The total 90\% uncertainty on the ${\rm log} N - {\rm
  log} S$ is then defined as the mean absolute deviation of the 10,000
simulated distributions combined in quadrature with the 90\%
Poissonian error on the main distribution, defined using the formalism
of Gehrels (1986). From our MC simulations, in Figure
\ref{fig:lognlogs} we show that the XDEEP2 ${\rm log} N - {\rm log} S$
is very well constrained ($\sim 0.12$~dex) in the flux range $f_X \sim
(0.2$--$5) \times 10^{-15} \ergpcmsqps$ owing to the large sensitive
area in Field 1 around the `knee' of the ${\rm log} N - {\rm log} S$
at $f_X \sim ($6--8$) \times 10^{-15} \ergpcmsqps$. However, we find
that the uncertainty on the distribution increases to $\sim 0.3$~dex
towards the bright flux tail ($f_X \gtrsim 10^{-14} \ergpcmsqps$) of
the ${\rm log} N - {\rm log} S$. We determined that this is caused by
the decrease in the space-density of the far rarer bright sources,
combined with the relatively large uncertainties on the fluxes for
those sources identified in the more shallow exposure Fields 2--4. For
these particular sources, which dominate the distribution within this
moderate--high flux regime, the majority are detected with relatively
few counts ($\sim 6$--15) and hence, $1 \sigma$ flux errors are $\sim
25$--50\% of the overall flux. In turn, these relatively large flux
uncertainties cause significant scatter of the sources within the
simulated distributions.

In Figure~\ref{fig:lognlogs}, we additionally compare the ${\rm log} N
- {\rm log} S$ derived from XDEEP2 to the distributions found in
previous wide and deep {\it Chandra} surveys [CDF-N (Bauer et
al. 2004); Extended-CDF-S (Lehmer et al. 2005); {\it Chandra}-COSMOS
(\citealt{puccetti09})]. In the flux range $f_X \sim (0.09$--$20)
\times 10^{-15} \ergpcmsqps$, we find excellent agreement with these
previous surveys. We confirm previous results (e.g., Luo et al. 2008),
that the CDF-N field may be subject to mild cosmic variance, as it
appears to over-estimate (a factor $\sim 1.5$--4) the number count
distribution of sources with $f_X \gtrsim 10^{-14}
\ergpcmsqps$. Furthermore, using the X-ray background (XRB) synthesis
models of \cite{gilli07}, we have simulated the expected ${\rm log} N
- {\rm log} S$ distribution of both obscured and unobscured
populations of AGN with $N_H \sim 10^{20}$--$10^{25}$~cm$^{-2}$, $L_X
\sim 10^{38}$--$10^{46} \Lsun$ in the redshift range $z \sim 0$--8. In
accordance with previous surveys, we consistently underestimate the
number counts of AGN with $f_X \lesssim 8 \times 10^{-15} \ergpcmsqps$
in comparison to that expected from the XRB (see upper panel of
Figure~\ref{fig:lognlogs}), suggesting that many heavily obscured
sources are still being missed in even the most sensitive
surveys. Indeed, multi-wavelength studies of deep and wide field X-ray
surveys find a large population of seemingly obscured AGN which remain
undetected using X-ray data alone (e.g.,
\citealt{alonso06,donley07,daddi07,melendez08a,fiore09,brusa10,goulding11,georgantopoulos11,dma11}).
However, for XDEEP2 sources with $f_X \gtrsim 2.5 \times 10^{-14}
\ergpcmsqps$, we find a mild ($\approx 30$--50\%) systematic offset
from previous X-ray surveys (e.g., E-CDFS; C-COSMOS), resulting in
number counts closer to those predicted by XRB models, although the
results from each of these surveys are all consistent at the 90\%
significance level.

\section{Optical DEEP2 \& X-ray XDEEP2 source identification} \label{sec:XOPT}

By design, the XDEEP2 {\it Chandra} survey is within the same spatial
region as the DEEP2 Galaxy Spectroscopic Redshift survey fields. In
this section we identify optical counterparts to the sources in the
XDEEP2 catalog using a custom Bayesian style analysis. For DEEP2,
optical $B$, $R$ and $I$-band photometry was obtained with the
Canada-France-Hawaii Telescope (CFHT) 12k camera. The main photometric
catalog contains over $>710$,000 sources with a typical absolute
astrometric accuracy of $\sim 0.2$ arc-seconds and is complete to
$R_{\rm AB} \sim 25.2$ (see \citealt{coil04}). In Table
\ref{tbl_opt_counter} we show the breakdown for the approximate number
of optical sources within the XDEEP2 survey fields. In DEEP2 Field 1,
all galaxies which have magnitudes of $R_{\rm AB} < 24.1$ were
targeted for spectroscopy using the DEIMOS spectrograph on Keck (see
\citealt{davis03} for further information on the observational setup
of DEEP2). However in Fields 2--4, only those galaxies which meet both
a simple $BRI$ color-cut threshold and have magnitudes of $R_{\rm AB}
< 24.1$ were targeted. The 4th data release of the DEEP2 spectroscopic
catalog contains 50,319 unique sources \citep{newman12}.

\subsection{A Bayesian optical--X-ray matching routine}

\begin{figure}[tb]
\includegraphics[width=0.97\linewidth]{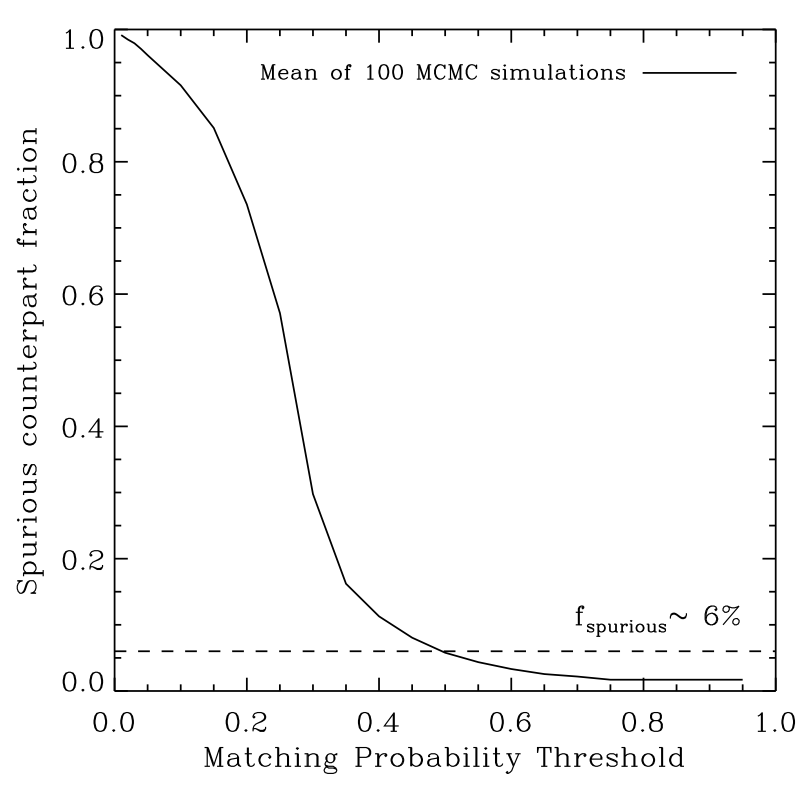}
\caption{Fraction of counterparts found between the DEEP2 optical
  catalog and 100 randomly simulated X-ray catalogs. We find that the
  fraction of spurious matches decreases rapidly as a function of the
  probability threshold ($P({\rm match})$) calculated in our
  Bayesian-style matching algorithm. For $P({\rm match}) = 0.46$, we
  expect a spurious matching fraction of $\lesssim 6$\% (dashed-line).}
\label{fig:optXspurious}
\end{figure}

Given the unique observational construction of the combined XDEEP2
survey, in that it is both relatively shallow in wide areas, while
simultaneously being extremely deep in smaller regions across the
fields, we require a method of source matching which will account for
changes in both the optical and X-ray source densities and
statistically associate bright X-ray sources in the shallow fields
with optical counterparts which are likely to contain bright AGN
(i.e., QSOs). To this end, we have extended the Bayesian
source-matching algorithm of Brand et al. (2006) to now include the
X-ray source density and the properties of the candidate optical
counterparts. Briefly, this method uses Bayesian-style statistics to
calculate the probability of a random association occurring between
two counterparts given the angular and magnitude distributions of the
optical sources in a specific region of the sky. Simultaneously this
algorithm accounts for the distribution of matching radii appropriate
for a given off-axis position of the X-ray source in a {\it Chandra}
observation. Furthermore, we allow modifications to the optical source
positions, assuming a Gaussian probability based on the centroid and
astrometric error of the DEEP2 data. As stated previously, median
offsets between DEEP2 and XDEEP2 have been removed a-priori (see
\S~\ref{sec:astrocalib}). We use a Gaussian prior based on the
characteristics of the {\it Chandra} PSF for the positional
uncertainty of the X-ray source to derive the probability, $f$ of an
X-ray source having an optical counterpart within the catalog (i.e., the
survey mean completeness). We combine these posterior assumptions with
information specific to the X-ray source (total counts, background
level, proximity to other X-ray sources) and the optical properties
(star, normal galaxy, quasar etc.) of possible counterparts to assign
likelihood association probabilities between pairs of sources. In our
new implementation of the algorithm, the probability of identifying an
X-ray source $i$ with optical source $k$ is then,

\begin{equation}
P_{ik,match} = f \frac{M_{ik}}{B_{k}} F_{ik} O_{ik} \left[ (1-f) + f \sum \limits_{l=1}^{n_{i}} \sum \limits_{k=1}^{n_{j}}  \frac{M_{il}}{B_l} F_{jk} \right]^{-1}
\end{equation}

where $M_{ik}$ is the simple Gaussian probability of associating an
X-ray source $i$ with an optical counterpart $k$ at a given separation
including the X-ray and optical positional uncertainties; $B_{k}$ are
the Poisson-idealized number counts as a function of optical magnitude
within a region encompassing the X-ray position, in effect, $B_{k}$
accounts for both the changing $R$-band magnitude depth and source
density within the optical DEEP2 catalog; $F_{ik}$ is the probability
that an X-ray source of a given flux and flux limit has an optical
association which is then marginalized over the $R$-band magnitude of
the proposed optical counterpart; and $O_{ik}$ is the probability
function containing the optical classification of the source, and is
essentially a weighting based on the probabilistic galaxy
classification of the source ($P({\rm gal})$ of 0 ($=$star) to 1
($=$galaxy) defined in Coil et~al. 2004) derived from the optical
photometry and SED fitting. We determine the priors for $F_{ik}$ by
randomly selecting from a cumulatively summed set of Poisson
distributions in Markov-Chain simulations of the X-ray and optical
catalogs. For computation speed, we limit the counterpart selection to
only optical sources detected in the $R$-band. This also conforms with
the selection method used to determine targets for optical
spectroscopy. We note here, that while this method increases our
ability to include optical sources with R-band magnitudes fainter than
the completeness limit of the DEEP2 survey ($R \sim 25.2$ mags), the
identification of X-ray sources with optically-faint counterparts is
still incomplete at $R \gtrsim 25.2$ (e.g., \citealt{dma01,brusa10}).

\begin{figure}[tb]
\includegraphics[width=0.97\linewidth]{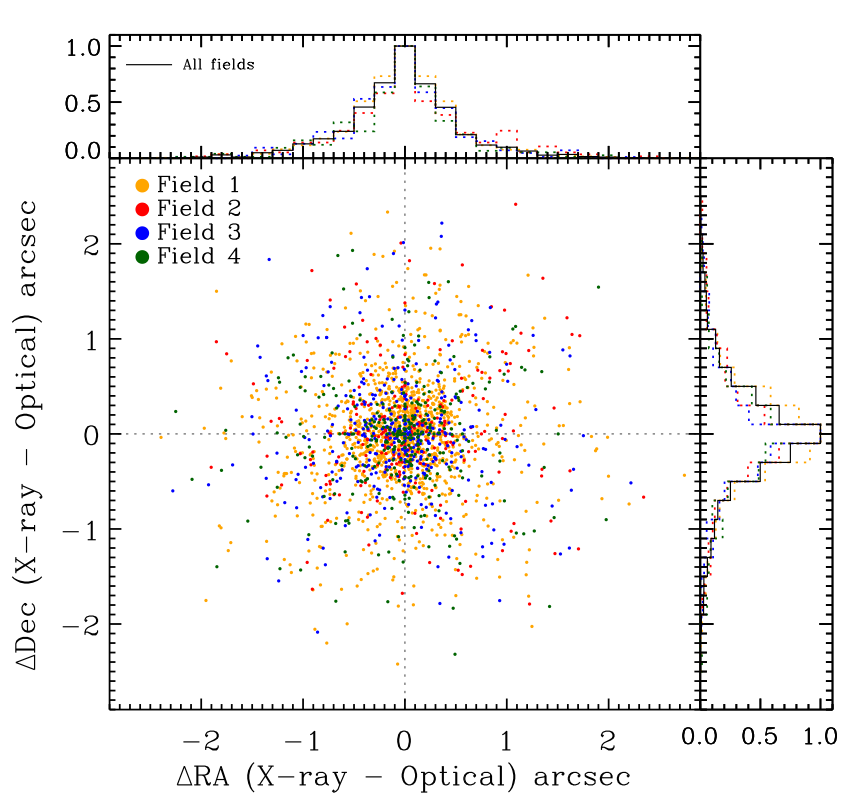}
\caption{Positional offset between optical and X-ray positions for the
  2126 XDEEP2 X-ray sources with secure DEEP2 optical counterparts
  found using our Bayesian-style matching algorithm. The spread in
  residuals is approximately Gaussian across all four DEEP2 fields
  with a mean positional offset of $\Delta_{\alpha,\delta} < 0.45$
  arc-seconds between the X-ray and optical source catalogs.}
\label{fig:radecoffset}
\end{figure}

To compute the probability threshold required to accept the optical
source as a counterpart to the X-ray source and to quantitatively
assess the false association fraction, we simulated mock XDEEP2
catalogs and compared them to the optical DEEP2 catalog. Following
Brand et al. (2006), we randomized the positions of the XDEEP2 sources
by $\pm 30''$ offsets and compared the number of false matches
produced. In Figure~\ref{fig:optXspurious} we show the behaviour of
the fraction of spurious counterparts for a given matching probability
threshold ($P_{\rm match}$) produced by our association routine. We
find that using $P_{\rm match} = 0.46$ produces one spurious optical
counterpart for $\lesssim 6$\% of the X-ray sources in the randomized
catalogs (see Figure \ref{fig:optXspurious}). The spurious counterpart
fraction of $\lesssim 6$\% is chosen specifically to be consistent
with that found for the previous {\it AEGIS-X} catalog which was
matched using the Maximum Likelihood technique (see
\citealt{civano12}, and references there-in); in turn, this also
allows for further comparison between the catalogs. In 100
Markov-Chain Monte-Carlo (MCMC) simulations, we find that the spurious
fraction remains relatively constant for $P_{\rm match} > 0.46$ across
all four XDEEP2 fields with an overall dispersion of $<1$\% within the
MCMC simulations.

\begin{table*}
\begin{center}
\caption{X-ray sources with optical counterparts\label{tbl_opt_counter}}
\begin{tabular}{ccccccc}
\tableline\tableline
\multicolumn{1}{c}{\textbf{Field \#}\tablenotemark{a}} &
\multicolumn{1}{c}{$N_{\rm X-ray,XDEEP2}$\tablenotemark{b}} &
\multicolumn{1}{c}{$N_{\rm opt, DEEP2}$\tablenotemark{c}} &
\multicolumn{1}{c}{$N_{\rm X-ray,opt}$\tablenotemark{d}} &
\multicolumn{1}{c}{$f_{\rm X-ray,opt}$\tablenotemark{e}} &
\multicolumn{1}{c}{Median $\Delta_{\rm RA}$\tablenotemark{f}} &
\multicolumn{1}{c}{Median $\Delta_{\rm Dec}$\tablenotemark{f}} \\
\tableline
1 & 1720 & $\sim 100,200$ & 1183 & 68.8 & -0.02 & 0.01 \\
2 &  342 & $\sim 119,400$ & 254  & 74.3 & 0.05 & 0.02 \\
3 &  528 & $\sim 146,100$ & 381  & 72.2 & 0.04 & 0.04 \\
4 &  386 & $\sim 145,300$ & 308  & 79.8 & 0.04 & -0.04 \\
\tableline
\end{tabular}
\end{center}
\footnotesize
{\sc Notes}-- \\
$^{a}$XDEEP2 field number \\
$^{b}$Number of X-ray sources in the XDEEP2 field \\
$^{c}$Approximate number of optical sources in the XDEEP2 field region\\
$^{d}$Number of X-ray sources with secure optical counterparts \\
$^{e}$Percentage fraction of X-ray sources with secure optical counterparts \\
$^{f}$Median positional offsets between the DEEP2 optical source co-ordinates and the X-ray source co-ordinates in arc-seconds 
\end{table*}

In Figure \ref{fig:radecoffset} we present the offset in astrometric
co-ordinates between the X-ray source position and that of the optical
counterpart from the XDEEP2 catalog. We find that the spread in
positional offsets is approximately Gaussian across all four DEEP2
fields with a mean positional offset of $\Delta_{\alpha,\delta} <
0.45$ arc-seconds with an approximately zero systematic offset between
the two catalogs. This mean offset is consistent with that found in
previous deep-wide surveys (e.g., C-COSMOS with 0.81'' for 90\% of the
sources; \citealt{elvis09,civano12}) Furthermore, we find that the
positional offset between the X-ray source and optical counterpart
appears to be a moderately-strong function of the ACIS-I off-axis
position with on-axis ($< 1.5'$) and off-axis ($> 6'$) X-ray sources
having median offsets of $\sim 0.28''$ and $\sim 0.96''$,
respectively.

\begin{figure}[tb]
\includegraphics[width=0.97\linewidth]{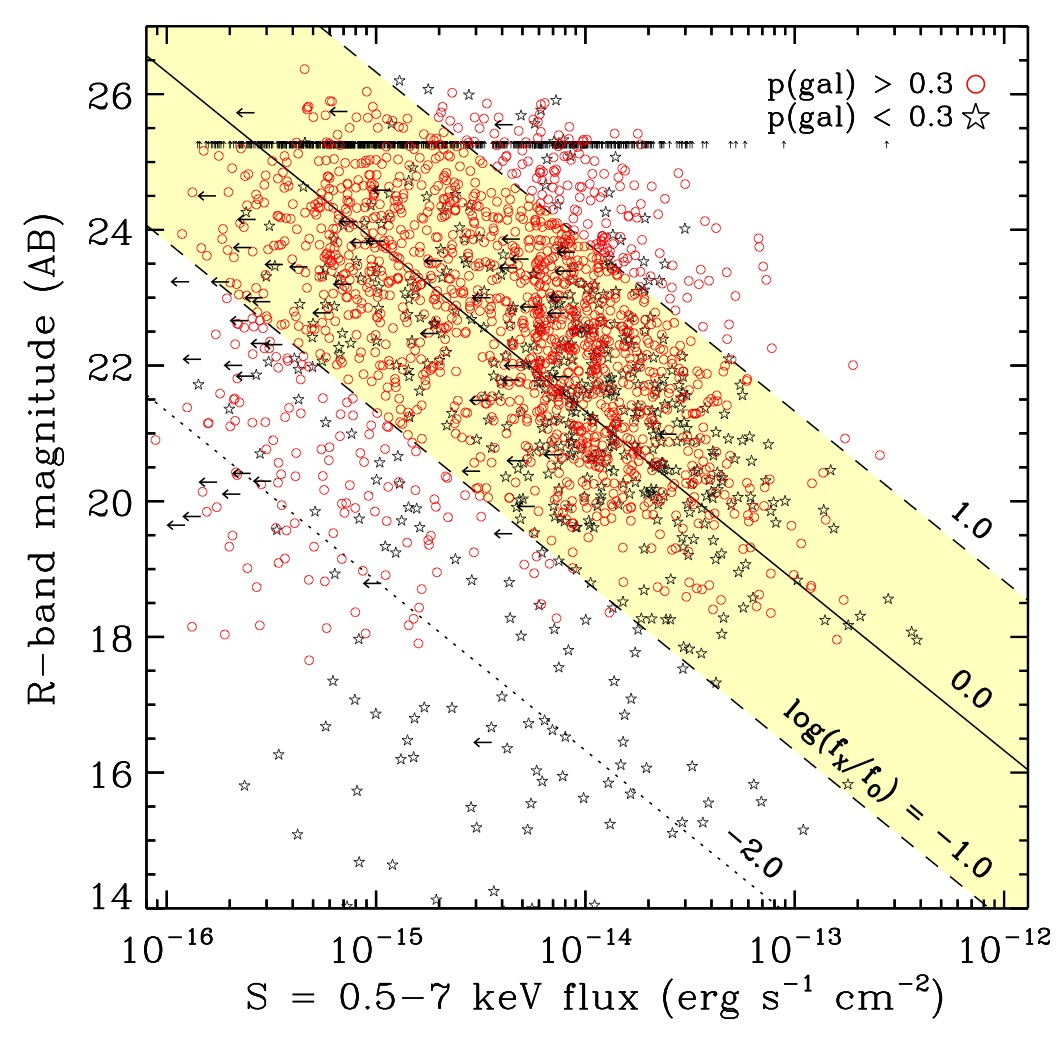}
\caption{R-band (AB) magnitude versus full-band (0.5--7.0 keV) flux
  for all XDEEP2 sources. X-ray sources are divided between those with
  galaxy probabilities ($P(gal)$) $> 0.3$ (i.e., optically extended
  sources; open circles) and $< 0.3$ (i.e., point-like sources; open
  stars). X-ray sources which lack optical counterparts are shown with
  upper-limits at $R=25.2$, i.e., the magnitude-limit of the DEEP2
  catalog.  Additionally, constant X-ray--optical flux ratios
  ($f_X/f_O$) are shown for log$(f_X/f_O) = \{-2.0;-1.0;0.0;1.0\}$,
  calculated using the relation of McHardy et al. (2003).}
\label{fig:fx_rmag}
\end{figure}

\begin{table*}
\begin{center}
\setlength{\tabcolsep}{1.5mm}
\caption{DEEP2 X-ray--optical counterpart catalog\label{tbl_xopt_counter}}
\begin{tabular}{ccccccccccccc}
\tableline\tableline
\multicolumn{1}{c}{} &
\multicolumn{1}{c}{} &
\multicolumn{2}{c}{X-ray} &
\multicolumn{1}{c}{} &
\multicolumn{2}{c}{Optical} &
\multicolumn{3}{c}{} &
\multicolumn{3}{c}{Photometry\tablenotemark{i}} \\
\multicolumn{1}{c}{XDEEP2\tablenotemark{a}} &
\multicolumn{1}{c}{Field\tablenotemark{b}} & 
\multicolumn{1}{c}{$\alpha_{\rm J2000}$\tablenotemark{c}} & 
\multicolumn{1}{c}{$\delta_{\rm J2000}$\tablenotemark{c}} & 
\multicolumn{1}{c}{DEEP2\tablenotemark{d}} & 
\multicolumn{1}{c}{$\alpha_{\rm J2000}$\tablenotemark{e}} & 
\multicolumn{1}{c}{$\delta_{\rm J2000}$\tablenotemark{e}} & 
\multicolumn{1}{c}{$d_{\rm OX}$\tablenotemark{f}} & 
\multicolumn{1}{c}{$z$\tablenotemark{g}} & 
\multicolumn{1}{c}{P(gal)\tablenotemark{h}} & 
\multicolumn{1}{c}{$B$} & 
\multicolumn{1}{c}{$R$} & 
\multicolumn{1}{c}{$I$} \\
\multicolumn{1}{c}{Name} &
\multicolumn{1}{c}{} &
\multicolumn{1}{c}{($^o$)} &
\multicolumn{1}{c}{($^o$)} &
\multicolumn{1}{c}{Objno} &
\multicolumn{1}{c}{($^o$)} &
\multicolumn{1}{c}{($^o$)} &
\multicolumn{1}{c}{('')} &
\multicolumn{1}{c}{} &
\multicolumn{1}{c}{} &
\multicolumn{3}{c}{(AB mag)} \\
\tableline

aeg1\_001 & 1 & 214.78246 & 52.99710 & - & - & - & - & - & - & - & - & - \\
aeg1\_002 & 1 & 214.78334 & 53.00712 & 13036677 & 214.78314 & 53.00728 & 0.72 & 0.5646 & 3 & 21.62 & 20.62 & 20.10 \\
aeg1\_003 & 1 & 214.79521 & 52.98033 & 13027633 & 214.79494 & 52.97998 & 1.38 & 0.7309 & 0.55 & 26.23 & 23.32 & 21.93 \\
aeg1\_004 & 1 & 214.79699 & 53.05600 & 13036612 & 214.79712 & 53.05598 & 0.29 & - & 3 & 20.31 & 18.55 & 17.91 \\
aeg1\_005 & 1 & 214.83506 & 53.04790 & - & - & - & - & - & - & - & - & - \\
aeg1\_006 & 1 & 214.83600 & 53.00792 & 13036601 & 214.83597 & 53.00809 & 0.63 & - & -2 & 16.67 & 16.36 & 16.23 \\
aeg1\_007 & 1 & 214.84506 & 53.02555 & 13035495 & 214.84502 & 53.02559 & 0.20 & - & 3 & 24.37 & 24.62 & 23.94 \\
aeg1\_008 & 1 & 214.85376 & 52.99871 & 13027346 & 214.85321 & 52.99912 & 1.90 & - & 3 & 24.13 & 23.49 & 23.26 \\
aeg1\_009 & 1 & 214.85694 & 53.00549 & 13100779 & 214.85669 & 53.00582 & 1.28 & - & 3 & 24.69 & 24.47 & 24.27 \\
aeg1\_010 & 1 & 214.85765 & 53.01971 & 13035756 & 214.85777 & 53.02011 & 1.45 & - & -2 & 22.79 & 21.16 & 20.41 \\
aeg1\_011 & 1 & 214.86239 & 53.03122 & 13035995 & 214.86253 & 53.03141 & 0.73 & - & 3 & 23.59 & 21.68 & 21.09 \\
aeg1\_012 & 1 & 214.86615 & 53.02515 & - & - & - & - & - & - & - & - & - \\
aeg1\_013 & 1 & 214.86670 & 52.97822 & 13027372 & 214.86674 & 52.97823 & 0.10 & 0.5608 & 0.81 & 23.00 & 22.87 & 22.61 \\
aeg1\_014 & 1 & 214.87337 & 53.03977 & 13035981 & 214.87335 & 53.03982 & 0.19 & - & 3 & 25.15 & 23.66 & 22.84 \\
aeg1\_015 & 1 & 214.87634 & 53.04383 & 13035650 & 214.87601 & 53.04362 & 1.03 & 0.3722 & 3 & 26.80 & 23.31 & 21.95 \\
aeg1\_016 & 1 & 214.87842 & 53.00748 & 13035444 & 214.87830 & 53.00769 & 0.81 & - & 1.00 & 26.06 & 23.14 & 22.27 \\
aeg1\_017 & 1 & 214.87886 & 52.98781 & 13027475 & 214.87888 & 52.98786 & 0.20 & - & 0.00 & 23.15 & 20.26 & 18.03 \\
aeg1\_018 & 1 & 214.88704 & 53.04167 & - & - & - & - & - & - & - & - & - \\
aeg1\_019 & 1 & 214.88706 & 52.99963 & 13027149 & 214.88704 & 52.99970 & 0.25 & - & 3 & 24.25 & 23.77 & 22.93 \\
aeg1\_020 & 1 & 214.88917 & 53.09005 & - & - & - & - & - & - & - & - & - \\

\tableline
\end{tabular}
\end{center}
\footnotesize
{\sc Notes}-- \\
$^{a}$XDEEP2 unique source identifier \\
$^{b}$XDEEP2 field number \\
$^{c}$X-ray source position in J2000 co-ordinates (degrees) \\
$^{d}$DEEP2 optical source identifier (Coil et al. 2004)  \\
$^{e}$Optical source position in J2000 co-ordinates (degrees) \\
$^{f}$Angular separation between optical and X-ray source positions (arc-seconds) \\
$^{g}$Redshift of optical counterpart \\
$^{h}$Bayesian probability of being a galaxy based on $R$-band image ($\leq 0$: star/compact; $\geq 1$: galaxy/extended; see Coil et al. 2004) \\
$^{i}$Optical photometry in the $B$, $R$ and $I$-bands (AB-magnitude)
\end{table*}

\subsection{X-ray--optical source properties}

Of the 2976 X-ray sources in XDEEP2, we find that 2126 ($\approx 71.4
\pm 2.8$\%) have at least one secure optical counterpart in the DEEP2
optical catalog. Multiple candidate counterparts are found for
$\approx 11$\% of the X-ray sources in XDEEP2. When multiple optical
counterparts are associated with one X-ray source, we accept the DEEP2
optical counterpart with the largest $P_{\rm match}$. Given the
cumulative distribution of $P_{\rm match}$ found for the XDEEP2
counterpart catalog, we expect a final spurious counterpart fraction
of $\approx 4$\%. In Table \ref{tbl_opt_counter} we show the breakdown
by field of the number of X-ray sources with optical counterparts, the
percentage identified and the median positional offset between the
optical and X-ray source positions. We find that 943 ($\approx
75.1$\%) of the X-ray sources in Fields 2--4 have secure optical
counterparts compared with 1183 ($\approx 68.8$\%) in Field 1. This
higher fraction of secure counterparts in Fields 2--4 is, in all
likelihood, due to the relatively shallow exposure of the {\it
  Chandra} observations in Fields 2--4 compared to those in Field 1,
and hence, brighter X-ray sources tending towards bright optical host
galaxies (i.e., X-ray-to-optical flux ratios $\sim 1$--10) which has
been found previously in very shallow wide-field X-ray surveys (e.g.,
\citealt{maccacaro88,stocke91,akiyama00,lehmann01,murray05}). Indeed,
AGN and QSOs are typically found to have similar ratios of $-1
<$~log$(f_X/f_O) < +1$ (e.g.,
\citealt{schmidt98,akiyama00,lehmann01}). In Figure~\ref{fig:fx_rmag}
we show the full-band X-ray flux versus the DEEP2 $R$-band magnitude
for the sources with secure optical counterparts. We illustrate
approximate X-ray-to-optical flux ratios for the sources assuming the
relation of \citet{mchardy03}, and we divide the sample between those
optical sources identified in DEEP2 to be extended/galaxy ($P({\rm
  gal}) > 0.3$; see \citealt{coil04,newman12}) and point-like sources
(stellar or QSO; $P({\rm gal}) < 0.3$). Of the 1559 optically extended
X-ray-optical sources, $\approx 90$\% (1425) have log$(f_X/f_O) > -1$,
suggesting a significant fraction are bright AGN. We also find that 77
X-ray sources are also detected with very low X-ray-to-optical flux
ratios (i.e., log$(f_X/f_O) < -2$. These X-ray sources generally
include normal galaxies, stars, and low-luminosity AGN, and as we show
in Figure~\ref{fig:fx_rmag}, all 77 X-ray--optical sources with
$R_{\rm AB} < 18$ are point-like suggesting a stellar origin for the
X-ray emission. As is clearly evident from the distribution of
galaxies in Figure~\ref{fig:fx_rmag}, our X-ray--optical source
matching becomes incomplete towards optically-faint ($R \gtrsim 25$)
systems for $f_X \lesssim 6 \times 10^{-15} \ergpcmsqps$ due to the
flux-limit of the optical DEEP2 data when compared to the depth of the
X-ray observations within Field 1.

\begin{figure}[tb]
\includegraphics[width=0.97\linewidth]{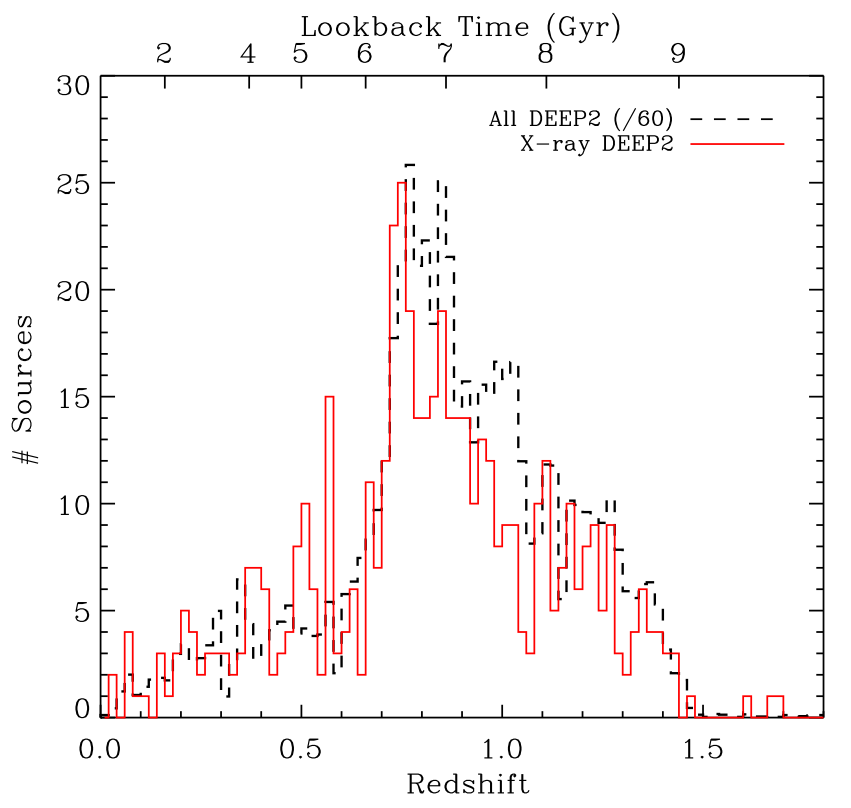}
\caption{Redshift histograms for all 510 XDEEP2 galaxies with optical
  spectroscopic counterparts (solid-line) and all optical DEEP2
  galaxies (dashed-line). The optical DEEP2 distribution is divided by
  a factor for 60 for ease of comparison to the X-ray sources. On the
  top-axis we show the present-day look-back times as a function of
  redshift, with $z=0$ equivalent to $\tau_{lb} = 0$.}
\label{fig:redshift_hist}
\end{figure}

We also consider the $\approx 450,000$ optical galaxies identified in
the CFHT Legacy Survey Deep 3 (CFHTLSD-3) field, which covers an area
roughly coincident with the AEGIS 1--3 sub-fields and is complete to
$i'_{\rm AB} < 27.0$ with sources detected down to $i'_{\rm AB} \sim
28.6$ (\citealt{ilbert06}; i.e., complete to $\sim 2$ magnitudes
deeper than DEEP2). We find that 1009 X-ray sources have optical
counterparts in the CFHTLSD-3 and 228/1009 were not previously
identified in the DEEP2 catalog. Of the 228 X-ray sources, which were
not previously identified to have optical DEEP2 counterparts, 163/228
have $i'_{\rm AB} > 24.4$ and 118 have $i'_{\rm AB} > 25.0$. With the
subsequent inclusion of the CFHTLSD-3, we find an X-ray--optical
counterpart fraction of $\approx 82$\% within Field 1.

In Table~\ref{tbl_xopt_counter} we provide the matching optical DEEP2
counterpart information for the entire XDEEP2 catalog (e.g., X-ray
name; X-ray position; optical DEEP2 counterpart; positional offset;
basic optical properties). Furthermore, to guide future
multi-wavelength surveys, we additionally include the spectroscopic
redshift information from the recently released DEEP2 DR4 catalog
(Newman et al. 2012). Of the 2126 X-ray sources with optical
counterparts, 700 are included as part of the spectroscopic catalog
and 510/700 have secure extragalactic redshifts, with the majority in
the range $0.3 \lesssim z \lesssim 1.4$ and the highest redshift
source at $z \sim 3.04$. We show in Figure~\ref{fig:redshift_hist}
that the X-ray sub-sample follows a similar redshift distribution to
the main parent DEEP2 redshift catalog. Hence, in redshift terms, the
X-ray sources may be considered a representative sample of the overall
galaxy population in DEEP2. Future in-depth analyses of the AGN and
galaxy redshift populations will allow us to understand the AGN
clustering properties and possible correlations of AGN presence and
large-scale structures.

\section{Summary} \label{sec:summary}

We have presented the X-ray source catalog and basic analyses of
sources detected in the $\approx 10$ks--1.1~Ms {\it Chandra} ACIS-I
observations of the four X-ray DEEP2 (XDEEP2) survey fields. The total
area of XDEEP2 is $\sim 3.2$~deg$^2$, and to date is the largest
medium-deep {\it Chandra} X-ray survey constructed. Using wavelet
decomposition software ({\tt wvdecomp}), we detected X-ray point
sources in the individual (non-merged) events and overlapping merged
images in the 0.5--2~keV (soft-band [SB]), 2--7~keV (hard-band [HB])
and 0.5--7~keV (full-band [FB]) energy ranges, complete to a
false-source probability threshold of $1 \times 10^{-6}$. When
considering the survey regions where at least 10\% of the area is
sensitive, the flux limits in the merged observations are $f_{X,FB} >
2.8 \times 10^{-16} \ergpcmsqps$, $f_{X,FB} > 4.5 \times 10^{-15}
\ergpcmsqps$, $f_{X,FB} > 4.6 \times 10^{-15} \ergpcmsqps$ and
$f_{X,FB} > 4.6 \times 10^{-15} \ergpcmsqps$ in XDEEP2 Fields 1, 2, 3
and 4, respectively. The full XDEEP2 point source catalog contains
2976 sources, with 1720, 342, 528 and 386 sources in Fields 1--4. For
the detected sources, we have presented the flux band ratio ($f_{\rm
  HB} / f_{\rm SB}$) distributions. Consistent with previous results,
we confirm that low flux X-ray sources tend towards higher flux ratios
($f_{\rm HB} / f_{\rm SB} \sim 2$--10), consistent with that expected
for flatter spectral slopes with $\Gamma \sim 1.2$--1.4.

We have performed a rigorous comparison between our new catalog of
Field 1 and that previously presented in Laird et al. (2009). Our new
catalog now contains the more recent 600~ks observations of three
sub-fields within Field 1. We find excellent agreement between the two
catalogs, and show that 96\% of the sources identified in the previous
catalog, using a substantially different detection technique, are also
identified in the new catalog of Field 1. Through extensive source
detection simulations, we suggest that the small $\approx 4$\%
discrepancy between the catalogs can be mainly attributed to our
conservative removal of low-significance and possibly spurious
sources. Indeed, with the inclusion of the low significance X-ray
sources, we show that $\sim 99$\% of the sources identified by Laird
et al. would be identified here. Furthermore, we present a comparison
between the {\it Chandra} Source Catalog (CSC) and the X-ray sources
identified in the more shallow 10ks Fields 2, 3 and 4. We find that
$\sim 41.9 \pm 2.2$\% of the XDEEP2 sources within these fields are
included in the CSC. The vast majority ($\approx 90$\%) of the XDEEP2
sources not identified in the CSC fall below their conservative
detection threshold. We have presented the combined log N -- log S
distribution of soft-band detected sources identified across the
XDEEP2 fields; the distribution shows excellent agreement with the
Extended {\it Chandra} Deep Field and {\it Chandra}-COSMOS fields to
$f_{\rm X,0.5-2keV} \sim 2 \times 10^{-16} \ergpcmsqps$. Given the
large survey area of XDEEP2, we additionally place relatively strong
constraints on the log N -- log S distribution at high fluxes ($f_{\rm
  X,0.5-2keV} > 2 \times 10^{-14} \ergpcmsqps$), and find a small
systematic offset (a factor $\sim 1.5$) towards lower source numbers
in the high-flux regime than observed previously in smaller area
surveys. The number counts for sources with $f_{\rm 0.5-2keV} > 2
\times 10^{-14} \ergpcmsqps$ are in close agreement with the X-ray
background synthesis models of Gilli et al. (2007). However, based on
our careful analyses of the uncertainty associated with the log N --
log S distribution, derived through the use of a Monte-Carlo
simulation, we find that at the 90\% level we cannot reject the number
count distribution predicted by the previous surveys.

We have additionally built upon a previous Bayesian-style method for
associating the X-ray sources with their optical counterparts
\citep{brand06} in the DEEP2 photometric catalog (complete to $R_{\rm
  AB} < 25.2$; Coil et al. 2004), and find that 2126 of the X-ray
sources presented here ($\approx 71.4 \pm 2.8$\%) have at least one
secure optical counterpart. However, due to the much deeper X-ray
exposure regions, we find a lower fraction of optical counterparts in
Field 1 ($\approx 68.8$\%) compared with Fields 2--4 ($\approx
75.1$\%). We have additionally presented the optical photometric
properties of the X-ray sources, the X-ray-to-optical ratios and find
that the XDEEP2 sample have a similar redshift distribution to the
main optical DEEP2 parent catalog, in the range $0 < z < 3$.

\acknowledgments

We would like to thank the anonymous referee for their considered and
comprehensive report, which has allowed us to greatly improve and
qualify many aspects of the X-ray catalog and analysis. We are
thankful to K. Nandra, F. Civano and N. Wright for helpful discussions
that have allowed us to clarify our analyses throughout the
manuscript. We are also grateful to B. Lehmer for kindly providing
data from the Extended Chandra Deep Field. This research has made use
of data obtained from the Chandra Source Catalog, provided by the
Chandra X-ray Center (CXC) as part of the Chandra Data Archive.

{\it Facilities:} \facility{CXO (ACIS)}.




\bibliography{bibtex1}

\begin{thebibliography}{91}
\expandafter\ifx\csname natexlab\endcsname\relax\def\natexlab#1{#1}\fi

\bibitem[{{Akiyama} {et~al.}(2000){Akiyama}, {Ohta}, {Yamada}, {Kashikawa},
  {Yagi}, {Kawasaki}, {Sakano}, {Tsuru}, {Ueda}, {Takahashi}, {Lehmann},
  {Hasinger}, \& {Voges}}]{akiyama00}
{Akiyama}, M., {Ohta}, K., {Yamada}, T., {et~al.} 2000, \apj, 532, 700

\bibitem[{{Alexander} {et~al.}(2001){Alexander}, {Brandt}, {Hornschemeier},
  {Garmire}, {Schneider}, {Bauer}, \& {Griffiths}}]{dma01}
{Alexander}, D.~M., {Brandt}, W.~N., {Hornschemeier}, A.~E., {et~al.} 2001,
  \aj, 122, 2156

\bibitem[{{Alexander} {et~al.}(2003{\natexlab{a}}){Alexander}, {Bauer},
  {Brandt}, {Schneider}, {Hornschemeier}, {Vignali}, {Barger}, {Broos},
  {Cowie}, {Garmire}, {Townsley}, {Bautz}, {Chartas}, \& {Sargent}}]{dma03a}
{Alexander}, D.~M., {Bauer}, F.~E., {Brandt}, W.~N., {et~al.}
  2003{\natexlab{a}}, \aj, 126, 539

\bibitem[{{Alexander} {et~al.}(2003{\natexlab{b}}){Alexander}, {Bauer},
  {Brandt}, {Hornschemeier}, {Vignali}, {Garmire}, {Schneider}, {Chartas}, \&
  {Gallagher}}]{dma03b}
---. 2003{\natexlab{b}}, \aj, 125, 383

\bibitem[{{Alexander} {et~al.}(2011){Alexander}, {Bauer}, {Brandt}, {Daddi},
  {Hickox}, {Lehmer}, {Luo}, {Xue}, {Young}, {Comastri}, {Del Moro}, {Fabian},
  {Gilli}, {Goulding}, {Mainieri}, {Mullaney}, {Paolillo}, {Rafferty},
  {Schneider}, {Shemmer}, \& {Vignali}}]{dma11}
---. 2011, \apj, 738, 44

\bibitem[{{Alonso-Herrero} {et~al.}(2006){Alonso-Herrero},
  {P{\'e}rez-Gonz{\'a}lez}, {Alexander}, {Rieke}, {Rigopoulou}, {Le Floc'h},
  {Barmby}, {Papovich}, {Rigby}, {Bauer}, {Brandt}, {Egami}, {Willner}, {Dole},
  \& {Huang}}]{alonso06}
{Alonso-Herrero}, A., {P{\'e}rez-Gonz{\'a}lez}, P.~G., {Alexander}, D.~M.,
  {et~al.} 2006, \apj, 640, 167

\bibitem[{{Barger} {et~al.}(2005){Barger}, {Cowie}, {Mushotzky}, {Yang},
  {Wang}, {Steffen}, \& {Capak}}]{barger05}
{Barger}, A.~J., {Cowie}, L.~L., {Mushotzky}, R.~F., {et~al.} 2005, \aj, 129,
  578

\bibitem[{{Bauer} {et~al.}(2011){Bauer}, {Gr{\"u}tzbauch}, {J{\o}rgensen},
  {Varela}, \& {Bergmann}}]{bauer11}
{Bauer}, A.~E., {Gr{\"u}tzbauch}, R., {J{\o}rgensen}, I., {Varela}, J., \&
  {Bergmann}, M. 2011, \mnras, 411, 2009

\bibitem[{{Bauer} {et~al.}(2002){Bauer}, {Alexander}, {Brandt},
  {Hornschemeier}, {Miyaji}, {Garmire}, {Schneider}, {Bautz}, {Chartas},
  {Griffiths}, \& {Sargent}}]{bauer02}
{Bauer}, F.~E., {Alexander}, D.~M., {Brandt}, W.~N., {et~al.} 2002, \aj, 123,
  1163

\bibitem[{{Bell} {et~al.}(2004){Bell}, {Wolf}, {Meisenheimer}, {Rix}, {Borch},
  {Dye}, {Kleinheinrich}, {Wisotzki}, \& {McIntosh}}]{bell04}
{Bell}, E.~F., {Wolf}, C., {Meisenheimer}, K., {et~al.} 2004, \apj, 608, 752

\bibitem[{{Boyle} \& {Terlevich}(1998)}]{boyle98}
{Boyle}, B.~J., \& {Terlevich}, R.~J. 1998, \mnras, 293, L49

\bibitem[{{Brand} {et~al.}(2005){Brand}, {Dey}, {Brown}, {Watson}, {Jannuzi},
  {Najita}, {Kochanek}, {Shields}, {Fazio}, {Forman}, {Green}, {Jones},
  {Kenter}, {McNamara}, {Murray}, {Rieke}, \& {Vikhlinin}}]{brand05}
{Brand}, K., {Dey}, A., {Brown}, M.~J.~I., {et~al.} 2005, \apj, 626, 723

\bibitem[{{Brand} {et~al.}(2006){Brand}, {Brown}, {Dey}, {Jannuzi}, {Kochanek},
  {Kenter}, {Fabricant}, {Fazio}, {Forman}, {Green}, {Jones}, {McNamara},
  {Murray}, {Najita}, {Rieke}, {Shields}, \& {Vikhlinin}}]{brand06}
{Brand}, K., {Brown}, M.~J.~I., {Dey}, A., {et~al.} 2006, \apj, 641, 140

\bibitem[{{Brandt} \& {Hasinger}(2005)}]{brandt05}
{Brandt}, W.~N., \& {Hasinger}, G. 2005, \araa, 43, 827

\bibitem[{{Brusa} {et~al.}(2010){Brusa}, {Civano}, {Comastri}, {Miyaji},
  {Salvato}, {Zamorani}, {Cappelluti}, {Fiore}, {Hasinger}, {Mainieri},
  {Merloni}, {Bongiorno}, {Capak}, {Elvis}, {Gilli}, {Hao}, {Jahnke},
  {Koekemoer}, {Ilbert}, {Le Floc'h}, {Lusso}, {Mignoli}, {Schinnerer},
  {Silverman}, {Treister}, {Trump}, {Vignali}, {Zamojski}, {Aldcroft},
  {Aussel}, {Bardelli}, {Bolzonella}, {Cappi}, {Caputi}, {Contini},
  {Finoguenov}, {Fruscione}, {Garilli}, {Impey}, {Iovino}, {Iwasawa},
  {Kampczyk}, {Kartaltepe}, {Kneib}, {Knobel}, {Kovac}, {Lamareille},
  {Leborgne}, {Le Brun}, {Le Fevre}, {Lilly}, {Maier}, {McCracken}, {Pello},
  {Peng}, {Perez-Montero}, {de Ravel}, {Sanders}, {Scodeggio}, {Scoville},
  {Tanaka}, {Taniguchi}, {Tasca}, {de la Torre}, {Tresse}, {Vergani}, \&
  {Zucca}}]{brusa10}
{Brusa}, M., {Civano}, F., {Comastri}, A., {et~al.} 2010, \apj, 716, 348

\bibitem[{{Cappelluti} {et~al.}(2010){Cappelluti}, {Ajello}, {Burlon},
  {Krumpe}, {Miyaji}, {Bonoli}, \& {Greiner}}]{cappelluti10}
{Cappelluti}, N., {Ajello}, M., {Burlon}, D., {et~al.} 2010, \apjl, 716, L209

\bibitem[{{Civano} {et~al.}(2012){Civano}, {Elvis}, {Brusa}, {Comastri},
  {Salvato}, {Zamorani}, {Aldcroft}, {Bongiorno}, {Capak}, {Cappelluti},
  {Cisternas}, {Fiore}, {Fruscione}, {Hao}, {Kartaltepe}, {Koekemoer}, {Gilli},
  {Impey}, {Lanzuisi}, {Lusso}, {Mainieri}, {Miyaji}, {Lilly}, {Masters},
  {Puccetti}, {Schawinski}, {Scoville}, {Silverman}, {Trump}, {Urry},
  {Vignali}, \& {Wright}}]{civano12}
{Civano}, F., {Elvis}, M., {Brusa}, M., {et~al.} 2012, ArXiv 1205.5030

\bibitem[{{Coil} {et~al.}(2004){Coil}, {Newman}, {Kaiser}, {Davis}, {Ma},
  {Kocevski}, \& {Koo}}]{coil04}
{Coil}, A.~L., {Newman}, J.~A., {Kaiser}, N., {et~al.} 2004, \apj, 617, 765

\bibitem[{{Coil} {et~al.}(2009){Coil}, {Georgakakis}, {Newman}, {Cooper},
  {Croton}, {Davis}, {Koo}, {Laird}, {Nandra}, {Weiner}, {Willmer}, \&
  {Yan}}]{coil09}
{Coil}, A.~L., {Georgakakis}, A., {Newman}, J.~A., {et~al.} 2009, \apj, 701,
  1484

\bibitem[{{Cooper} {et~al.}(2005){Cooper}, {Newman}, {Madgwick}, {Gerke},
  {Yan}, \& {Davis}}]{cooper05}
{Cooper}, M.~C., {Newman}, J.~A., {Madgwick}, D.~S., {et~al.} 2005, \apj, 634,
  833

\bibitem[{{Cooper} {et~al.}(2006){Cooper}, {Newman}, {Croton}, {Weiner},
  {Willmer}, {Gerke}, {Madgwick}, {Faber}, {Davis}, {Coil}, {Finkbeiner},
  {Guhathakurta}, \& {Koo}}]{cooper06}
{Cooper}, M.~C., {Newman}, J.~A., {Croton}, D.~J., {et~al.} 2006, \mnras, 370,
  198

\bibitem[{{Daddi} {et~al.}(2007){Daddi}, {Alexander}, {Dickinson}, {Gilli},
  {Renzini}, {Elbaz}, {Cimatti}, {Chary}, {Frayer}, {Bauer}, {Brandt},
  {Giavalisco}, {Grogin}, {Huynh}, {Kurk}, {Mignoli}, {Morrison}, {Pope}, \&
  {Ravindranath}}]{daddi07}
{Daddi}, E., {Alexander}, D.~M., {Dickinson}, M., {et~al.} 2007, \apj, 670, 173

\bibitem[{{Davis} {et~al.}(2003){Davis}, {Faber}, {Newman}, {Phillips},
  {Ellis}, {Steidel}, {Conselice}, {Coil}, {Finkbeiner}, {Koo}, {Guhathakurta},
  {Weiner}, {Schiavon}, {Willmer}, {Kaiser}, {Luppino}, {Wirth}, {Connolly},
  {Eisenhardt}, {Cooper}, \& {Gerke}}]{davis03}
{Davis}, M., {Faber}, S.~M., {Newman}, J., {et~al.} 2003, in Society of
  Photo-Optical Instrumentation Engineers (SPIE) Conference Series, Vol. 4834,
  Society of Photo-Optical Instrumentation Engineers (SPIE) Conference Series,
  ed. {P.~Guhathakurta}, 161--172

\bibitem[{{Dom{\'{\i}}nguez S{\'a}nchez} {et~al.}(2011){Dom{\'{\i}}nguez
  S{\'a}nchez}, {Pozzi}, {Gruppioni}, {Cimatti}, {Ilbert}, {Pozzetti},
  {McCracken}, {Capak}, {Le Floch}, {Salvato}, {Zamorani}, {Carollo},
  {Contini}, {Kneib}, {Le F{\`e}vre}, {Lilly}, {Mainieri}, {Renzini},
  {Scodeggio}, {Bardelli}, {Bolzonella}, {Bongiorno}, {Caputi}, {Coppa},
  {Cucciati}, {de la Torre}, {de Ravel}, {Franzetti}, {Garilli}, {Iovino},
  {Kampczyk}, {Knobel}, {Kova{\v c}}, {Lamareille}, {Le Borgne}, {Le Brun},
  {Maier}, {Mignoli}, {Pell{\'o}}, {Peng}, {Perez-Montero}, {Ricciardelli},
  {Silverman}, {Tanaka}, {Tasca}, {Tresse}, {Vergani}, \&
  {Zucca}}]{dominguez11}
{Dom{\'{\i}}nguez S{\'a}nchez}, H., {Pozzi}, F., {Gruppioni}, C., {et~al.}
  2011, \mnras, 417, 900

\bibitem[{{Done} {et~al.}(1996){Done}, {Madejski}, \& {Smith}}]{done96}
{Done}, C., {Madejski}, G.~M., \& {Smith}, D.~A. 1996, \apjl, 463, L63+

\bibitem[{{Donley} {et~al.}(2007){Donley}, {Rieke}, {P{\'e}rez-Gonz{\'a}lez},
  {Rigby}, \& {Alonso-Herrero}}]{donley07}
{Donley}, J.~L., {Rieke}, G.~H., {P{\'e}rez-Gonz{\'a}lez}, P.~G., {Rigby},
  J.~R., \& {Alonso-Herrero}, A. 2007, \apj, 660, 167

\bibitem[{{Ebeling} {et~al.}(2006){Ebeling}, {White}, \&
  {Rangarajan}}]{ebeling06}
{Ebeling}, H., {White}, D.~A., \& {Rangarajan}, F.~V.~N. 2006, \mnras, 368, 65

\bibitem[{{Elvis} {et~al.}(2009){Elvis}, {Civano}, {Vignali}, {Puccetti},
  {Fiore}, {Cappelluti}, {Aldcroft}, {Fruscione}, {Zamorani}, {Comastri},
  {Brusa}, {Gilli}, {Miyaji}, {Damiani}, {Koekemoer}, {Finoguenov}, {Brunner},
  {Urry}, {Silverman}, {Mainieri}, {Hasinger}, {Griffiths}, {Carollo}, {Hao},
  {Guzzo}, {Blain}, {Calzetti}, {Carilli}, {Capak}, {Ettori}, {Fabbiano},
  {Impey}, {Lilly}, {Mobasher}, {Rich}, {Salvato}, {Sanders}, {Schinnerer},
  {Scoville}, {Shopbell}, {Taylor}, {Taniguchi}, \& {Volonteri}}]{elvis09}
{Elvis}, M., {Civano}, F., {Vignali}, C., {et~al.} 2009, \apjs, 184, 158

\bibitem[{{Evans} {et~al.}(2010){Evans}, {Primini}, {Glotfelty}, {Anderson},
  {Bonaventura}, {Chen}, {Davis}, {Doe}, {Evans}, {Fabbiano}, {Galle}, {Gibbs},
  {Grier}, {Hain}, {Hall}, {Harbo}, {(Helen He}, {Houck}, {Karovska},
  {Kashyap}, {Lauer}, {McCollough}, {McDowell}, {Miller}, {Mitschang},
  {Morgan}, {Mossman}, {Nichols}, {Nowak}, {Plummer}, {Refsdal}, {Rots},
  {Siemiginowska}, {Sundheim}, {Tibbetts}, {Van Stone}, {Winkelman}, \&
  {Zografou}}]{evans10}
{Evans}, I.~N., {Primini}, F.~A., {Glotfelty}, K.~J., {et~al.} 2010, \apjs,
  189, 37

\bibitem[{{Faber} {et~al.}(2007){Faber}, {Willmer}, {Wolf}, {Koo}, {Weiner},
  {Newman}, {Im}, {Coil}, {Conroy}, {Cooper}, {Davis}, {Finkbeiner}, {Gerke},
  {Gebhardt}, {Groth}, {Guhathakurta}, {Harker}, {Kaiser}, {Kassin},
  {Kleinheinrich}, {Konidaris}, {Kron}, {Lin}, {Luppino}, {Madgwick},
  {Meisenheimer}, {Noeske}, {Phillips}, {Sarajedini}, {Schiavon}, {Simard},
  {Szalay}, {Vogt}, \& {Yan}}]{faber07}
{Faber}, S.~M., {Willmer}, C.~N.~A., {Wolf}, C., {et~al.} 2007, \apj, 665, 265

\bibitem[{{Fassbender} {et~al.}(2011){Fassbender}, {Nastasi}, {B{\"o}hringer},
  {{\v S}uhada}, {Santos}, {Rosati}, {Pierini}, {M{\"u}hlegger}, {Quintana},
  {Schwope}, {Lamer}, {de Hoon}, {Kohnert}, {Pratt}, \& {Mohr}}]{fassbender11}
{Fassbender}, R., {Nastasi}, A., {B{\"o}hringer}, H., {et~al.} 2011, \aap, 527,
  L10+

\bibitem[{{Fiore} {et~al.}(2009){Fiore}, {Puccetti}, {Brusa}, {Salvato},
  {Zamorani}, \& {et~al.}}]{fiore09}
{Fiore}, F., {Puccetti}, S., {Brusa}, M., {et~al.} 2009, \apj, 693, 447

\bibitem[{{Franceschini} {et~al.}(1999){Franceschini}, {Hasinger}, {Miyaji}, \&
  {Malquori}}]{Franceschini99}
{Franceschini}, A., {Hasinger}, G., {Miyaji}, T., \& {Malquori}, D. 1999,
  \mnras, 310, L5

\bibitem[{{Fukazawa} {et~al.}(2001){Fukazawa}, {Iyomoto}, {Kubota},
  {Matsumoto}, \& {Makishima}}]{fukazawa01}
{Fukazawa}, Y., {Iyomoto}, N., {Kubota}, A., {Matsumoto}, Y., \& {Makishima},
  K. 2001, \aap, 374, 73

\bibitem[{{Gehrels}(1986)}]{gehrels86}
{Gehrels}, N. 1986, \apj, 303, 336

\bibitem[{{Georgakakis} {et~al.}(2008){Georgakakis}, {Nandra}, {Laird}, {Aird},
  \& {Trichas}}]{georgakakis08}
{Georgakakis}, A., {Nandra}, K., {Laird}, E.~S., {Aird}, J., \& {Trichas}, M.
  2008, \mnras, 388, 1205

\bibitem[{{Georgantopoulos} {et~al.}(2009){Georgantopoulos}, {Akylas},
  {Georgakakis}, \& {Rowan-Robinson}}]{georgantopoulos09}
{Georgantopoulos}, I., {Akylas}, A., {Georgakakis}, A., \& {Rowan-Robinson}, M.
  2009, \aap, 507, 747

\bibitem[{{Georgantopoulos} {et~al.}(2011){Georgantopoulos}, {Rovilos},
  {Xilouris}, {Comastri}, \& {Akylas}}]{georgantopoulos11}
{Georgantopoulos}, I., {Rovilos}, E., {Xilouris}, E.~M., {Comastri}, A., \&
  {Akylas}, A. 2011, \aap, 526, A86+

\bibitem[{{Giacconi} {et~al.}(2002){Giacconi}, {Zirm}, {Wang}, {Rosati},
  {Nonino}, {Tozzi}, {Gilli}, {Mainieri}, {Hasinger}, {Kewley}, {Bergeron},
  {Borgani}, {Gilmozzi}, {Grogin}, {Koekemoer}, {Schreier}, {Zheng}, \&
  {Norman}}]{giacconi02}
{Giacconi}, R., {Zirm}, A., {Wang}, J., {et~al.} 2002, \apjs, 139, 369

\bibitem[{{Gilli} {et~al.}(2007){Gilli}, {Comastri}, \& {Hasinger}}]{gilli07}
{Gilli}, R., {Comastri}, A., \& {Hasinger}, G. 2007, \aap, 463, 79

\bibitem[{{Gilli} {et~al.}(2009){Gilli}, {Zamorani}, {Miyaji}, {Silverman},
  {Brusa}, {Mainieri}, {Cappelluti}, {Daddi}, {Porciani}, {Pozzetti}, {Civano},
  {Comastri}, {Finoguenov}, {Fiore}, {Salvato}, {Vignali}, {Hasinger}, {Lilly},
  {Impey}, {Trump}, {Capak}, {McCracken}, {Scoville}, {Taniguchi}, {Carollo},
  {Contini}, {Kneib}, {Le Fevre}, {Renzini}, {Scodeggio}, {Bardelli},
  {Bolzonella}, {Bongiorno}, {Caputi}, {Cimatti}, {Coppa}, {Cucciati}, {de La
  Torre}, {de Ravel}, {Franzetti}, {Garilli}, {Iovino}, {Kampczyk}, {Knobel},
  {Kova{\v c}}, {Lamareille}, {Le Borgne}, {Le Brun}, {Maier}, {Mignoli},
  {Pell{\`o}}, {Peng}, {Perez Montero}, {Ricciardelli}, {Tanaka}, {Tasca},
  {Tresse}, {Vergani}, {Zucca}, {Abbas}, {Bottini}, {Cappi}, {Cassata},
  {Fumana}, {Guzzo}, {Leauthaud}, {Maccagni}, {Marinoni}, {Memeo}, {Meneux},
  {Oesch}, {Scaramella}, \& {Walcher}}]{gilli09}
{Gilli}, R., {Zamorani}, G., {Miyaji}, T., {et~al.} 2009, \aap, 494, 33

\bibitem[{{Goulding} {et~al.}(2011){Goulding}, {Alexander}, {Mullaney},
  {Gelbord}, {Hickox}, {Ward}, \& {Watson}}]{goulding11}
{Goulding}, A.~D., {Alexander}, D.~M., {Mullaney}, J.~R., {et~al.} 2011,
  \mnras, 411, 1231

\bibitem[{{Hasinger} {et~al.}(1993){Hasinger}, {Burg}, {Giacconi}, {Hartner},
  {Schmidt}, {Trumper}, \& {Zamorani}}]{hasinger93}
{Hasinger}, G., {Burg}, R., {Giacconi}, R., {et~al.} 1993, \aap, 275, 1

\bibitem[{{Hasinger} {et~al.}(2007){Hasinger}, {Cappelluti}, {Brunner},
  {Brusa}, \& {et al.}}]{hasinger07}
{Hasinger}, G., {Cappelluti}, N., {Brunner}, H., {Brusa}, M., \& {et al.} 2007,
  \apjs, 172, 29

\bibitem[{{Hasinger} {et~al.}(2005){Hasinger}, {Miyaji}, \&
  {Schmidt}}]{hasinger05}
{Hasinger}, G., {Miyaji}, T., \& {Schmidt}, M. 2005, \aap, 441, 417

\bibitem[{{Hickox} \& {Markevitch}(2006)}]{hickox06}
{Hickox}, R.~C., \& {Markevitch}, M. 2006, \apj, 645, 95

\bibitem[{{Hickox} {et~al.}(2009){Hickox}, {Jones}, {Forman}, {Murray},
  {Kochanek}, {Eisenstein}, {Jannuzi}, {Dey}, {Brown}, {Stern}, {Eisenhardt},
  {Gorjian}, {Brodwin}, {Narayan}, {Cool}, {Kenter}, {Caldwell}, \&
  {Anderson}}]{hickox09}
{Hickox}, R.~C., {Jones}, C., {Forman}, W.~R., {et~al.} 2009, \apj, 696, 891

\bibitem[{{Hilton} {et~al.}(2009){Hilton}, {Stanford}, {Stott}, {Collins},
  {Hoyle}, {Davidson}, {Hosmer}, {Kay}, {Liddle}, {Lloyd-Davies}, {Mann},
  {Mehrtens}, {Miller}, {Nichol}, {Romer}, {Sabirli}, {Sahl{\'e}n}, {Viana},
  {West}, {Barbary}, {Dawson}, {Meyers}, {Perlmutter}, {Rubin}, \&
  {Suzuki}}]{hilton09}
{Hilton}, M., {Stanford}, S.~A., {Stott}, J.~P., {et~al.} 2009, \apj, 697, 436

\bibitem[{{Hopkins} {et~al.}(2006){Hopkins}, {Hernquist}, {Cox}, {Di Matteo},
  {Robertson}, \& {Springel}}]{hopkins06}
{Hopkins}, P.~F., {Hernquist}, L., {Cox}, T.~J., {et~al.} 2006, \apjs, 163, 1

\bibitem[{{Hopkins} {et~al.}(2008){Hopkins}, {Hernquist}, {Cox}, \& {Kere{\v
  s}}}]{Hopkins08}
{Hopkins}, P.~F., {Hernquist}, L., {Cox}, T.~J., \& {Kere{\v s}}, D. 2008,
  \apjs, 175, 356

\bibitem[{{Hopkins} {et~al.}(2007){Hopkins}, {Richards}, \&
  {Hernquist}}]{hopkins07}
{Hopkins}, P.~F., {Richards}, G.~T., \& {Hernquist}, L. 2007, \apj, 654, 731

\bibitem[{{Ilbert} {et~al.}(2006){Ilbert}, {Arnouts}, {McCracken},
  {Bolzonella}, {Bertin}, {Le F{\`e}vre}, {Mellier}, {Zamorani}, {Pell{\`o}},
  {Iovino}, {Tresse}, {Le Brun}, {Bottini}, {Garilli}, {Maccagni}, {Picat},
  {Scaramella}, {Scodeggio}, {Vettolani}, {Zanichelli}, {Adami}, {Bardelli},
  {Cappi}, {Charlot}, {Ciliegi}, {Contini}, {Cucciati}, {Foucaud}, {Franzetti},
  {Gavignaud}, {Guzzo}, {Marano}, {Marinoni}, {Mazure}, {Meneux}, {Merighi},
  {Paltani}, {Pollo}, {Pozzetti}, {Radovich}, {Zucca}, {Bondi}, {Bongiorno},
  {Busarello}, {de La Torre}, {Gregorini}, {Lamareille}, {Mathez}, {Merluzzi},
  {Ripepi}, {Rizzo}, \& {Vergani}}]{ilbert06}
{Ilbert}, O., {Arnouts}, S., {McCracken}, H.~J., {et~al.} 2006, \aap, 457, 841

\bibitem[{{Kenter} {et~al.}(2005){Kenter}, {Murray}, {Forman}, {Jones},
  {Green}, {Kochanek}, {Vikhlinin}, {Fabricant}, {Fazio}, {Brand}, {Brown},
  {Dey}, {Jannuzi}, {Najita}, {McNamara}, {Shields}, \& {Rieke}}]{kenter05}
{Kenter}, A., {Murray}, S.~S., {Forman}, W.~R., {et~al.} 2005, \apjs, 161, 9

\bibitem[{{La Franca} {et~al.}(2005){La Franca}, {Fiore}, {Comastri}, {Perola},
  {Sacchi}, {Brusa}, {Cocchia}, {Feruglio}, {Matt}, {Vignali}, {Carangelo},
  {Ciliegi}, {Lamastra}, {Maiolino}, {Mignoli}, {Molendi}, \&
  {Puccetti}}]{lafranca05}
{La Franca}, F., {Fiore}, F., {Comastri}, A., {et~al.} 2005, \apj, 635, 864

\bibitem[{{Laird} {et~al.}(2009){Laird}, {Nandra}, {Georgakakis}, {Aird},
  {Barmby}, {Conselice}, {Coil}, {Davis}, {Faber}, {Fazio}, {Guhathakurta},
  {Koo}, {Sarajedini}, \& {Willmer}}]{laird09}
{Laird}, E.~S., {Nandra}, K., {Georgakakis}, A., {et~al.} 2009, \apjs, 180, 102

\bibitem[{{Lehmann} {et~al.}(2001){Lehmann}, {Hasinger}, {Schmidt}, {Giacconi},
  {Tr{\"u}mper}, {Zamorani}, {Gunn}, {Pozzetti}, {Schneider}, {Stanke},
  {Szokoly}, {Thompson}, \& {Wilson}}]{lehmann01}
{Lehmann}, I., {Hasinger}, G., {Schmidt}, M., {et~al.} 2001, \aap, 371, 833

\bibitem[{{Lehmer} {et~al.}(2005){Lehmer}, {Brandt}, {Alexander}, {Bauer},
  {Schneider}, \& {et al.}}]{lehmer05}
{Lehmer}, B.~D., {Brandt}, W.~N., {Alexander}, D.~M., {et~al.} 2005, \apjs,
  161, 21

\bibitem[{{Lidman} {et~al.}(2008){Lidman}, {Rosati}, {Tanaka}, {Strazzullo},
  {Demarco}, {Mullis}, {Ageorges}, {Kissler-Patig}, {Petr-Gotzens}, \&
  {Selman}}]{lidman08}
{Lidman}, C., {Rosati}, P., {Tanaka}, M., {et~al.} 2008, \aap, 489, 981

\bibitem[{{Lira} {et~al.}(2002){Lira}, {Ward}, {Zezas}, {Alonso-Herrero}, \&
  {Ueno}}]{lira02}
{Lira}, P., {Ward}, M., {Zezas}, A., {Alonso-Herrero}, A., \& {Ueno}, S. 2002,
  \mnras, 330, 259

\bibitem[{{Luo} {et~al.}(2008){Luo}, {Bauer}, {Brandt}, {Alexander}, {Lehmer},
  {Schneider}, {Brusa}, {Comastri}, {Fabian}, \& {et~al.}}]{Luo08}
{Luo}, B., {Bauer}, F.~E., {Brandt}, W.~N., {et~al.} 2008, \apjs, 179, 19

\bibitem[{{Maccacaro} {et~al.}(1988){Maccacaro}, {Gioia}, {Wolter}, {Zamorani},
  \& {Stocke}}]{maccacaro88}
{Maccacaro}, T., {Gioia}, I.~M., {Wolter}, A., {Zamorani}, G., \& {Stocke},
  J.~T. 1988, \apj, 326, 680

\bibitem[{{Madgwick} {et~al.}(2003){Madgwick}, {Coil}, {Conselice}, {Cooper},
  {Davis}, {Ellis}, {Faber}, {Finkbeiner}, {Gerke}, {Guhathakurta}, {Kaiser},
  {Koo}, {Newman}, {Phillips}, {Steidel}, {Weiner}, {Willmer}, {Yan}, \& {Deep2
  Survey Team}}]{madgwick03}
{Madgwick}, D.~S., {Coil}, A.~L., {Conselice}, C.~J., {et~al.} 2003, \apj, 599,
  997

\bibitem[{{Maiolino} {et~al.}(1998){Maiolino}, {Salvati}, {Bassani}, {Dadina},
  {della Ceca}, {Matt}, {Risaliti}, \& {Zamorani}}]{maiolino98}
{Maiolino}, R., {Salvati}, M., {Bassani}, L., {et~al.} 1998, \aap, 338, 781

\bibitem[{{Matt} {et~al.}(1996){Matt}, {Brandt}, \& {Fabian}}]{matt96}
{Matt}, G., {Brandt}, W.~N., \& {Fabian}, A.~C. 1996, \mnras, 280, 823

\bibitem[{{McHardy} {et~al.}(2003){McHardy}, {Gunn}, {Newsam}, {Mason}, {Page},
  {Takata}, {Sekiguchi}, {Sasseen}, {Cordova}, {Jones}, \&
  {Loaring}}]{mchardy03}
{McHardy}, I.~M., {Gunn}, K.~F., {Newsam}, A.~M., {et~al.} 2003, \mnras, 342,
  802

\bibitem[{{Mehrtens} {et~al.}(2012){Mehrtens}, {Romer}, {Hilton},
  {Lloyd-Davies}, {Miller}, {Stanford}, {Hosmer}, {Hoyle}, {Collins}, {Liddle},
  {Viana}, {Nichol}, {Stott}, {Dubois}, {Kay}, {Sahl{\'e}n}, {Young}, {Short},
  {Christodoulou}, {Watson}, {Davidson}, {Harrison}, {Baruah}, {Smith},
  {Burke}, {Mayers}, {Deadman}, {Rooney}, {Edmondson}, {West}, {Campbell},
  {Edge}, {Mann}, {Sabirli}, {Wake}, {Benoist}, {da Costa}, {Maia}, \&
  {Ogando}}]{mehrtens12}
{Mehrtens}, N., {Romer}, A.~K., {Hilton}, M., {et~al.} 2012, \mnras, 2912

\bibitem[{{Mel{\'e}ndez} {et~al.}(2008){Mel{\'e}ndez}, {Kraemer}, {Schmitt},
  {Crenshaw}, {Deo}, {Mushotzky}, \& {Bruhweiler}}]{melendez08a}
{Mel{\'e}ndez}, M., {Kraemer}, S.~B., {Schmitt}, H.~R., {et~al.} 2008, \apj,
  689, 95

\bibitem[{{Moran} {et~al.}(1999){Moran}, {Lehnert}, \& {Helfand}}]{moran99}
{Moran}, E.~C., {Lehnert}, M.~D., \& {Helfand}, D.~J. 1999, \apj, 526, 649

\bibitem[{{Murray} {et~al.}(2005){Murray}, {Kenter}, {Forman}, {Jones},
  {Green}, {Kochanek}, {Vikhlinin}, {Fabricant}, {Fazio}, {Brand}, {Brown},
  {Dey}, {Jannuzi}, {Najita}, {McNamara}, {Shields}, \& {Rieke}}]{murray05}
{Murray}, S.~S., {Kenter}, A., {Forman}, W.~R., {et~al.} 2005, \apjs, 161, 1

\bibitem[{{Nandra} {et~al.}(2005){Nandra}, {Laird}, {Adelberger}, {Gardner},
  {Mushotzky}, {Rhodes}, {Steidel}, {Teplitz}, \& {Arnaud}}]{nandra05}
{Nandra}, K., {Laird}, E.~S., {Adelberger}, K., {et~al.} 2005, \mnras, 356, 568

\bibitem[{{Nastasi} {et~al.}(2011){Nastasi}, {Fassbender}, {B{\"o}hringer},
  {{\v S}uhada}, {Rosati}, {Pierini}, {Verdugo}, {Santos}, {Schwope}, {de
  Hoon}, {Kohnert}, {Lamer}, {M{\"u}hlegger}, \& {Quintana}}]{nastasi11}
{Nastasi}, A., {Fassbender}, R., {B{\"o}hringer}, H., {et~al.} 2011, \aap, 532,
  L6+

\bibitem[{{Newman} {et~al.}(2012){Newman}, {Cooper}, {Davis}, {Faber}, {Coil},
  {Guhathakurta}, {Koo}, {Phillips}, {Conroy}, {Dutton}, {Finkbeiner}, {Gerke},
  {Rosario}, {Weiner}, {Willmer}, {Yan}, {Harker}, {Kassin}, {Konidaris},
  {Lai}, {Madgwick}, {Noeske}, {Wirth}, {Connolly}, {Kaiser}, {Kirby},
  {Lemaux}, {Lin}, {Lotz}, {Luppino}, {Marinoni}, {Matthews}, {Metevier}, \&
  {Schiavon}}]{newman12}
{Newman}, J.~A., {Cooper}, M.~C., {Davis}, M., {et~al.} 2012, ArXiv 1203.3192

\bibitem[{{Papovich} {et~al.}(2010){Papovich}, {Momcheva}, {Willmer},
  {Finkelstein}, {Finkelstein}, {Tran}, {Brodwin}, {Dunlop}, {Farrah}, {Khan},
  {Lotz}, {McCarthy}, {McLure}, {Rieke}, {Rudnick}, {Sivanandam}, {Pacaud}, \&
  {Pierre}}]{papovich10}
{Papovich}, C., {Momcheva}, I., {Willmer}, C.~N.~A., {et~al.} 2010, \apj, 716,
  1503

\bibitem[{{Park} {et~al.}(2006){Park}, {Kashyap}, {Siemiginowska}, {van Dyk},
  {Zezas}, {Heinke}, \& {Wargelin}}]{park06}
{Park}, T., {Kashyap}, V.~L., {Siemiginowska}, A., {et~al.} 2006, \apj, 652,
  610

\bibitem[{{Puccetti} {et~al.}(2009){Puccetti}, {Vignali}, {Cappelluti},
  {Fiore}, {Zamorani}, {Aldcroft}, {Elvis}, {Gilli}, {Miyaji}, {Brunner},
  {Brusa}, {Civano}, {Comastri}, {Damiani}, {Fruscione}, {Finoguenov},
  {Koekemoer}, \& {Mainieri}}]{puccetti09}
{Puccetti}, S., {Vignali}, C., {Cappelluti}, N., {et~al.} 2009, \apjs, 185, 586

\bibitem[{{Richards} {et~al.}(2006){Richards}, {Strauss}, {Fan}, {Hall},
  {Jester}, {Schneider}, {Vanden Berk}, {Stoughton}, {Anderson}, {Brunner},
  {Gray}, {Gunn}, {Ivezi{\'c}}, {Kirkland}, {Knapp}, {Loveday}, {Meiksin},
  {Pope}, {Szalay}, {Thakar}, {Yanny}, {York}, {Barentine}, {Brewington},
  {Brinkmann}, {Fukugita}, {Harvanek}, {Kent}, {Kleinman}, {Krzesi{\'n}ski},
  {Long}, {Lupton}, {Nash}, {Neilsen}, {Nitta}, {Schlegel}, \&
  {Snedden}}]{Richards06}
{Richards}, G.~T., {Strauss}, M.~A., {Fan}, X., {et~al.} 2006, \aj, 131, 2766

\bibitem[{{Risaliti} {et~al.}(1999){Risaliti}, {Maiolino}, \&
  {Salvati}}]{risaliti99}
{Risaliti}, G., {Maiolino}, R., \& {Salvati}, M. 1999, \apj, 522, 157

\bibitem[{{Schmidt} {et~al.}(1998){Schmidt}, {Hasinger}, {Gunn}, {Schneider},
  {Burg}, {Giacconi}, {Lehmann}, {MacKenty}, {Trumper}, \&
  {Zamorani}}]{schmidt98}
{Schmidt}, M., {Hasinger}, G., {Gunn}, J., {et~al.} 1998, \aap, 329, 495

\bibitem[{{Serjeant} {et~al.}(2010){Serjeant}, {Bertoldi}, {Blain}, {Clements},
  \& {et~al.}}]{Serjeant10}
{Serjeant}, S., {Bertoldi}, F., {Blain}, A.~W., {Clements}, D.~L., \& {et~al.}
  2010, \aap, 518, L7+

\bibitem[{{Silverman} {et~al.}(2009){Silverman}, {Lamareille}, {Maier},
  {Lilly}, {Mainieri}, {Brusa}, {Cappelluti}, {Hasinger}, {Zamorani},
  {Scodeggio}, {Bolzonella}, {Contini}, {Carollo}, {Jahnke}, {Kneib}, {Le
  F{\`e}vre}, {Merloni}, {Bardelli}, {Bongiorno}, {Brunner}, {Caputi},
  {Civano}, {Comastri}, {Coppa}, {Cucciati}, {de la Torre}, {de Ravel},
  {Elvis}, {Finoguenov}, {Fiore}, {Franzetti}, {Garilli}, {Gilli}, {Iovino},
  {Kampczyk}, {Knobel}, {Kova{\v c}}, {Le Borgne}, {Le Brun}, {Mignoli},
  {Pello}, {Peng}, {Perez Montero}, {Ricciardelli}, {Tanaka}, {Tasca},
  {Tresse}, {Vergani}, {Vignali}, {Zucca}, {Bottini}, {Cappi}, {Cassata},
  {Fumana}, {Griffiths}, {Kartaltepe}, {Koekemoer}, {Marinoni}, {McCracken},
  {Memeo}, {Meneux}, {Oesch}, {Porciani}, \& {Salvato}}]{silverman09}
{Silverman}, J.~D., {Lamareille}, F., {Maier}, C., {et~al.} 2009, \apj, 696,
  396

\bibitem[{{Smol{\v c}i{\'c}} {et~al.}(2009){Smol{\v c}i{\'c}}, {Zamorani},
  {Schinnerer}, {Bardelli}, {Bondi}, {B{\^\i}rzan}, {Carilli}, {Ciliegi},
  {Elvis}, {Impey}, {Koekemoer}, {Merloni}, {Paglione}, {Salvato}, {Scodeggio},
  {Scoville}, \& {Trump}}]{smolic09}
{Smol{\v c}i{\'c}}, V., {Zamorani}, G., {Schinnerer}, E., {et~al.} 2009, \apj,
  696, 24

\bibitem[{{Stark} {et~al.}(1992){Stark}, {Gammie}, {Wilson}, {Bally}, {Linke},
  {Heiles}, \& {Hurwitz}}]{stark92}
{Stark}, A.~A., {Gammie}, C.~F., {Wilson}, R.~W., {et~al.} 1992, \apjs, 79, 77

\bibitem[{{Stocke} {et~al.}(1991){Stocke}, {Morris}, {Gioia}, {Maccacaro},
  {Schild}, {Wolter}, {Fleming}, \& {Henry}}]{stocke91}
{Stocke}, J.~T., {Morris}, S.~L., {Gioia}, I.~M., {et~al.} 1991, \apjs, 76, 813

\bibitem[{{Tozzi} {et~al.}(2006){Tozzi}, {Gilli}, {Mainieri}, {Norman},
  {Risaliti}, {Rosati}, {Bergeron}, {Borgani}, {Giacconi}, {Hasinger},
  {Nonino}, {Streblyanska}, {Szokoly}, {Wang}, \& {Zheng}}]{tozzi06}
{Tozzi}, P., {Gilli}, R., {Mainieri}, V., {et~al.} 2006, \aap, 451, 457

\bibitem[{{Ueda} {et~al.}(2003){Ueda}, {Akiyama}, {Ohta}, \& {Miyaji}}]{ueda03}
{Ueda}, Y., {Akiyama}, M., {Ohta}, K., \& {Miyaji}, T. 2003, \apj, 598, 886

\bibitem[{{Vikhlinin} {et~al.}(1995){Vikhlinin}, {Forman}, {Jones}, \&
  {Murray}}]{vikhlinin95}
{Vikhlinin}, A., {Forman}, W., {Jones}, C., \& {Murray}, S. 1995, \apj, 451,
  564

\bibitem[{{Vikhlinin} {et~al.}(1998){Vikhlinin}, {McNamara}, {Forman}, {Jones},
  {Quintana}, \& {Hornstrup}}]{vikhlinin98}
{Vikhlinin}, A., {McNamara}, B.~R., {Forman}, W., {et~al.} 1998, \apj, 502, 558

\bibitem[{{Willmer} {et~al.}(2006){Willmer}, {Faber}, {Koo}, {Weiner},
  {Newman}, {Coil}, {Connolly}, {Conroy}, {Cooper}, {Davis}, {Finkbeiner},
  {Gerke}, {Guhathakurta}, {Harker}, {Kaiser}, {Kassin}, {Konidaris}, {Lin},
  {Luppino}, {Madgwick}, {Noeske}, {Phillips}, \& {Yan}}]{willmer06}
{Willmer}, C.~N.~A., {Faber}, S.~M., {Koo}, D.~C., {et~al.} 2006, \apj, 647,
  853

\bibitem[{{Worsley} {et~al.}(2005){Worsley}, {Fabian}, {Bauer}, {Alexander},
  {Hasinger}, {Mateos}, {Brunner}, {Brandt}, \& {Schneider}}]{worsley05}
{Worsley}, M.~A., {Fabian}, A.~C., {Bauer}, F.~E., {et~al.} 2005, \mnras, 357,
  1281

\bibitem[{{Xue} {et~al.}(2011){Xue}, {Luo}, {Brandt}, {Bauer}, {Lehmer},
  {Broos}, {Schneider}, {Alexander}, {Brusa}, {Comastri}, {Fabian}, {Gilli},
  {Hasinger}, {Hornschemeier}, {Koekemoer}, {Liu}, {Mainieri}, {Paolillo},
  {Rafferty}, {Rosati}, {Shemmer}, {Silverman}, {Smail}, {Tozzi}, \&
  {Vignali}}]{xue11}
{Xue}, Y.~Q., {Luo}, B., {Brandt}, W.~N., {et~al.} 2011, \apjs, 195, 10

\bibitem[{{Zheng} {et~al.}(2009){Zheng}, {Bell}, {Somerville}, {Rix}, {Jahnke},
  {Fontanot}, {Rieke}, {Schiminovich}, \& {Meisenheimer}}]{Zheng09}
{Zheng}, X.~Z., {Bell}, E.~F., {Somerville}, R.~S., {et~al.} 2009, \apj, 707,
  1566

\end{thebibliography}

\end{document}